\documentclass[11pt,a4paper]{article}
\pdfoutput=1
\usepackage{jheppub}

\usepackage{multirow, graphicx,amssymb,url,mathrsfs,amsmath}
\usepackage{wrapfig,boxedminipage,setspace,subfigure,epsfig}
\usepackage{amsxtra,amstext,latexsym,dsfont,amsfonts}
\usepackage{color,eucal}
\usepackage[dvipsnames]{xcolor}
\usepackage{float}
\usepackage{slashed,comment}
\usepackage{kotex}
\usepackage{colortbl}
\usepackage{multirow}
\usepackage{tikz}
\usepackage{amsmath}
\usetikzlibrary{tikzmark}
\usetikzlibrary{arrows}

\usepackage{ulem}
 
\usepackage{tabularx, array}
\newcolumntype{L}[1]{>{\raggedright\arraybackslash}p{#1}}
\newcolumntype{C}[1]{>{\centering\arraybackslash}p{#1}}
\newcolumntype{R}[1]{>{\raggedleft\arraybackslash}p{#1}}

\usepackage{pdflscape}

\newcommand{\dd}{\mathrm{d}}

\newcommand{\emu}{{\left(\frac{\partial{\epsilon}}{\partial{\mu}}\right)_{T, B}}}
\newcommand{\eT}{{\left(\frac{\partial{\epsilon}}{\partial{T}}\right)_{\mu, B}}}
\newcommand{\eB}{{\left(\frac{\partial{\epsilon}}{\partial{B}}\right)_{T, \mu}}}

\newcommand{\Pmu}{{\left(\frac{\partial{p}}{\partial{\mu}}\right)_{T, B}}}
\newcommand{\PT}{{\left(\frac{\partial{p}}{\partial{T}}\right)_{\mu, B} }}
\newcommand{\PB}{{\left(\frac{\partial{p}}{\partial{B}}\right)_{T, \mu} }}

\newcommand{\rhomu}{{\left(\frac{\partial{\rho}}{\partial{\mu}}\right)_{T, B}}}
\newcommand{\rhoT}{{\left(\frac{\partial{\rho}}{\partial{T}}\right)_{\mu, B}}}
\newcommand{\rhoB}{{\left(\frac{\partial{\rho}}{\partial{B}}\right)_{T, \mu}}}

\title{Holography and magnetohydrodynamics with dynamical gauge fields}

\author[a,b,c,d,\dagger]{Yongjun Ahn,}
\author[c,d]{Matteo Baggioli,}
\author[a,b,\dagger]{Kyoung-Bum Huh,}
\author[e,f,g,h,\dagger]{Hyun-Sik Jeong,}
\author[a,b]{Keun-Young Kim,}
\author[e,f]{and Ya-Wen Sun}

\emailAdd{yongjunahn619@gmail.com}
\emailAdd{b.matteo@sjtu.edu.cn}
\emailAdd{hkabell1689@gm.gist.ac.kr}
\emailAdd{sicobysico@gmail.com}
\emailAdd{fortoe@gist.ac.kr}
\emailAdd{yawen.sun@ucas.ac.cn}

\affiliation[a]{Department of Physics and Photon Science, Gwangju Institute of Science and Technology, \\
123 Cheomdan-gwagiro, Gwangju 61005, Korea}
\affiliation[b]{Research Center for Photon Science Technology, Gwangju Institute of Science and Technology, \\
123 Cheomdan-gwagiro, Gwangju 61005, Korea}
\affiliation[c]{Wilczek Quantum Center, School of Physics and Astronomy, Shanghai Jiao Tong University, Shanghai 200240, China}
\affiliation[d]{Shanghai Research Center for Quantum Sciences, Shanghai 201315, China}
\affiliation[e]{School of physics $\&$ CAS Center for Excellence in Topological Quantum Computation, University of Chinese Academy of Sciences, Zhongguancun east road 80, Beijing 100049, China}
\affiliation[f]{Kavli Institute for Theoretical Sciences, University of Chinese Academy of Sciences, \\ Zhongguancun east road 80, Beijing 100049, China}
\affiliation[g]{Instituto de Física Teórica UAM/CSIC, Calle Nicolás Cabrera 13-15, 28049 Madrid, Spain}
\affiliation[h]{Departamento de Física Teórica, Universidad Autónoma de Madrid, Campus de Cantoblanco, 28049 Madrid, Spain}

\affiliation[\dagger]{These authors contributed equally to this paper and should be considered as co-first authors}

\abstract{Within the framework of holography, the Einstein-Maxwell action with Dirichlet boundary conditions corresponds to a dual conformal field theory in presence of an external gauge field. Nevertheless, in many real-world applications, e.g., magnetohydrodynamics, plasma physics, superconductors, etc. dynamical gauge fields and Coulomb interactions are fundamental. In this work, we consider bottom-up holographic models at finite magnetic field and (free) charge density in presence of dynamical boundary gauge fields which are introduced using mixed boundary conditions. We numerically study the spectrum of the lowest quasi-normal modes and successfully compare the obtained results to magnetohydrodynamics theory in $2+1$ dimensions. Surprisingly, as far as the electromagnetic coupling is small enough, we find perfect agreement even in the large magnetic field limit. Our results prove that a holographic description of magnetohydrodynamics does not necessarily need higher-form bulk fields but can be consistently derived using mixed boundary conditions for standard gauge fields.}

\begin{document}
\maketitle

\section{Introduction}

The essence of the holographic duality lies in the correspondence between a gravitational theory in $d+1$ dimensions and a ``dual" field theory in $d$ dimensions which is formally defined via the so-called GPKW (Gubser, Polyakov, Klebanov, Witten) master rule \cite{Witten:1998qj,Gubser:1998bc}. Importantly, the bulk gravitational action does not uniquely define the boundary dual field theory but, for that scope, needs to be supplemented with boundary conditions for all the bulk fields. 

Let us consider the example of a bulk gauge field $A_\mu$. Formally, in the $(d+1)$-dimensional bulk, one usually writes a gravitational action of the form:
\begin{equation}\label{eq1}
    S=\int d^{d+1}x \sqrt{-g}\,\left[R-2\Lambda+\mathcal{L}\left(F_{\mu\nu}\right)+\dots\right] \,,
\end{equation}
where $R$ is the Ricci scalar, $\Lambda$ a negative cosmological constant and $\mathcal{L}\left(F_{\mu\nu}\right)$ a generic Lagrangian for the bulk field $A_\mu$ which is written in terms of its field strength $F:=dA$, as imposed by gauge invariance in the bulk. The simplest example possible is a Maxwell kinetic term, $\mathcal{L}\left(F_{\mu\nu}\right)=-F^2/4$ with $F^2:=F_{\mu\nu}F^{\mu\nu}$, which gives rise to the so-called Einstein-Maxwell action. The ellipsis in Eq.\eqref{eq1} indicates the presence of other possible bulk fields which are irrelevant for the present discussion and therefore not shown. Also, for simplicity, let us neglect possible couplings between the field strength $F_{\mu\nu}$ and other bulk fields, e.g., dilaton couplings. The asymptotic solution for a massless gauge field in an asymptotically Anti-de Sitter bulk geometry is then generically given by:
\begin{equation}\label{exp}
    A_\mu\left(r,t,\Vec{x}\right)\,\underset{r\rightarrow \infty}{\sim}\,A^{(0)}_\mu\left(t,\Vec{x}\right)\,+\,A^{(1)}_\mu\left(t,\Vec{x}\right)\,r^{2-d} \,,
\end{equation}
where $r$ is the radial holographic direction, $r=\infty$ the location of the AdS boundary. At this point, $A^{(0),(1)}_\mu\left(t,\Vec{x}\right)$ are two undetermined functions, usually denoted as the \textit{leading/subleading} terms, which must be fixed by the choice of boundary conditions at $r=\infty$. The canonical (but not unique) procedure, which goes under the name of \textit{standard quantization}, is to fix the value of $A^{(0)}_\mu$ at the boundary and dynamically determine the value of $A^{(1)}_\mu$. From a dual field theory perspective, this corresponds to a CFT deformation of the type:
\begin{equation}
    \mathcal{L}_{CFT}\,\,\longrightarrow\,\,\mathcal{L}_{CFT}\,+\,\int d^dx\,A^{(0)}_\mu J^\mu \,,
\end{equation}
where $J^\mu$ is a U(1) current operator in the dual CFT with conformal dimension $\Delta=d-1$ whose vacuum expectation value is determined by $A^{(1)}_\mu$ (see \cite{Hartnoll:2016apf} for details). Gauge invariance in the bulk, $A_\mu \rightarrow A_\mu+\partial_\mu \xi$, implies that the dual U(1) current is conserved, $\partial_\mu J^\mu=0$. Additionally, the current operator two-point function $\langle J J \rangle$ (whose spatial component is related to the electric conductivity $\sigma$) can be easily computed in linear response theory by looking at the ratio $A^{(1)}/A^{(0)}$. Within this picture, it is often said that a local symmetry in the bulk (in this case, a local U(1) gauge symmetry) corresponds to a global symmetry in the dual boundary theory. In other words, following the prescription just described and using standard quantization, the dual field theory does not possess a local U(1) symmetry and, as a consequence, no electromagnetism nor long-range Coulomb interactions are present therein.

Having in mind possible applications of the holographic duality to realistic systems, this outcome might appear rather disappointing and limiting. There are indeed several circumstances in which the role of dynamical gauge fields and electromagnetic interactions are fundamental for the correct physical description and cannot be neglected. Plasma physics and condensed matter systems are clear examples of this sort. Even if too often ignored, within the holographic framework, the solution to this problem is more than ten years old. Starting from the seminal works by Witten \cite{Witten:1998qj,Witten:2003ya} (see also \cite{Klebanov:1999tb,Leigh:2003ez,Yee:2004ju,Breitenlohner:1982jf}), the existence and meaning of different boundary conditions for bulk vector fields were analyzed in detail by Marolf and Ross in 2006 \cite{Marolf:2006nd}\footnote{As noted in \cite{Cottrell:2017gkb}, the b.c.s of \cite{Marolf:2006nd} are actually different from the original ones in \cite{Witten:2003ya}.} and few years later applied for the first time in the context of holographic superconductors \cite{Montull:2009fe,Maeda:2010br,Domenech:2010nf}, to be contrasted with the more famous holographic superfluid model of Hartnoll, Herzog and Horowitz \cite{Hartnoll:2008kx}.\footnote{This model is often improperly labelled as ``holographic superconductor''. This is imprecise since the $U(1)$ symmetry at the boundary is not dynamical and the gauge field is external. In this sense, the dual of the HHH model \cite{Hartnoll:2008kx} is a superfuid state and not a superconducting one.}

Without going into details, the main idea of this program is to promote the b.c.s. for the gauge field at the AdS boundary to the most general form:
\begin{equation}
    \alpha\,A^{(0)}_\mu\left(t,\Vec{x}\right)\,+\,\beta\,A^{(1)}_\mu\left(t,\Vec{x}\right)\,=\,\text{fixed} \,,
\end{equation}
which takes the name of \textit{mixed boundary conditions}.\footnote{See also \cite{Compere:2008us,Ecker:2021cvz} for the possibility and meaning of implementing the same procedure for the boundary metric $g_{\mu\nu}$.} The standard quantization is recovered by setting $\beta=0$, while the \textit{alternative quantization} by setting $\alpha=0$. It was early realized (mostly for the simpler case of a bulk scalar field) \cite{Witten:2003ya,Leigh:2003ez,Witten:2001ua,Berkooz:2002ug,Hartman:2006dy} that these most general b.c.s. are connected to double trace deformations in the boundary field theory and specific SL(2,$\mathbb{R}$) transformations in the moduli space of the dual CFTs.\\

As already mentioned, one of the first concrete realizations of these modified boundary conditions in the context of applied holography has appeared in the construction of a ``real" holographic superconductor model \cite{Domenech:2010nf,Maeda:2010br,Silva:2011zzc} which has turned out to be fundamental for the study of certain specific properties of superconductors such as vortices \cite{Montull:2009fe,Albash:2009iq,Rozali:2012ry,Gao:2012yw,Salvio:2012at,Salvio:2013ja,Dias:2013bwa,Montull:2011im,delCampo:2021rak,Zeng:2019yhi} or the Meissner effect \cite{Natsuume:2022kic}. Similar types of mixed b.c.s. have been considered in the study of anyons physics in \cite{Jokela:2013hta,Brattan:2013wya,Brattan:2014moa}. A more recent explosion of efforts to incorporate, understand and utilise the effects of dynamical electromagnetism in the boundary field theory is connected to the study of magnetohydrodynamics and plasmons physics. From one side, the latter has been initiated by Gran, Torns\"{o} and Zingg \cite{Gran:2017jht} using mixed b.c.s. for the bulk gauge field and has been investigated in several directions \cite{Gran:2018iie,Gran:2018vdn,Gran:2018jnt,Baggioli:2019aqf,Gran:2019djz,Baggioli:2019sio,Baggioli:2021ujk,Romero-Bermudez:2019lzz}. The connection of this framework to the diagrammatic Random Phase Approximation (RPA) \cite{pines2018theory} and double trace deformations in the dual field theory has been explained in \cite{Mauri:2018pzq,Romero-Bermudez:2018etn}. On the contrary, the aspects related to magnetohydrodynamics have been so far dominated by the usage of higher-form symmetry structures as proposed in the original work by Hofman, Grozdanov and Iqbal \cite{Grozdanov:2016tdf} which has been implemented within the holographic framework in \cite{Grozdanov:2017kyl,Poovuttikul:2021fdi,Das:2022auy} and discussed from a hydrodynamic perspective in \cite{Hernandez:2017mch,Grozdanov:2018fic,Armas:2018ibg,Benenowski:2019ule}. In particular, Grozdanov and Poovuttikul \cite{Grozdanov:2017kyl} have demonstrated a complete match between the magnetohydrodynamic expectations \cite{Hernandez:2017mch} and the bulk higher-form picture. Moreover, in this class of theories, a \textit{bona fide} photon has been identified \cite{Hofman:2017vwr}\footnote{See \cite{Gao:2012yw} for an earlier identification of a propagating photon using alternative boundary conditions in AdS$_3$. We will comment again on the results of \cite{Gao:2012yw} at the end of this work proposing a slightly different interpretation.} and interestingly described as a Goldstone mode of the emergent higher-form symmetry \cite{Hofman:2018lfz}.\\

Is the higher-form bulk picture really necessary to have dynamical electromagnetism at the boundary? What is its relation with the mixed boundary conditions discussed so far? Can we recover magnetohydrodynamics in the boundary field theory without using higher-form symmetries? Most of the answers to these questions have been already addressed in a beautiful work by Higginbotham and DeWolfe \cite{DeWolfe:2020uzb} (see Fig. 1 therein for a nice summary of their results). In a nutshell, the Hodge dual operation performed in the bulk, and needed to pass from the standard Maxwell picture to the higher-form description (which in AdS$_5$ appears as a Maxwell action for a 3-form field strength), does not leave the boundary conditions unchanged. On the contrary, it hiddenly modifies the standard Dirichlet b.c.s. into mixed b.c.s. rendering the original global U(1) symmetry in the boundary dynamical. As a consequence, the higher-form formalism introduces non-trivial physics in the boundary field theory because it corresponds to deforming the original Dirichlet boundary conditions for the Maxwell bulk gauge field $A_\mu$. In other words, despite the bulk physics is unchanged because of the harmless Hodge dual, the field theory interpretation of the two scenarios is completely different.\footnote{This is happening also for the simpler case of bulk massless scalars in the context of broken translations. See for example \cite{Grozdanov:2018ewh,Armas:2019sbe}.} Therefore, the naive expectation is that one could be able to obtain the same results, i.e., to obtain dynamical electromagnetism and magnetohydrodynamics in the boundary, by sticking to the maybe less elegant but more direct gauge field picture and deforming its asymptotic b.c.s. without the need of any higher-form structure. This possibility and its outcomes are the subject of this paper.\\

More concretely, we are asking whether a standard Einstein-Maxwell action:
\begin{equation}
    S=\int d^{d+1}x\,\sqrt{g}\,\left[R-2\Lambda-\frac{1}{4}F^2\right] \,,
\end{equation}
implemented with the ``right'' (and indeed not Dirichlet) boundary conditions for the gauge field $A_\mu$ is able to provide the physics of a dual system exhibiting a dynamical U(1) symmetry with EM (electromagnetic) long-range interactions.\\

In the case of standard Dirichlet b.c.s., the dual field theory is a finite temperature CFT with a conserved U(1) current $J^\mu$ in presence of an external, and not dynamical, gauge field $A_\mu$ (and therefore an external magnetic field $B$ as well).  This scenario has been studied in several works~\cite{Hartnoll:2007ih,Jensen:2011xb,Kovtun:2016lfw,Jeong:2022luo,Hartnoll:2007ip} and the complete consistency between the holographic picture and the dual hydrodynamic framework has been recently verified in~\cite{Jeong:2022luo}. This same system has also been studied in presence of explicit and/or spontaneous breaking of translations \cite{Amoretti:2021fch,Amoretti:2020mkp,Baggioli:2020edn,Blake:2015hxa,donos2016dc,Amoretti:2014zha,Amoretti:2015gna} and anomalies \cite{Ammon:2020rvg}. Here, in analogy with the higher-symmetry analysis in \cite{Grozdanov:2017kyl}, we aim at running a parallel program for the case in which the U(1) symmetry, and correspondingly the magnetic field $B$, in the boundary are dynamical. Our holographic results will be compared to the magnetohydrodynamics derived in Ref.\cite{Hernandez:2017mch} reduced, for simplicity, to two spatial dimensions. We will study the system at finite charge density and finite dynamical magnetic field and explore the regime of strong magnetic field.\\

Finally, we will discuss the more speculative possibility of modifying the nature of the dual field theory not by using boundary conditions nor by performing a Hodge duality in the bulk but rather by substituting the original Maxwell term $F_{\mu\nu}F^{\mu\nu}$ in the bulk with a non-canonical higher derivative action of the form $\left(F_{\mu\nu}F^{\mu\nu}\right)^{N/2}$.
\begin{equation}\label{eq2}
    S=\int d^{d+1}x\,\sqrt{g}\,\left[R-2\Lambda-\left(F_{\mu\nu}F^{\mu\nu}\right)^{N/2}\right]\quad \text{with}\quad N>2\,.
\end{equation}
Aware of the issues of ``naturalness" in the effective field theory sense, we will consider this case as a toy model to understand better the implementation of symmetries and boundary conditions in bottom-up holography. The idea for an action as in \eqref{eq2} is borrowed from the holographic axions model \cite{Baggioli:2021xuv} in which this type of higher order kinetic terms \cite{Baggioli:2014roa} has been employed to realize the spontaneous symmetry breaking of translations in the dual field theory \cite{Alberte:2017oqx}. Therein, this procedure turned out to be equivalent to modifying appropriately the boundary conditions for the axion fields responsible for the breaking of translations \cite{Armas:2019sbe}. Despite its odd nature, the bulk action written in terms of non-canonical kinetic terms exactly reproduces the structure and dynamics of viscoelasticity theory \cite{Ammon:2020xyv} (see \cite{Baggioli:2022pyb} for a review on the topic) proving its validity as an effective bulk description. Here, we will perform the same analysis for a bulk gauge field with non-canonical kinetic term as in Eq.\eqref{eq2}. As we will explore in detail, the deformation of the bulk action as in \eqref{eq2} automatically modifies the nature of the coefficient $A^{(0)}_\mu\left(t,\Vec{x}\right)$ in the asymptotic expansion of the gauge field \eqref{exp} from \textit{leading} to \textit{subleading}. This implies that, assuming standard quantization for the theory in \eqref{eq2}, the coefficient $A^{(0)}_\mu\left(t,\Vec{x}\right)$ is not anymore a source for an external field $A_\mu$ but rather the expectation value of the current $J^\mu$. This is exactly what would happen by considering the standard Maxwell action, $N=2$, but with alternative b.c.s. for the bulk gauge field $A_\mu$. Indeed, the two frameworks will give analogous results. \\ \\

\noindent \textbf{Structure of the paper} -- In Section \ref{SECMHDDIS222}, we revisit the the magnetohydrodynamic framework of \cite{Hernandez:2017mch} in $2+1$ dimensions and obtain the dispersion relation of the low-energy modes at finite charge density and dynamical magnetic field; in Section \ref{SECMHDDIS333}, we introduce our holographic setup and the precise boundary conditions used: in Section \ref{newsec}, we present all the main results of our work with modified mixed b.c.s. for the bulk gauge field $A_\mu$; in Section \ref{SECnew1}, we discuss the features of a higher-derivative bulk model and its meaning from the point of view of the dual field theory; finally, in Section \ref{SECMHDDIS444}, we conclude and discuss a few points for future investigation. Appendix \ref{lala} discusses some interesting outcomes regarding the regime of validity of the hydrodynamic framework.

%
\section{Hydrodynamics with dynamical gauge fields}\label{SECMHDDIS222}
In this section, we consider relativistic magnetohydrodynamics in (2+1) dimensions, including the effects of finite charge density and magnetic field. With the term \textit{relativistic magnetohydrodynamics} we refer to the hydrodynamic description of a relativistic charged fluid in presence of long-range EM interactions mediated by a dynamical gauge field (see \cite{Hattori:2022hyo} for a recent review on the topic). This is very different from the situation (which is often loosely labelled in the same way) in which the magnetic and electric fields are external and non dynamical (see \cite{andreanew} for a review). From a practical perspective, the computations presented in this section are identical to those in ~\cite{Hernandez:2017mch} but in the simpler situation of (2+1) dimensions. The main simplification with respect to \cite{Hernandez:2017mch} arises from the fact that a magnetic field in two spatial dimensions cannot be associated to a proper vector field and therefore one cannot define an angle $\theta$ between the magnetic field $B$ and the wave-vector $k$. This avoids several complications related to the anisotropy of the system in (3+1) dimensions. As a downside, the dynamics in $(2+1)$ is less richer than that in $(3+1)$ dimensions. For example, it does not include the so-called Alfv\'en waves nor the separation between fast and slow magnetosonic waves.

\subsection{Setup}\label{SECSETUP}
Let us start by considering the generating functional $Z[g_{\mu\nu},A_\mu]$:
\begin{equation}\label{EXTTH}
  Z[g_{\mu\nu},A_\mu]= \int \mathcal{D}\Phi \exp \left[i S_0\left(\Phi\right)+ i \int d^3 x \sqrt{-g} \left(A_\mu J^\mu\left(\Phi\right)+\frac{1}{2}g_{\mu\nu}T^{\mu\nu}\left(\Phi\right)\right)\right] \,,
\end{equation}
where $\Phi$ denotes a set of dynamical fields, $A_\mu$ an external gauge field coupled to the field theory current $J^\mu\left(\Phi\right)$ and $g_{\mu\nu}$ a fixed external metric coupled to the stress tensor $T^{\mu\nu}\left(\Phi\right)$.
Using Eq.\eqref{EXTTH}, and the standard functional derivative prescription, the n-point functions of the corresponding conserved current operator $J^\mu$ (and not only) can be computed. Within the holographic business, this would correspond to imposing the standard Dirichlet boundary conditions on the bulk gauge field (see more details below).

From the generating functional in Eq.\eqref{EXTTH}, we can define an effective action as
\begin{equation}\label{EXTTH2}
    S[g_{\mu\nu}, A_\mu] := -i \, \ln Z[g_{\mu\nu}, A_\mu] = \int \dd^3 x \, \sqrt{-g} \,\, \mathcal{F}\,,
\end{equation}
where $\mathcal{F}$ denotes the free energy density. The derivative expansion of $\mathcal{F}$,
\begin{equation}\label{EXTTHnew}
    \mathcal{F} = p(T,\mu,B^2) \,+\, \mathcal{O}(\partial) \,,
\end{equation}
gives the thermodynamic pressure $p$ at the leading order in fluctuations, i.e., at equilibrium.
Here, $T$ is the temperature, $\mu$ the chemical potential, and $B$ the magnetic field. {Moreover, we have assumed that $T,\mu,B$ are $\mathcal{O}(1)$ in derivatives while $E$ is order $\mathcal{O}(\partial)$. This is the correct assumption in case of magnetohydrodynamics (MHD) \cite{Hernandez:2017mch}.} Using Eq.\eqref{EXTTH2}, one can further define the stress-energy tensor $T^{\mu\nu}$ and the U(1) conserved current $J^{\mu}$ as
\begin{equation} \label{CONJ2}
\begin{split}
\delta_{g_{\mu\nu}} S \,=\, \frac{1}{2} \int \dd^3 x \,\, \sqrt{-g} \,\,  T^{\mu\nu} \, \delta g_{\mu\nu} \,,\quad  \delta_{A_\mu} S \,=\, \int \dd^3 x \,\, \sqrt{-g} \,\, J^{\mu} \, \delta A_{\mu} \,.
\end{split}
\end{equation}

So far, all the quantities discussed are defined in terms of the external fields $g_{\mu\nu},A_\mu$ which are not dynamical. In order to promote the external gauge field to be dynamical, one considers the following Legendre-transformed action
\begin{align}\label{DYNATH}
\begin{split}
    S_{\text{tot}} \,&=\, S[g_{\mu\nu}, A_\mu] \,+\, \int \dd^3 x\sqrt{-g} \,\, A_\mu \, J^{\mu}_{\text{ext}}\,,\\
    & = S_{\text{m}}[g_{\mu\nu}, A_\mu] \,+\, \int \dd^3 x\sqrt{-g} \,\left[\,-\frac{1}{4\lambda}F^2\, + A_\mu \, J^{\mu}_{\text{ext}}\right]\,.
\end{split}
\end{align}
{In the second line of Eq.\eqref{DYNATH} we separate $S$ into two pieces which correspond to the ``matter contribution" $S_{\text{m}}$, and the Maxwell kinetic term for the dynamical gauge field. The Maxwell kinetic term is defined using the field strength $F:=dA$. The last term in Eq.\eqref{DYNATH} represents a coupling of the dynamical gauge field to an external current $J^{\mu}_{\text{ext}}$.} Here, $\lambda$ is the square of the electromagnetic coupling. {For convenience, we define the matter contribution to the stress-energy tensor $T^{\mu\nu}_{\text{m}}$ and to the U(1) conserved current $J^{\mu}_{\text{m}}$ as
\begin{equation} \label{CONJ3}
\begin{split}
\delta_{g_{\mu\nu}} S_{\text{m}} \,=\, \frac{1}{2} \int \dd^3 x \,\, \sqrt{-g} \,\,  T^{\mu\nu}_{\text{m}} \, \delta g_{\mu\nu} \,,\quad  \delta_{A_\mu} S_{\text{m}} \,=\, \int \dd^3 x \,\, \sqrt{-g} \,\, J^{\mu}_{\text{m}} \, \delta A_{\mu} \,.
\end{split}
\end{equation}
}
In order to show the physical meaning of the action in Eq.\eqref{DYNATH}, it is convenient to vary it with respect to $A_\mu$ obtaining
\begin{equation}\label{SUMBDY333}
\delta_{A_\mu} S_{\text{tot}} \,=\, \int \dd^3x \, \sqrt{-g} \,\, \left[ J^{\mu}_{\text{m}} \,-\, \frac{1}{\lambda} \nabla_{\nu} F^{\mu\nu} + J_{\text{ext}}^{\mu}  \right]  \delta A_{\mu}   \,.\,
\end{equation}
The vanishing of  Eq.\eqref{SUMBDY333} corresponds to the standard Maxwell equations, i.e.,
\begin{equation} \label{BMEQS}
\begin{split}
\nabla_{\nu} F^{\mu\nu}  = \lambda\left(J^{\mu}_{\text{m}}+ J_{\text{ext}}^{\mu}\right)\,,
\end{split}
\end{equation}
implying that the gauge field $A_{\mu}$ is now dynamical and coupled to the external current $J^{\mu}_{\text{ext}}$ as in standard electromagnetism through the coupling $\lambda$. Using the action in Eq.\eqref{DYNATH}, the dynamical equations of motion can be summarized as 
\begin{align}\label{}
&\nabla_\mu \left(T^{\mu \nu}_{\text{m}} \,+\, T^{\mu \nu}_{\text{EM}}\right) \,\,=\, F^{\lambda \nu}J_{\text{ext} \lambda} \,, \qquad\qquad\,\,\,  \nabla_\mu J^\mu_{\text{m}} \,=\, 0 \,,  \label{Eq:MHDdyn1} \\
&J^{\mu}_{\text{m}} \,+\, J^{\mu}_{\text{EM}} \,+\, J_{\text{ext}}^{\mu} = 0\,, \qquad \,\,\,\,   \qquad \quad \epsilon^{\alpha \beta \gamma} \, \nabla_\alpha F_{\beta \gamma} \,=\, 0 \,, \label{Eq:MHDdyn2}
\end{align}
{where the Levi-Civita symbol is taken following the notation $\epsilon^{012} = 1/\sqrt{-g}$, and $T^{\mu\nu}_{\text{EM}}$ is the stress-energy tensor of the the Maxwell kinetic term given by
\begin{equation}
    T^{\mu\nu}_{\text{EM}}=\frac{1}{\lambda}F^{\mu\sigma}F^{\nu}_{\,\,\,\,\sigma}-\frac{1}{4\lambda}F^2 g^{\mu \nu}\,.
\end{equation}
Likewise, we can define the contribution of the Maxwell kinetic term to the current as:
\begin{equation}
    J^{\mu}_{\text{EM}}=-\frac{1}{\lambda}\nabla_\nu F^{\mu\nu}\,.
\end{equation}}
{With these notations, the total stress tensor and current are given by:
\begin{equation}
   T^{\mu\nu}= T^{\mu\nu}_{\text{m}}+ T^{\mu\nu}_{\text{EM}}  \,,\qquad  J^{\mu}= J^{\mu}_{\text{m}} +J^{\mu}_{\text{EM}} \,,
\end{equation}
and, in absence of external sources, $J_{\text{ext}}=0$, are both conserved. In terms of the total stress tensor and total current, the EOMs in \eqref{Eq:MHDdyn1}-\eqref{Eq:MHDdyn2} become simply:
\begin{equation}
    \nabla_\mu T^{\mu\nu} \,=\, F^{\lambda \nu}J_{\text{ext} \lambda} \,, \qquad J^{\mu}+ J_{\text{ext}}^{\mu}=0\,,
\end{equation}
as reported in \cite{Hernandez:2017mch}.}
Notice that the first equation in \eqref{Eq:MHDdyn2} implies the independent conservation of the external current $J^\mu_{\text{ext}}$ as well. 
The first two equations \eqref{Eq:MHDdyn1}, can be obtained by utilising the diffeomorphism invariance (and the gauge invariance) of the action in Eq.\eqref{DYNATH}.
The other equations \eqref{Eq:MHDdyn2}, are the Maxwell equation and the electromagnetic Bianchi identity, respectively.
Note that the Maxwell equation determines the evolution of the dynamical gauge field $A_\mu$ and the Bianchi identity is used to ensure that the electric/magnetic fields can be derived from a scalar/vector potential.

In order to solve the equations of motion \eqref{Eq:MHDdyn1}-\eqref{Eq:MHDdyn2}, {we need to further specify the constitutive relations for either $T^{\mu\nu}_{\text{m}}$ and $J^{\mu}_{\text{m}}$, or equivalently $T^{\mu\nu}$ and $J^{\mu}$. Following the standard procedure to construct hydrodynamic theories, for this purpose, we will use a gradient expansion (more details about this procedure and its validity will be provided below).}
In the Landau (or energy) frame, the constitutive relations at first order in derivatives are
\begin{align}\label{Eq:consext}
    \begin{split}
        T^{\mu\nu} &\,=\, \epsilon \, u^{\mu}u^\nu \,+\, p \, \Delta^{\mu \nu} \,+\, {\mathcal{H}^{\mu \gamma} }\, F^{\nu}_{\,\,\gamma} \,+\, \Pi^{\mu \nu}\,,\\
        J^\mu &\,=\, \rho \, u^\mu \,{-\, \nabla_\nu \mathcal{H}^{\mu \nu}} \,+\, \nu^\mu\,,
    \end{split}
\end{align}
where $\epsilon$ is the energy density, $\rho$ is the charge density, $p$ is the pressure given in \eqref{EXTTHnew}, and $\Delta^{\mu\nu} :=  g^{\mu \nu}+u^\mu u^\nu$ the projection tensor in terms of the fluid velocity $u^\mu$. {In addition, we have defined the $\mathcal{H}$ tensor:
\begin{equation}
    \mathcal{H}^{\mu\nu}:= \frac{1}{\lambda}F^{\mu\nu}-M^{\mu\nu}_{\text{m}}\,,
\end{equation}
where $M^{\mu\nu}_{\text{m}}$ is the polarization tensor \eqref{POMEQ}.} Finally, ($\Pi^{\mu \nu}, \, \nu^{\mu}$) are the first order in derivatives dissipative corrections \eqref{Eq:consextdiss}. Then the constitutive relation of matter part of the stress-energy tensor and $U(1)$ current density are given by

\begin{align}\label{Eq:consext2}
    \begin{split}
        T^{\mu\nu}_{\text{m}} &\,=\, T^{\mu\nu}-T^{\mu\nu}_{\text{EM}}\,=\, \epsilon_{\text{m}} \, u^{\mu}u^\nu \,+\, p_{\text{m}}  \, \Delta^{\mu \nu} \,-\, {M^{\mu \gamma}_{\text{m}} }\, F^{\nu}_{\,\,\gamma} \,+\, \Pi^{\mu \nu} \,, \\
        J^\mu_{\text{m}} &\,=\, J^{\mu}-J^{\mu}_{\text{EM}}\,\,\,\,=\, \rho_{\text{m}}  \, u^\mu \,{+\, \nabla_\nu M^{\mu \nu}_{\text{m}} } \,+\, \nu^\mu\,,
    \end{split}
\end{align}
where $\epsilon_{\text{m}}=\epsilon-\frac{1}{4\lambda}F^2$, $p_{\text{m}}=p+\frac{1}{4\lambda}F^2$, and $\rho_{\text{m}}=\rho$. {All the quantities indicated with a sub-index $_m$ relates to the matter part of the total action in Eq.\eqref{DYNATH}.}
Using the constitutive relation for the current $J^{\mu}_{\text{m}}$ in \eqref{Eq:consext2}, one can notice that Eq.\eqref{BMEQS} corresponds to the standard Maxwell equation in matter:
\begin{equation}
    \label{MBEQS2}
    \frac{1}{\lambda} \nabla_\nu F^{\mu\nu} \,=\, J^\mu_{\text{free}}+J^\mu_{\text{bound}} + J^\mu_{\text{ext}} \,,
\end{equation}
in which $J^\mu_{\text{free}} : = \rho_{\text{m}} u^\mu + \nu^{\mu}$ and  $J^\mu_{\text{bound}} : = \nabla_\nu M^{\mu\nu}_{\text{m}}$. {$J^\mu_{\text{free}}$ refers to the current of free charges while $J^\mu_{\text{bound}}$ incorporates the polarization effects.} {We can decompose the polarization tensor $M^{\mu \nu}_{\text{m}}$ and $\mathcal{H}^{\mu\nu}$ with respect to fluid velocity as}
\begin{equation}\label{POMEQ}
\begin{split}
M^{\mu \nu}_{\text{m}} \,&=\, P^\mu \, u^\nu - P^\nu \, u^\mu - \epsilon^{\mu\nu\rho} \, u_\rho M \,,\\
 \mathcal{H}^{\mu\nu}\, &=\, u^{\mu}D^{\nu}-u^{\nu}D^{\mu}-\epsilon^{\mu\nu\rho}u_{\rho}H\,,
\end{split}
\end{equation}
and can also be identified with $M^{\mu \nu}_{\text{m}}= 2 \partial p_{\text{m}}/\partial F_{\mu\nu}$, $\mathcal{H}^{\mu \nu}= 2 \partial p/\partial F_{\mu\nu}$. {The objects $D^\mu$ and $H$ are respectively the displacement vector and the magnetic $H$-field, also known as the magnetic field strength \cite{griffiths2014introduction}.} In (3+1) dimensions~\cite{Hernandez:2017mch}, the magnetization $M$ in Eq.\eqref{POMEQ} becomes the magnetic polarization vector $M^\mu$.
The electric polarization vector $P^\mu$ and the magnetization $M$ are associated with the electric field $E^\mu$ and magnetic field $B$ via the susceptibilities ($\chi_{EE}\,, \chi_{BB}$), i.e.,
\begin{equation}\label{COMPHYHOLO}
P^\mu = \chi_{EE}E^\mu\,, \qquad M = \chi_{BB} B\,,
\end{equation}
with
\begin{equation}\label{SUSEMU}
    \chi_{EE} \,=\, 2\frac{\partial p_{\text{m}}}{\partial E^2} \,, \qquad \chi_{BB} \,=\, 2\frac{\partial p_{\text{m}}}{\partial B^2} \,.
\end{equation}
%

%
{To understand better the physical meaning of $D^{\mu}$ and $H$, it is convenient to re-write Eq.\eqref{MBEQS2} in terms of $\mathcal{H}^{\mu\nu}$}
\begin{equation}
    \nabla_{\nu}\mathcal{H}^{\mu\nu}=J^{\mu}_{\text{free}}+J^\mu_{\text{ext}}\,.
\end{equation}
Eq.\eqref{COMPHYHOLO} implies that $ D^{\mu}$ and $H$ are also proportional to the electric and magnetic field $E^{\mu}$ and $B$ via the following relations
\begin{equation}
    D^{\mu} =\frac{1}{\lambda}E^{\mu} + P^{\mu}= \epsilon_{\text{e}}E^{\mu}\,, \qquad H = \frac{1}{\lambda}B - M = \frac{1}{\mu_{\text{m}}}B\,,
\end{equation}
in which we have defined the electric permittivity $\epsilon_{\text{e}}$ and the magnetic permeability $\mu_{\text{m}}$. Using all the previous identities and definitions, we finally arrive at the following identities
\begin{equation}\label{SUSEMU2}
    \chi_{EE} =\, \epsilon_{\text{e}}-\frac{1}{\lambda} \,, \qquad \chi_{BB} =\, \frac{1}{\lambda} -\frac{1}{\mu_{\text{m}}} \,,
\end{equation}
which connect the susceptibilities to the electric permittivity and the magnetic permeability.

Continuing with the hydrodynamic construction, the dissipative terms $\Pi^{\mu \nu}$ and $\nu^\mu$ are given by
\begin{align}\label{Eq:consextdiss}
\begin{split}
        \Pi^{\mu \nu} &\,=\, -\eta \left[\Delta^{\mu \alpha}\Delta^{\nu \beta}\left(\partial_\alpha u_\beta + \partial_\beta u_\alpha \right)-\Delta^{\mu \nu}\partial_\gamma u^\gamma\right]\,,\\
        \nu^\mu &\,=\, \sigma \Delta^{\mu \nu}\left(-\partial_\nu \mu + F_{\nu \alpha}u^\alpha+\frac{\mu}{T}\partial_\nu T \right)\,,
\end{split}
\end{align}
where $\eta$ is the shear viscosity and $\sigma$ the conductivity. Here, conformal symmetry has been assumed (e.g., the bulk viscosity is not appearing therein).

One can now solve the equations of motion \eqref{Eq:MHDdyn1}-\eqref{Eq:MHDdyn2} together with the constitutive relations \eqref{Eq:consext} and obtain the low-energy excitations of the system. For this purpose, we consider the following set of fluctuations ($\delta T, \delta \mu$, $\delta u^{i=x,y}, \delta E^{i=x,y}, \delta B$) around the equilibrium configuration
\begin{align}\label{}
\begin{split}
u^\mu = (1,0,0) \,, \quad T = T_{\text{eq}} \,, \quad \mu = \mu_\text{eq} \,, \quad B = B_\text{eq} \,,
\end{split}
\end{align}
with {$(T_\text{eq},\mu_\text{eq},B_\text{eq})$} being the equilibrium values for the corresponding thermodynamic quantities. In order to simplify the notations, we will drop the subfix $_\text{eq}$ in the following.
Note that the fluctuations of the electric field ($\delta E^i$) and magnetic field ($\delta B$) appear explicitly in this case as a direct manifestation of the dynamical, rather than external, gauge field. Indeed, this is one of the major differences with the hydrodynamics with external gauge fields considered for example in~\cite{Hartnoll:2007ih,Jensen:2011xb,Kovtun:2016lfw,Jeong:2022luo}.

In Fourier space, assuming the spacetime dependence of the fluctuations to be proportional to $e^{-i\omega t + i k x}$, the linearised equations of motion can be rewritten in matrix form as
\begin{equation}\label{MMATRIX}
    \mathcal{M}(\omega, k) \cdot s_A = 0\,,
\end{equation}
where $\mathcal{M}(\omega, k)$ is a $6\times6$ matrix and $s_A = \{\delta T, \delta u^{i=x,y}, \delta E^{i=x,y}, \delta B\}$.
The matrix $\mathcal{M}(\omega, k)$ is a function of the thermodynamical susceptibilities
\begin{align}\label{THDQSE}
\begin{split}
   & \rhomu \,,\,\, \rhoT \,,\,\, \rhoB \,,\,\,\,\,  \emu \,,\,\, \eT \,,\,\, \eB \,, \\ 
   & \Pmu \,,\,\,    \PT \,,\,\,  \PB \,,
\end{split}
\end{align}
and the various thermodynamical parameters ($\epsilon\,,\, p \,,\, T \,,\, \mu\,,\, \rho \,,\, B \,,\, \sigma \,,\, \eta \,,\, \epsilon_\text{e} \,,\, \mu_\text{m}$).

Then, the dispersion relations (or eigenmodes), $\omega=\omega(k)$, can be obtained by solving the condition
\begin{equation}\label{MMATRIX2}
    \text{det} \, \mathcal{M}(\omega, k) = 0\,.
\end{equation}
The complete analytic expressions of Eq.\eqref{MMATRIX2} and of the dynamical matrix $\mathcal{M}(\omega, k)$ itself are rather lengthy and therefore not made explicit. We will show the dispersion relations of the low-energy modes only in the low $\omega,k$ expansion in the next sections.
Furthermore, in the main text, to simplify our formulas and avoid clutter, we will only show the formulas in the low $B$ expansion. {The small $B$ expansion is sometimes called the ``weak field" limit and it is somehow equivalent to assuming the magnetic field to be of order $\mathcal{O}(\partial)$ \cite{Hernandez:2017mch}. When expanding the expressions in this limit, all the thermodynamic quantities, such as $\epsilon,p$, will be independent of $B$. To avoid clutter, we will still denote them in the same way, avoiding additional subscripts.}\\
Readers interested in the complete expressions are referred to the GitHub repository available \href{https://github.com/sicobysico/MHD_HOLO}{here}.

\subsection{Zero density}\label{SECZD}
We first study the hydrodynamics of a neutral plasma, at zero charge density, $\rho=0$, or equivalently at zero chemical potential, $\mu=0$. Moreover, we may further set the following thermodynamic susceptibilities
\begin{align}\label{ZDLIT}
\begin{split}
\rhoT= \emu = \Pmu = 0 \,,
\end{split}
\end{align} 
for the neutral state. This assumption will be verified a posteriori using the holographic computations in the next section.  Using Eq.\eqref{ZDLIT} together with $\rho=0$, the spectrum of low-energy excitations exhibits six modes: four gapless modes and two gapped modes.

Let us first discuss the simplest case, i.e., the case with a vanishing magnetic field, $B=0$. For such a system, we have a pair of longitudinal sound waves together with one transverse shear diffusion mode
\begin{align}\label{SWSHW}
\begin{split}
\omega = \pm v_s \, k - i \frac{\Gamma_s}{2} \, k^2\,, \qquad \omega = -i \frac{\eta}{\epsilon+p} \, k^2 \,,
\end{split}
\end{align} 
where $v_s^2 = \partial p/\partial \epsilon$ and $\Gamma_s = {\eta}/{\left(\epsilon+p\right)}$. In this case, $(\epsilon+p)$ is just the momentum susceptibility. The modes above are decoupled from the others and simply follow from the conservation of energy and momentum as in simple relativistic fluids.
In addition, the remaining gapless mode can be determined from the ``telegrapher equation" for electromagnetic (EM) waves \cite{doi:10.1080/14786447608639176,Baggioli:2019jcm}:
\begin{align}\label{EMWAVE}
\begin{split}
\omega \left(\omega + i \, \frac{\sigma}{\epsilon_{\text{e}}}\right) = \frac{k^2}{\epsilon_{\text{e}} \, \mu_{\text{m}}} \,,
\end{split}
\end{align} 
where $\epsilon_{\text{e}}$ is the electric permittivity, $\mu_{\text{m}}$ magnetic permeability. In vacuum, $\sigma=0$, Eq.\eqref{EMWAVE} would just give rise to standard electromagnetic waves with light speed equal to $c^2\equiv 1/\left(\epsilon_{\text{e}} \, \mu_{\text{m}}\right)$. Screening effects are introduced when $\sigma \neq 0$ and relate to the well-known skin effect (for which, differently from here, one usually assumes $\omega \in \mathbb{R}$ and $k \in \mathbb{C}$). Notice, that in our holographic setup, which does not preserve Galilean symmetry, the conductivity $\sigma$ is finite even at zero charge density.

Eq.\eqref{EMWAVE} gives rise to the following solution:
\begin{equation}\label{cce}
    \omega=\pm \frac{1}{2}\sqrt{4 c^2 k^2-\frac{\sigma^2}{\epsilon_\text{e}^2}}-i\frac{\sigma}{2 \epsilon_\text{e}} \,,
\end{equation}
where we have identified $c^2=1/(\epsilon_\text{e}\mu_\text{m})$. The dynamics in Eq.\eqref{cce} is sometimes labelled as k-gap \cite{Baggioli:2019jcm} since the dispersion relation of the modes acquires a finite real part only above a certain critical wave-vector. This happens via a collision between a diffusive hydrodynamic mode and a relaxational non-hydrodynamic mode. As we will prove explicitly, Eq.\eqref{cce} is an accurate description only when the k-gap is small (in this case $\sigma/(2 \,c\, \epsilon_\text{e})\ll T$), or also in the so-called quasihydrodynamic regime \cite{Grozdanov:2018fic}.

In the small wave-vector limit, Eq.\eqref{EMWAVE} gives rise to two modes with dispersion
\begin{align}\label{EMWAVE2}
\begin{split}
\omega = -i \frac{k^2}{\sigma \, \mu_{\text{m}}} +\dots\,, \qquad \omega = -i \, \frac{\sigma}{\epsilon_{\text{e}}} + i \, \frac{k^2}{\sigma \, \mu_{\text{m}}} +\dots\,\,.
\end{split}
\end{align} 
EM waves do not propagate anymore at long distances (small $k$). Conversely, for $\sigma/\epsilon_\text{e} \neq 0$, the magnetic field diffuses with a diffusion constant $D_B := 1/(\sigma \mu_\text{m})$ while the electric field relaxes with a rate $\tau_e^{-1}:= \sigma/\epsilon_\text{e}$. In the language of global higher-form symmetries \cite{Grozdanov:2016tdf}, the electric U(1) symmetry is explicitly broken while the magnetic one is preserved implying the conservation of the magnetic flux and the presence of magnetic diffusion.

The last low energy mode is a longitudinal damped charge diffusion mode
\begin{align}\label{DCDD}
\begin{split}
\omega = -i \, \frac{\sigma}{\epsilon_{\text{e}}} - i \, \frac{\sigma}{\chi_{\rho\rho}} \, k^2 \,.
\end{split}
\end{align} 
Here, we have defined the charge susceptibility $\chi_{\rho\rho}=\partial \rho/\partial \mu$.
Charge fluctuations do not diffuse anymore but they are rather relaxing with a rate equal to $\tau_e^{-1}$, and identical to that for the electric field $E$.\\

Next, let us turn on the magnetic field and discuss its effects on the low energy modes. We focus on the small $B$ limit, {defined as $B/T^2\ll 1$ (where the $_{\text{eq}}$ subscripts are neglected for simplicity),}   and use the following identities:{
\begin{equation}
\label{Eq:preandener}
    \begin{split}
    \PB&= \left(\frac{\partial p_{\text{m}}}{\partial B}\right)_{T,\mu}-\frac{B}{\lambda}=\chi_{BB} B -\frac{B}{\lambda}\,,\\ 
    \eB &= \left(\frac{\partial \epsilon_{\text{m}}}{\partial B}\right)_{T,\mu}+\frac{B}{\lambda} = -2\chi_{BB} B+\frac{B}{\lambda}\,,
    \end{split}
\end{equation}}
which are valid in that regime.

The longitudinal sound waves and transverse shear mode in \eqref{SWSHW} are now modified into 
\begin{align}\label{SWSHW2}
\begin{split}
\omega = \pm v_{\text{ms}} \, k - i \frac{\Gamma_\text{ms}}{2} \, k^2\,, \qquad \omega = -i \left( \frac{\eta}{\epsilon+p} - \frac{\eta B^2}{\mu_{\text{m}}(\epsilon+p)^2} \right) \, k^2 \,.
\end{split}
\end{align} 
The sound modes in \eqref{SWSHW2} are known as \textit{magnetosonic waves}. Their velocity $v_{\text{ms}}$ and attenuation constant $\Gamma_\text{ms}$ are given by
\begin{align}\label{MSWEQ}
\begin{split}
v_{\text{ms}} &\,=\, v_s - \delta v_s \, B^2 \,,  \qquad  \delta v_s := \frac{\lambda \left(2 v_s^2 - 1 \right) - v_s^2 \mu_{\text{m}}}{2 \, \lambda \, \mu_{\text{m}} \, v_s \left(\epsilon+p\right)} \,,  \\
\Gamma_\text{ms} &\,=\,   \frac{\eta}{\epsilon+p}  - \left[\frac{\eta}{\mu_{\text{m}}(\epsilon+p)^2}  -  \frac{(v_s^2 - 1)(v_s (\epsilon_{\text{e}} \, \mu_{\text{m}} -1)  + 2 \mu_{\text{m}} \delta v_s (\epsilon+p))}{\mu_{\text{m}}^2 \, v_s \, \sigma \, (\epsilon+p)}  \right] B^2  \,.
\end{split}
\end{align} 
The EM waves still follow the same dynamics as in Eq.\eqref{EMWAVE} but their dispersion relations at small wave-vector are corrected by the presence of a finite magnetic field. In particular, the expressions in Eq.\eqref{EMWAVE2} are now modified into
\begin{align}\label{EMWAVE3}
\begin{split}
\omega &\,=\, -i \left(\frac{1}{\sigma \, \mu_{\text{m}}}  -  \frac{v_s - 2 \mu_{\text{m}} \delta v_s (\epsilon+p)}{\mu_{\text{m}}^2 v_s \sigma (\epsilon+p)} B^2 \right) k^2\,, \\ 
\omega &\,=\, -i \, \left(\frac{\sigma}{\epsilon_{\text{e}}} + \frac{\sigma B^2}{\epsilon_{\text{e}} \, \mu_{\text{m}} (\epsilon+p)} \right) \\
 & \qquad + i \, \left[\frac{1}{\sigma \, \mu_{\text{m}}}  + \left(\frac{v_s^2 (\epsilon_{\text{e}} \mu_{\text{m}} -1 ) - \epsilon_{\text{e}} \mu_{\text{m}} }{\mu_{\text{m}}^2 \sigma (\epsilon+p)} + \frac{2 v_s \delta v_s}{\mu_{\text{m}} \sigma} - \frac{\eta}{\mu_{\text{m}} (\epsilon+p)^2} \right) B^2  \right]k^2 \,.
\end{split}
\end{align} 
Finally, the dispersion relation of the damped charge diffusion mode \eqref{DCDD} becomes
\begin{align}\label{DCDD2}
\begin{split}
\omega = -i \, \left(\frac{\sigma}{\epsilon_{\text{e}}} + \frac{\sigma B^2}{\epsilon_{\text{e}} \, \mu_{\text{m}} (\epsilon+p)} \right) - i \, \left(\frac{\sigma}{\chi_{\rho\rho}} + \frac{\eta B^2}{\mu_{\text{m}}(\epsilon+p)^2} \right)\, k^2 \,,
\end{split}
\end{align} 
where, once again, the limit of small magnetic field is assumed. {We remind the Reader that in this limit, all the thermodynamic quantities appearing in the expressions above are not functions of the magnetic field $B$. This is equivalent to the ``weak field" limit in \cite{Hernandez:2017mch}.}
Note that all the dispersion relations presented in \eqref{SWSHW2}-\eqref{DCDD2} are consistent with the ones in (3+1) dimensions derived in~\cite{Hernandez:2017mch}. A main qualitative difference is that in (2+1), one does not have Alfv\'en waves because the magnetic field is always perpendicular to the wave-vector.\footnote{To be more precise, in (2+1) dimensions, the magnetic field does not relate to a well-defined vector but rather to a pseudo-scalar. Therefore, formally, one cannot define any angle between the magnetic field and the wave-vector in (2+1) dimensions.} Indeed, Alfv\'en waves disappear even in (3+1) dimensions when the direction of propagation is perpendicular to the magnetic field~\cite{Hernandez:2017mch} ($\theta=\pi/2$ in their notations, see also \cite{Grozdanov:2017kyl}). Moreover, the distinction between fast and slow magnetosonic waves, which relies on the presence of a finite angle $\theta$, is also absent in (2+1).

\subsection{Finite density}\label{FDskl}
Next, we study the hydrodynamics at finite density ($\rho\neq0$). {To be precise, by ``finite density" we mean a finite density of free charges $\rho=J^t$. Because of Maxwell's equations, we necessarily have $J^\mu+J_{\text{ext}}^\mu=0$ at equilibrium. This implies that we also have $J^t_{\text{ext}}=-\rho$ and that the total charge density is zero, $J^t_{\text{tot}}\equiv J^t+J^t_{\text{ext}}=0$. Physically, we should think of this situation as a system with a finite density of free charges (e.g., electrons) together with an equal finite density of ions which render the total system neutral. In \cite{Hernandez:2017mch}, this state is labelled as ``charged state offset by background charge''. For simplicity, we will continue with the simpler notation ``finite density", well aware of the caveat discussed above.}

Solving Eq.\eqref{MMATRIX2} in the small wave-vector limit, one finds six low energy modes corresponding this time to two gapless modes and four gapped modes. Note that the number of the gapless modes (and gapped modes) is different from the zero density case. This is related to the fact that the small $\rho$ limit does not commute with the small $k$ limit when the gauge field is dynamical~\cite{Hernandez:2017mch}.

We now show the structure of the low-energy modes at finite charge density and finite magnetic field $B$, focusing on the small $B$ limit. We find one longitudinal diffusive mode and one transverse subdiffusive shear mode with dispersions
\begin{align}\label{FINITEDEN1}
\begin{split}
\omega =    -i \, \frac{ \, \rhomu (\epsilon+p)^2 \,\, \sigma}{T \, \left[ \eT\rhomu - \emu\rhoT \right] \left(\rho^2+B^2 \sigma\right)} \,\, k^2 \,, \qquad  \omega = -i \frac{\eta}{\mu_{\text{m}} \, \rho^2} \, k^4  \,.
\end{split}
\end{align} 
The fact that the dispersion relation of the subdiffusive mode is not well defined at $\rho \rightarrow 0$ is just a manifestation of the non-commutativity of limits. Note that the gapless modes in \eqref{FINITEDEN1} appear also in the case of external gauge fields~\cite{Jeong:2022luo}.
Interestingly, the dispersion of the diffusive mode in \eqref{FINITEDEN1} is exactly the same (at this order in $k$) as the one in the case of external gauge fields (see Eq.\eqref{AP3}). This observation is also consistent with the results in the higher dimensional (3+1) theory~\cite{Hernandez:2017mch}. On the contrary, the subdiffusive mode in \eqref{FINITEDEN1} displays a slightly different form (see Eq.\eqref{AP3} for the subdiffusive mode in the case of external gauge field). One could worry whether the subdiffusive mode in Eq.\eqref{FINITEDEN1} is a spurious mode whose dispersion is not robust under higher-order corrections. We will discuss this problem in detail in Section \ref{nn} and we will concretely show in appendix \ref{lala} that this is not the case.

The remaining four modes are all non-hydrodynamic and their dispersions can be found by solving the following equation:
\begin{align}\label{GAPEOMD}
\begin{split}
\left[ \omega \left( \omega + i \, \frac{\sigma}{\epsilon_{\text{e}}} \right) - \Omega_p^2 \right]^2 \,=\, \frac{B^2}{\epsilon_{\text{e}}^2 \, \mu_{\text{m}}^2(\epsilon+p)^2} \left[ \rho^2 - \mu_{\text{m}}^2 \sigma^2 (\rho^2-B^2)  +\omega^2\left( 2(\epsilon+p)(\sigma - i \epsilon_{\text{e}} \omega) \right)  \right]  \,,
\end{split}
\end{align} 
where $\Omega_p$ is the plasma frequency
\begin{align}\label{PSFOUR}
\begin{split}
\Omega_p^2 \,:=\, \frac{\rho^2}{\epsilon_{\text{e}}(\epsilon+p)}\,.
\end{split}
\end{align} 
In particular, in the small $B$ limit, the four modes are given as follows
\begin{align}\label{FINITEDEN2}
\begin{split}
\omega &\,=\, - \frac{i}{2}\left( \frac{\sigma}{\epsilon_\text{e}} \,-\, \sqrt{\frac{\sigma^2}{\epsilon_\text{e}^2} - 4 \, \Omega_p^2 } \right) \,\pm\, \left(1 - \frac{1+2 \epsilon_{\text{e}} \mu_{\text{m}}}{\sqrt{\frac{\sigma^2}{\epsilon_\text{e}^2} -4 \, \Omega_p^2 }} \, \frac{\sigma}{\epsilon_\text{e}} \right) \frac{\rho B}{2 \epsilon_{\text{e}} \mu_{\text{m}} (\epsilon+p)} \,, \\
\omega &\,=\, - \frac{i}{2}\left( \frac{\sigma}{\epsilon_\text{e}} \,+\, \sqrt{\frac{\sigma^2}{\epsilon_\text{e}^2} - 4 \, \Omega_p^2 } \right) \,\pm\, \left(1 + \frac{1+2 \epsilon_{\text{e}} \mu_{\text{m}}}{\sqrt{\frac{\sigma^2}{\epsilon_\text{e}^2} - 4 \, \Omega_p^2}} \, \frac{\sigma}{\epsilon_\text{e}} \right) \frac{\rho B}{2 \epsilon_{\text{e}} \mu_{\text{m}} (\epsilon+p)} \,.
\end{split}
\end{align} 
One interesting feature of the gaps at finite density \eqref{FINITEDEN2} can be observed in the limit of $B=0$. More precisely, depending on the value of the charge density $\rho$ (entering through the plasma frequency $\Omega_\rho$), the dispersions in Eq.\eqref{FINITEDEN2} can be purely imaginary or complex, i.e.,
\begin{align}\label{GAPSDENSI}
\begin{split}
\left( \sigma^2/\epsilon_\text{e}^2 \,\gg\, 4 \, \Omega_p^2 \,\,; \,\, \text{small density}\right):  \quad  \omega &\,=\, -\, i \, \frac{\epsilon_{\text{e}}}{\sigma} \, \Omega_p^2 \,, \qquad\, \omega  \,=\,  - i  \, \frac{\sigma}{\epsilon_{\text{e}}} \,+\, i \, \frac{\epsilon_{\text{e}}}{\sigma} \, \Omega_p^2   \,, \\
\left( \sigma^2/\epsilon_\text{e}^2 \,\ll\, 4 \, \Omega_p^2 \,\,; \,\, \text{large density}\right): \quad  \omega &\,=\,  \pm\,\, \Omega_p -\, i \, \frac{\sigma}{2 \epsilon_{\text{e}}}   \,.
\end{split}
\end{align} 
Finally, setting all the dissipative coefficients (e.g., $\sigma=0$) to zero, one finds
\begin{align}\label{GOE}
\begin{split}
\omega^2 \,=\,  \Omega_p^2 \,+\, v_s^2 \, k^2 \,, \qquad \omega^2 \,=\, \Omega_p^2 \,+\, \frac{k^2}{\epsilon_{\text{e}} \, \mu_{\text{m}}} \,.
\end{split}
\end{align} 
The plasma frequency $\Omega_p$ gaps out both the sound waves (the first equation in \eqref{GOE}) and the electromagnetic waves (the second equation in \eqref{GOE}).\footnote{The gapped sound waves, the first equation in \eqref{GOE}, are the relativistic analogues of Langmuir oscillations~\cite{Hernandez:2017mch}.}
As the sound waves become the magnetosonic waves \eqref{SWSHW2} at finite $B$, one may say that finite density gaps out the magnetosonic waves as well~\cite{Hernandez:2017mch}.

For later use, we summarize the small $B$ correction to \eqref{GAPSDENSI} as follows.
For small density, ($\sigma^2/\epsilon_\text{e}^2 \,\gg\, 4 \, \Omega_p^2$), we have 
\begin{align}\label{SMDG}
\begin{split}
\omega &\,=\, -\, i \, \frac{\epsilon_{\text{e}}}{\sigma} \, \Omega_p^2 \,\,\pm\,\,  \left[\frac{1}{\epsilon+p} + \frac{\epsilon_{\text{e}} (1+2 \epsilon_{\text{e}}\, \mu_{\text{m}} ) }{\mu_{\text{m}} \sigma^2 (\epsilon+p)} \Omega_p^2  \right] \, \rho \,B   \,, \\
\omega  &\,=\,  - i  \, \frac{\sigma}{\epsilon_{\text{e}}} \,+\, i \, \frac{\epsilon_{\text{e}}}{\sigma} \, \Omega_p^2 \,\,\pm\,\, \left[\frac{1+\epsilon_{\text{e}}\,\mu_{\text{m}}}{\epsilon_{\text{e}}\,\mu_{\text{m}} (\epsilon+p)} + \frac{\epsilon_{\text{e}} (1+2 \epsilon_{\text{e}}\, \mu_{\text{m}} ) }{\mu_{\text{m}} \sigma^2 (\epsilon+p)} \Omega_p^2\right] \, \rho\, B   \,, 
\end{split}
\end{align} 
where the magnetic field $B$ produces a real gap.
For large density ($\sigma^2/\epsilon_\text{e}^2 \,\ll\, 4 \, \Omega_p^2$), we have
\begin{align}\label{LAGDG}
\omega &\,=\,  \,\, \pm \,\Omega_p -\, i \, \frac{\sigma}{2 \epsilon_{\text{e}}}  \,\,\pm\,\, \left[\frac{1}{2  \epsilon_{\text{e}}\,\mu_{\text{m}} (\epsilon+p) } \,\mp\, i \, \frac{(1+ 2 \epsilon_{\text{e}}\,\mu_{\text{m}})\sigma}{4 \epsilon_{\text{e}}^2 \, \mu_{\text{m}} (\epsilon+p) \Omega_p } \right] \rho\,B \,, 
\end{align} 
where the magnetic field $B$ produces both a real and an imaginary gap.

\subsection{Dynamical vs. external gauge fields}
We finish this section by comparing the dispersion relations of the low-energy modes for the distinct cases of dynamical and external gauge fields. The results are summarized in tables \ref{STED1}-\ref{STED2}.  

Let us briefly remind the Reader of the results in the case of external gauge fields. In the neutral case, the spectrum displays a energy diffusion mode and a subdiffusive mode with dispersion:
\begin{align}
\omega = -i  \frac{\partial p}{\partial \epsilon}\, \frac{\epsilon + p}{\sigma B^2} \,k^2  \,, \qquad \omega = -i  \frac{\eta}{B^2 \chi_{\rho\rho}} \,k^4 \,, \label{AP1}
\end{align} 
together with the so-called cyclotron mode
\begin{align}
\qquad \omega = -i \frac{\sigma B^2}{\epsilon+p} \,. \label{AP2}
\end{align} 
At the finite density, the dispersion of these modes are modified into
\begin{align}
\omega &=    -i\,\frac{ \, \rhomu (\epsilon+p)^2 \,\, \sigma}{T\left[ \eT\rhomu - \emu\rhoT \right]\left(\rho^2 + B^2 \sigma^2 \right)} \,\, k^2  \,, \quad\,\,\,\, \omega = -i  \frac{\eta}{B^2 \left( \frac{\partial \rho}{\partial \mu} \right)_{T}} \,k^4 \,,  \label{AP3} \\
\omega &=  \pm \,\Omega_B -i \frac{\sigma B^2}{\epsilon+P}     \label{AP4}  \,,
\end{align} 
where the real gap, $\Omega_B := {B \rho}/{(\epsilon+p)}$, is the cyclotron frequency.

\begin{table}[]
\begin{tabular}{| C{2.68cm} | C{4.99cm} | C{6.38cm}  |}
\hline
             & \cellcolor{green!08} \textbf{External gauge fields}    &   \cellcolor{blue!08} \textbf{Dynamical gauge fields}  \\ 
 \hline
 \hline
                             &  Energy diffusion mode \eqref{AP1},             & Magnetosonic waves \eqref{SWSHW2}, \\
 Gappless modes    &  Subdiffusive mode \eqref{AP1}, &  Shear diffusion mode \eqref{SWSHW2}, \\ 
                           &    &  Magnetic diffusion mode \eqref{EMWAVE3}, \\ 
\hline
   Gapped modes    &   Cyclotron mode \eqref{AP2}.    &        
Damped diffusion mode \eqref{EMWAVE3},          \\
             &       &        Damped charge diffusion mode \eqref{DCDD2}.          \\
 \hline
\end{tabular}
\caption{The low energy modes at zero density and finite magnetic field.}\label{STED1}
\end{table}
\begin{table}[]
\begin{tabular}{| C{2.68cm} | C{4.99cm} | C{6.38cm}  |}
\hline
             &   \cellcolor{green!08} \textbf{External gauge fields}    &   \cellcolor{blue!08}\textbf{Dynamical gauge fields}  \\ 
 \hline
 \hline
Gappless modes      &  Diffusion mode \eqref{AP3},             & Diffusion mode \eqref{FINITEDEN1}, \\
                               &  Subdiffusive mode \eqref{AP3}, &  Subdiffusive mode \eqref{FINITEDEN1}, \\ 
\hline
     Gapped modes    &   Cyclotron mode \eqref{AP4}.    &        
Gapped plasma modes \eqref{FINITEDEN2}.          \\
 \hline
\end{tabular}
\caption{The low energy modes at finite density and finite magnetic field.}\label{STED2}
\end{table}

By looking at the summary tables \ref{STED1}-\ref{STED2}, one can notice that the dispersion relations in the two cases (dynamical vs. external gauge fields) are noticeably different. However, as described below Eq. \eqref{FINITEDEN1}, there is one exception: the diffusion mode at finite density (cfr. Eq.\eqref{AP3} vs. Eq.\eqref{FINITEDEN1}). At least at such order in the wave-vector $k$, the dispersions are identical.

\subsection{A note on the regime of validity of first-order magnetohydrodynamics}\label{nn}
Before concluding this section about the magnetohydrodynamic framework, we would like to clarify its regime of validity and the role of possible higher order corrections. Everything discussed so far is valid in the approximation of first-order linearised hydrodynamics. The latter is the statement that the constitutive relations for the stress tensor $T^{\mu\nu}$ and the U(1) current $J^\mu$ are expanded in dissipative corrections up to terms which are linear, or first-order, in the gradients. The first-order dissipative corrections, which have been indicated respectively as $\Pi^{\mu\nu}$ and $\nu^\mu$, are only the first of an infinite series expansion in gradients. In particular, both the stress tensor and the U(1) current can be expanded as:
\begin{equation}
    T^{\mu\nu}=T^{\mu\nu}_{\text{eq}}+ \Pi^{\mu\nu}+ T^{\mu\nu}_{(2)}+T^{\mu\nu}_{(3)}+\dots\,,\qquad J^\mu=J^\mu_{\text{eq}}+\nu^\mu+J^\mu_{(2)}+J^\mu_{(3)}+\dots \,,
\end{equation}
where the terms indicated with suffix $_{(n)}$ refers to dissipative corrections beyond equilibrium which are n-th order in gradients. Higher order terms correspond to shorter timescales and lengthscales and they expand the validity of the effective description towards the microscopic world.\\

In our analysis, the first terms which are neglected are second order in gradients, i.e., $T^{\mu\nu}_{(2)},J^{\mu}_{(2)}$. Since all the dynamical equations contain at least an extra derivative, these neglected corrections enter into the dynamical matrix $\mathcal{M}(\omega,k)$ as terms $\sim k^3$, where $k$ is the wave-vector. All in all, this means that the dispersion relations within the second order magnetohydrodynamic framework should be extracted from:
\begin{equation}
    \mathrm{det}\left[\mathcal{M}(\omega,k) + \mathcal{C} \, k^3\right] = 0 \,,
\end{equation}
with $\mathcal{C}$ a matrix of $k-$independent coefficients. In the worst case scenario, all the entries of the matrix $\mathcal{C}$ are nonzero. This is obviously a very conservative view but at this point, without knowing the precise form of $\mathcal{C}$, the safest. Following this argument, we can confidently trust the results from first-order magnetohydrodynamics only up to the order in which the corresponding hydrodynamic coefficients are not affected by the possible corrections appearing in $\mathcal{C}$. This is a well known problem in hydrodynamics which sometimes leads to the appearance of spurious poles as well. See for example Section 2.6 in \cite{Kovtun:2012rj}.
To make this point clearer, let us make an example. Let us assume that as a solution of $\mathrm{det}\left[\mathcal{M}(\omega,k)\right]=0$ we get a mode whose dispersion relation within the first-order approximation can be written as:
\begin{equation}
    \omega(k)=\sum_{i=0}^n\,a_i k^i \,,
\end{equation}
with $a_i$ random complex numbers. Now, let us assume, that the same dispersion relation extracted from the second-order formalism reads:
\begin{equation}
    \omega(k)=\sum_{i=0}^n\,\tilde a_i k^i\,.
\end{equation}
Then, we can trust the dispersion relation obtained from the first-order formalism only up to the order at which $a_i=\tilde a_i$ or, in other words, up to the order at which the higher order corrections do not play any role. In the rest of the manuscript, whenever we will refer to \textit{hydrodynamic predictions}, we will always have in mind this first order formalism truncated up to the terms in the $k$ expansion which can be trusted within this approximation.\\

Notice that all the dispersion relations written so far, apart from that of the subdiffusive mode, $\omega\sim k^4$ in Eq.\eqref{FINITEDEN1} are at most order $k^2$. It is then immediate to verify that such expressions would not be corrected by possible higher order terms $\sim k^3$. The case of the subdiffusive mode could be potentially different. In particular, both the $k^3$ coefficient, which is zero in the first-order approximation, and the $k^4$ one shown in the text might in principle be affected by higher-order corrections. Nevertheless, we have verified with an accurate numerical analysis that this is not the case. We refer the Reader to Appendix \ref{lala} for an extensive discussion on this point.\\

Finally, in Appendix \ref{lala}, we will investigate further the validity of the dispersion relations obtained from solving $\mathrm{det}\left[\mathcal{M}(\omega,k)\right]=0$ without worrying about possible higher-order corrections. In particular, we will show that the dispersion relations obtained in that way, by assuming somehow that no higher-order corrections appear, significantly extend the range of agreement between the numerical data and the hydrodynamic predictions. This is an a-posteriori proof that many of the higher-order corrections are either zero or negligible for the problem at hand. Of course, one is not guaranteed that this is generally the case.

\section{Holography with dynamical boundary gauge fields}\label{SECMHDDIS333}

In this section, we study the dynamics of a simple holographic system at finite charge density and finite magnetic field in which the gauge field is taken as dynamical in the boundary field theory using mixed boundary conditions.

\subsection{Holographic setup}
Let us consider the Einstein-Maxwell action in (3+1) dimensions,
\begin{equation}\label{ACTIONH}
\begin{split}
S_{\text{bulk}} = \int \dd^4x\, \sqrt{-g} \,\left( R \,+\, 6 \,-\, \frac{1}{4} F^2 \right) \,, \qquad F= \dd A \,,
\end{split}
\end{equation}
where we set $16\pi G=1$ and the AdS radius $L=1$. 
We use Latin indices $M,N,\dots$ for the 4-dimensional bulk spacetime coordinates and use Greek indices $\mu,\nu,\dots$ for the 3-dimensional boundary coordinates.
In addition, let us consider the background dyonic black-brane ansatz 
\begin{equation}\label{BGMET}
\begin{split}
\dd s^2 =  -f(r)\, \dd t^2 +  \frac{1}{f(r)} \, \dd r^2  + r^2 (\dd x^2 + \dd y^2) \,,\quad   A = A_t(r) \, \dd t -\frac{B}{2} y \,\dd x \,+\, \frac{B}{2} x \, \dd y \,,
\end{split}
\end{equation}
where $r=\infty$ is the location of the AdS boundary and $B$ the magnetic field. The action in Eq.\eqref{ACTIONH} allows for a simply dyonic black-brane analytic solution given by 
\begin{equation}\label{BCF}
\begin{split}
 f(r) &\,= r^2 - \frac{m_{0}}{r} \,+ \, \frac{\mu^2 r_{h}^2 + B^2}{4\,r^2} \,, \quad m_{0} = r_{h}^3\left( 1 +  \frac{\mu^2 r_{h}^2+B^2}{4\, r_{h}^4} \right) \,, \\
 A_{t}(r) &\,= \mu \left( 1- \frac{r_{h}}{r} \right)\,,
\end{split}
\end{equation}
where $\mu$ is the chemical potential, $r_{h}$ the horizon radius, and the black-brane mass $m_{0}$ is determined by the condition $f(r_{h})=0$.

The various thermodynamic parameters associated with such a solution can be derived as~\cite{Hartnoll:2007ih,Hartnoll:2007ip,Kim:2015wba,Blake:2015hxa}. {We identify the bulk on-shell action in Eq.\eqref{ACTIONH} with the matter controbution $S_{m}[g_{\mu\nu}, A_{\mu}]$ in Eq.\eqref{DYNATH}. Moreover, we add the following boundary terms
\begin{equation}\label{Action:bdry}
S_{\text{boundary}} = \int \dd^3x \, \left[ -\frac{1}{4\lambda}F_{\mu\nu}^2 \,+\,  A_{\mu} \, J_{\text{ext}}^{\mu}  \right] \,.
\end{equation}
The latter, together with the bulk part, Eq.\eqref{ACTIONH}, constitute the full boundary action which has to be compared with Eq.\eqref{DYNATH}.
As a consequence, the thermodynamic quantities for $T^{\mu\nu}$ and $J^{\mu}$, which include the contributions from the Maxwell kinetic term, are given by}
\begin{align}\label{HAWKINGT}
\begin{split}
 T &\,=\, \frac{1}{4\pi} \left( 3\,r_{h} \,-\, \frac{\mu^2 r_{h}^2 + B^2}{4\,r_{h}^3}  \right) \,,  \quad \rho \,=\, \mu \, r_{h} \,, \quad s \,=\, 4\pi \, r_{h}^2  \,, \\
\epsilon &\,=\,  2r_{h}^3 + \frac{\mu^2 r_{h}}{2} + \frac{B^2}{2 r_{h}} {+\frac{B^2}{2\lambda}}  \,, \qquad p \,=\, r_{h}^3 + \frac{\mu^2 r_{h}}{4} - \frac{3 B^2}{4 r_{h}} {-\frac{B^2}{2\lambda}}\,,   
\end{split}
\end{align}
where $(T,\rho,s, \epsilon, p)$ are the temperature, charge density, entropy, energy and pressure density, respectively. One can easily verify that these expressions satisfy the Smarr relation $\epsilon + p\,=\, s \, T + \mu \, \rho$.
Furthermore, using \eqref{HAWKINGT}, one can compute all the thermodynamic susceptibilities in \eqref{THDQSE} holographically. Doing so, we have verified that some of them vanish for this concrete solution at zero charge density as anticipated in Eq.\eqref{ZDLIT}. {Notice that while the trace of the matter contribution to the stress tensor vanishes, the trace of the total stress tensor does not. In particular, we have:
\begin{equation}
 {T^\mu}_\mu=\frac{1}{4\lambda}F^2 \,.
\end{equation}
This result is not surprising since it corresponds to the statement that Maxwell theory in $3+1$ dimensions is scale invariant but not conformal invariant \cite{El-Showk:2011xbs, Nakayama:2013is}.  As a consequence, the trace of its stress tensor does not vanish but it is equal to the total divergence of a virial current \cite{El-Showk:2011xbs}.} Moreover, in presence of the magnetic field, the mechanical pressure is not equal to the thermodynamic pressure. In particular, we have:{
\begin{equation}
    T_{xx}=\frac{\epsilon}{2}+\frac{B^2}{4\lambda}=\left(p+\frac{B^2}{\mu_m}\right) \neq p\,.
\end{equation}}
The other transport coefficients ($\sigma\,,\eta$) can also be computed and read
\begin{align}\label{etasigma}
\begin{split}
\sigma = \left(\frac{s T}{\epsilon+p}\right)^2  \,, \qquad \eta = \frac{s}{4\pi} \,,
\end{split}
\end{align}
where the conductivity $\sigma$ is given in \cite{Hartnoll:2007ih,Hartnoll:2007ip,Kim:2015wba,Blake:2015hxa}\footnote{\label{ft14}See also \cite{Amoretti:2021fch,Amoretti:2020mkp,Amoretti:2019buu} for the development of transport property where $B$ is no longer taken to be of order one in derivatives.} and the shear viscosity $\eta$ in \eqref{etasigma} is obtained from the fact that the KSS bound~\cite{Kovtun:2004de,Kovtun:2003wp} is not violated in the presence of both charge density and magnetic field in (3+1) dimensions.\footnote{On the contrary, in the higher dimensional case~\cite{Jain:2015txa,Finazzo:2016mhm,Rebhan:2011vd,Giataganas:2013lga,Mamo:2012sy}, the KSS bound is violated at finite $B$.}

{Finally, we observed that the term $\epsilon+p$ which frequently appears in the hydrodynamic expressions is not affected by the Maxwell kinetic term. In particular, as evident from Eq.\eqref{HAWKINGT}, we have that:
\begin{equation}
    \epsilon_m+p_m=\epsilon+p\,.
\end{equation}
}
\subsection{Boundary conditions for dynamical gauge fields}\label{DTMF}

In order to investigate the quasi-normal modes, we consider the fluctuations $\delta g_{MN}$ and $\delta A_{M}$ as
\begin{align}\label{}
\begin{split}
g_{MN} \,\rightarrow\, g_{MN} + \delta g_{MN} \,, \quad A_{M} \,\rightarrow\, A_{M} + \delta A_{M} \,,
\end{split}
\end{align} 
where $g_{MN}$ and $A_{M}$ are the background bulk fields given in Eq.\eqref{BGMET}. For convenience, we choose the radial gauge: $\delta g_{tr} = \delta g_{rr} =\delta g_{xr}=\delta g_{yr} = \delta A_{r} = 0$. Additionally, we write the fluctuations in Fourier form using the following notations
\begin{align}\label{FLUCOURSETUP}
\begin{split}
\delta g_{MN} &= h_{MN}(r) \,e^{-i  \omega  t \,+\, i  k  x} \,,\, \quad  \delta A_{M} = a_{M}(r) \,e^{-i \omega  t \,+\, i  k  x} \,,
\end{split}
\end{align} 
with the wave-vector $k$ aligned along the $x$ direction.
We then construct four gauge-invariant variables (see for example~\cite{Jeong:2022luo}) as
\begin{align}\label{GIVOUR}
\begin{split}
Z_{H_1} &\,:=\, k \,h_{t}^{y} \,+\, \omega \, h_{x}^{y}  \,, \\
Z_{H_2} &\,:=\, \frac{4 k}{\omega} \, h_{t}^{x} \,+\,  2 h_{x}^{x} - \left( 2 - \frac{k^2}{\omega^2}\frac{f'(r)}{r} \right) h_{y}^{y}  + \frac{2k^2}{\omega^2}\frac{f(r)}{r^2} h_{t}^{t}  \,, \\
Z_{A_1} &\,:=\, k \,a_{t} \,+\, \omega \, a_{x} \,-\, \frac{i B\,\omega}{k} h_{x}^{y} \,-\,\frac{k\,r}{2} \,A_{t}' \, h_{y}^{y} \,, \\
Z_{A_2} &\,:=\, a_{y}  \,+\, \frac{i B}{2k} \left(h_{x}^{x}-h_{y}^{y}\right) \,,
\end{split}
\end{align}
where the index of the metric fluctuation $h_{MN}$ is raised with the background metric \eqref{BGMET}. The number of gauge-invariant variables is related to the structure of the equations of motion at the linearized level. In our case, one can find nine second-order equations with five first-order constraints. This implies that there are four independent fluctuations and therefore four gauge-invariant variables.

Then, we can study the quasi-normal modes by employing the determinant method~\cite{Kaminski:2009dh} in which the source matrix, $\mathcal{S}$, is constructed with the AdS boundary ($r\rightarrow\infty$) expansion of the variables \eqref{GIVOUR}. Note that, at the AdS boundary, the gauge-invariant variables \eqref{GIVOUR} are expanded as
\begin{align}\label{BD2}
\begin{split}
&Z_{H_{i}} = Z_{H_{i}}^{(L)} \, r^{0} \,(1 \,+\, \dots) \,+\, Z_{H_{i}}^{(S)} \, r^{-3} \,(1 \,+\, \dots) \,, \\
&Z_{A_{i}} = Z_{A_{i}}^{(L)} \, r^{0} \,(1 \,+\, \dots) \,+\, Z_{A_{i}}^{(S)}\, r^{-1} \,(1 \,+\, \dots) \,,
\end{split}
\end{align}
where the superscripts $(L,S)$ denote the leading/subleading term in the asymptotic expansion.

At this point, it is fundamental to understand how to construct the matrix $\mathcal{S}$ \eqref{APPENSMATA3} appearing in the determinant method in the case of dynamical gauge field. For this purpose, we need to consider the boundary action \eqref{Action:bdry}. Then the variation of the total action $S_{\text{on-shell}}+S_{\text{boundary}}$ produces the following equation at the AdS boundary,
\begin{equation} \label{SOLEQ2C}
\begin{split}
\Pi^{\mu} \,-\, \frac{1}{\lambda} \partial_{\nu} F^{\mu\nu} + J_{\text{ext}}^{\mu} = 0 \,, \qquad \Pi^{\mu} \,=\, \frac{\delta S_{\text{on-shell}}}{\delta A_{\mu}} \,=\, - \sqrt{-g}\, F^{r \mu} \big|_{r\rightarrow\infty} \,,
\end{split}
\end{equation}
where $S_{\text{on-shell}}$ is the on-shell action from \eqref{ACTIONH} and $\Pi^{\mu}$ the radially conserved bulk current obtained from the Maxwell equation: $ \partial_r \left( \sqrt{-g} \, F^{r \mu} \right)  \,=\,  \partial_r \,\Pi^{\mu}=0$. Notice that $\lambda$ parametrizes the ratio between the bulk electromagnetic coupling and the boundary one. Since we have fixed the bulk one to unity, then $\lambda$ corresponds directly to the boundary coupling as in the hydrodynamic description of the previous sections.

The first order variation of each terms is given by
\begin{equation}\label{Eq:firstordervar2}
\begin{split}
&\delta \Pi^t=\frac{1}{\omega^2-k^2}\left(k Z_{A_1}^{(S)}-\frac{1}{2}\rho Z_{H_2}^{(L)}-\frac{1}{2}\rho(k^2-\omega^2)(h_{xx}^{(L)}-h_{yy}^{(L)})\right)\,,\\
&\delta \Pi^x = \frac{1}{\omega^2-k^2}\left(\omega   Z_{A_1}^{(S)} - \frac{\omega}{2k}\rho Z_{H_2}^{(L)}-\frac{\omega}{2k}\rho(k^2-\omega^2)(h_{xx}^{(L)}-h_{yy}^{(L)}) \right)\,,\\
&\delta \Pi^y = Z_{A_2}^{(S)}-\frac{\rho}{k}Z_{H_1}^{(L)}+\frac{\omega}{k}\rho h_{xy}^{(L)}\,,\\
&\delta\left(\frac{1}{\lambda} \partial_{\mu} F^{\mu t}\right) = \frac{k}{\lambda}Z_{A_1}^{(L)} +i\frac{\omega^2}{\lambda}B h_{xy}^{(L)}\,,\\
& \delta\left(\frac{1}{\lambda} \partial_{\mu} F^{\mu x}\right) = \frac{\omega}{\lambda}Z_{A_1}^{(L)}+i\frac{\omega}{k \lambda}B h_{xy}^{(L)}\,,\\
& \delta\left(\frac{1}{\lambda} \partial_{\mu} F^{\mu y}\right) = \frac{\omega^2-k^2}{\lambda}Z_{A_2}^{(L)}+i\frac{\omega^2-k^2}{2k \lambda}B\left(h_{xx}^{(L)}-h_{yy}^{(L)}\right)\,,
\end{split}
\end{equation}
where, for convenience, gauge-invariant variables~\eqref{GIVOUR} are used. Using the Maxwell equations, we then get:
\begin{equation}
\label{Eq:source}
\begin{split}
&\delta J_{\text{ext}}^{\,t \,\,(L)} \,=  - \frac{k}{\lambda}Z_{A_1}^{(L)} - \frac{k}{\omega^2 - k^2}Z_{A_1}^{(S)} + \rho \left(\frac{1}{2(\omega^2-k^2)}Z_{H_2}^{(L)}-\left(h_{xx}^{(L)}-h_{yy}^{(L)}\right)\right)-iB\frac{\omega}{\lambda} h_{xy}^{(L)}\,, \\
&\delta J_{\text{ext}}^{\,x \,\,(L)} \,=  - \frac{\omega}{\lambda}Z_{A_1}^{(L)} - \frac{\omega}{\omega^2 - k^2}Z_{A_1}^{(S)} + \frac{\rho \omega}{k} \left(\frac{1}{2(\omega^2-k^2)}Z_{H_2}^{(L)}-\left(h_{xx}^{(L)}-h_{yy}^{(L)}\right)\right)-iB\frac{\omega^2}{k \lambda} h_{xy}^{(L)}\,, \\
&\delta J_{\text{ext}}^{\,y \,\,(L)} \,=  - \frac{\omega^2-k^2}{\lambda}Z_{A_2}^{(L)} - Z_{A_2}^{(S)}+\frac{\rho}{k}\left(Z_{H_1}^{(L)}-\omega h_{xy}^{(L)}\right)+iB\frac{\omega^2-k^2}{2k \lambda}\left(h_{xx}^{(L)}-h_{yy}^{(L)}\right)\,.
\end{split}
\end{equation}
In the following, we will set $h_{xx}^{(L)}\,, h_{xy}^{(L)}\,, h_{yy}^{(L)}$ to zero. As shown in \cite{Davison:2011uk}, those terms would contribute only to finite local counterterms in the on-shell action and therefore they would not modify the structure of the poles that we are interested in. Doing that, Eq.\eqref{Eq:source} becomes
\begin{equation}
\label{Eq:sourcess}
\begin{split}
&\delta J_{\text{ext}}^{\,t \,\,(L)} \,=  - \frac{k}{\lambda}Z_{A_1}^{(L)} - \frac{k}{\omega^2 - k^2}Z_{A_1}^{(S)} + \frac{\rho}{2(\omega^2-k^2)}Z_{H_2}^{(L)}\,, \\
&\delta J_{\text{ext}}^{\,x \,\,(L)} \,=  - \frac{\omega}{\lambda}Z_{A_1}^{(L)} - \frac{\omega}{\omega^2 - k^2}Z_{A_1}^{(S)} + \frac{\rho \omega}{2k(\omega^2-k^2)}Z_{H_2}^{(L)}\,, \\
&\delta J_{\text{ext}}^{\,y \,\,(L)} \,=  - \frac{\omega^2-k^2}{\lambda}Z_{A_2}^{(L)} - Z_{A_2}^{(S)}+\frac{\rho}{k}Z_{H_1}^{(L)}\,,
\end{split}
\end{equation}
and the conservation equation $\nabla_\mu J_{\text{ext}}^{\mu}=0$ is trivially satisfied. Then, we choose $\delta J_{\text{ext}}^{\,x \,\,(L)}\,, \delta J_{\text{ext}}^{\,y \,\,(L)}$ as our independent external sources.

To compute the quasi-normal modes, we are interested in the determinant of the source matrix. This is given by
\begin{align}\label{APPENSMATA3}
\begin{split}
& \det \mathcal{S} = \left|
\begin{array}{cccc} 
Z_{H_1}^{(L)(I)} & Z_{H_1}^{(L)(II)} & Z_{H_1}^{(L)(III)} & Z_{H_1}^{(L)(IV)} \\ [8pt]
Z_{H_2}^{(L)(I)} & Z_{H_2}^{(L)(II)} & Z_{H_2}^{(L)(III)} & Z_{H_2}^{(L)(IV)} \\ [8pt]
\delta J_{\text{ext}}^{\,x \,\,(L)(I)} & \delta J_{\text{ext}}^{\,x \,\,(L)(II)} & \delta J_{\text{ext}}^{\,x \,\,(L)(III)} & \delta J_{\text{ext}}^{\,x \,\,(L)(IV)} \\ [8pt]
\delta J_{\text{ext}}^{\,y \,\,(L)(I)} & \delta J_{\text{ext}}^{\,y \,\,(L)(II)} & \delta J_{\text{ext}}^{\,y \,\,(L)(III)} & \delta J_{\text{ext}}^{\,y \,\,(L)(IV)}
\end{array}  \right| \,, \\
\end{split}
\end{align}
where $(I),(II),(III),(IV)$ indicate four linearly independent solutions of the equations of motion for the fluctuations.
\\

{Before continuing, one remark is in order. In principle, the electric and magnetic susceptibilities in Eq.\eqref{SUSEMU} could be computed directly knowing the expression for the matter pressure $p_m$. In the case of holography, we are able to easily compute $\chi_{BB}$ since we dispose of a background solution at finite magnetic field. Nevertheless, we do not know how to compute the electric susceptibility $\chi_{EE}$ because introducing a background electric field will inevitably make the full solution time dependent. Therefore, in order to proceed, we will assume that $\chi_{EE}=0$. As we will see, this assumption will turn out to be a good approximation in the limit of small EM coupling, $\lambda/T \ll 1$, but not in general (see Fig. \ref{PFSLDEP} below). We cannot exclude that this might be one of the reasons behind the disagreement between the hydrodynamic predictions and the holographic results in the concomitant limit of large $B$ and large $\lambda$. Given this clarification, within this assumption}, the electric permittivity $(\epsilon_{\text{e}})$ and magnetic permeability $(\mu_{\text{m}})$ satisfy 
\begin{align}\label{HOLOPER}
\begin{split}
\epsilon_{\text{e}} \,=\, \frac{1}{\lambda} \,, \qquad \mu_\text{m} \,=\, \frac{\lambda}{1-\lambda \,\chi_{BB}} \,,
\end{split}
\end{align}
where $1/\mu_m = -2{\partial p}/{\partial B^2}$ can be computed via \eqref{HAWKINGT}. Interestingly, for the simple Reissner-Nordstrom solution considered, we find:
\begin{equation}\label{HOLOCHIFOR}
    \chi_{BB}\,=\,-\frac{1}{r_h}\,<\,0\,.
\end{equation}
This is tantamount to saying that the dual field theory is the avatar of a diamagnetic material as already noted in~\cite{Hartnoll:2008kx,Denef:2009yy,Donos:2012yu}.

\section{Results at finite electromagnetic coupling} \label{newsec}

By following the method just outlined, we are now ready to compute the low-energy excitations in our holographic model. The phase space of our system is defined by three scale invariant parameters ($\mu/T\,,B/T^2\,,\lambda/T$). 
For the moment, we mainly focus on the case of small EM interactions, $\lambda/T=0.1$. We will discuss in detail the effects of dialing the EM coupling $\lambda$ in Section~\ref{appena}. We study the quasi-normal modes at zero density ($\mu/T=0$) and finite density ($\mu/T\neq0$), separately. Moreover, to avoid clutter in the presentation of the results, for the pure imaginary dispersion relations (e.g., the shear diffusion mode in Eq.\eqref{SWSHW}), we only display the imaginary part in all the figures.\\

Unless indicated otherwise, in all the figures of this manuscript solid lines will refer to the hydrodynamic predictions as explained in Section \ref{nn}. On the contrary, colored dots will represent the numerical results obtained from the quasi-normal modes (QNMs) analysis.

\subsection{Zero density}
\begin{figure}[]
\centering
     {\includegraphics[width=7.3cm]{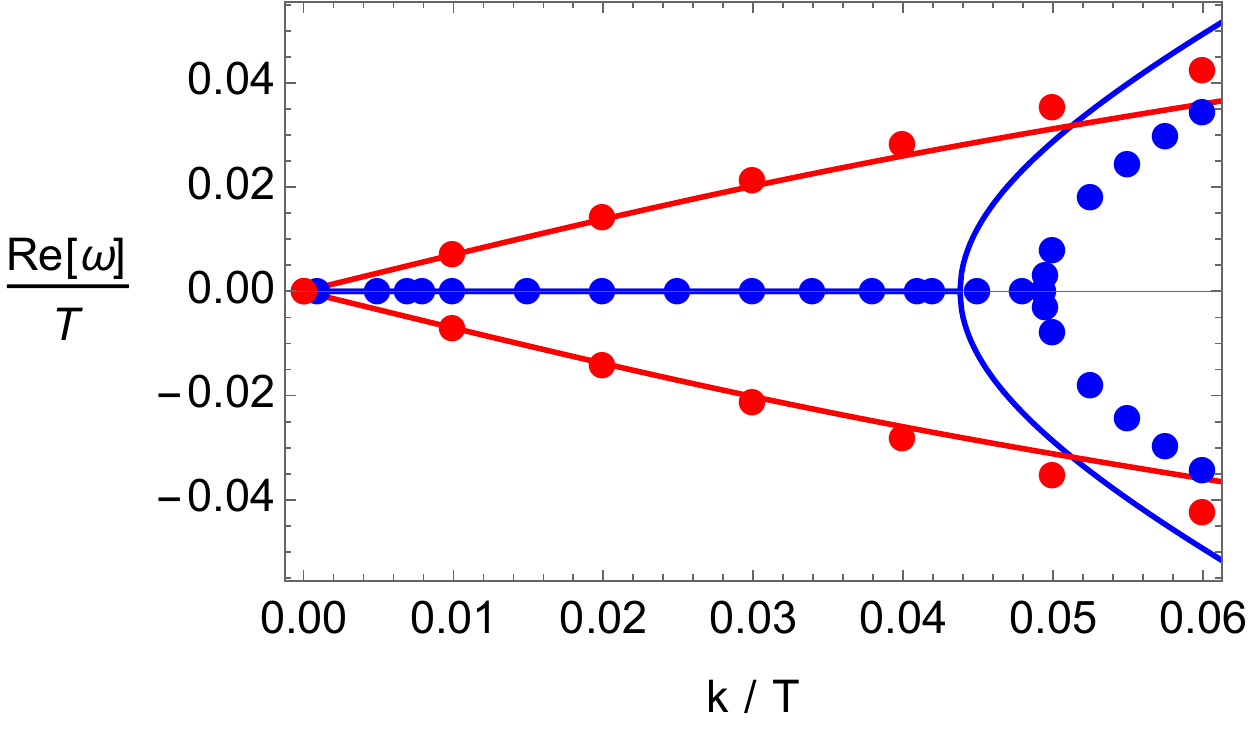} \label{zeroden1a}}
     {\includegraphics[width=7.3cm]{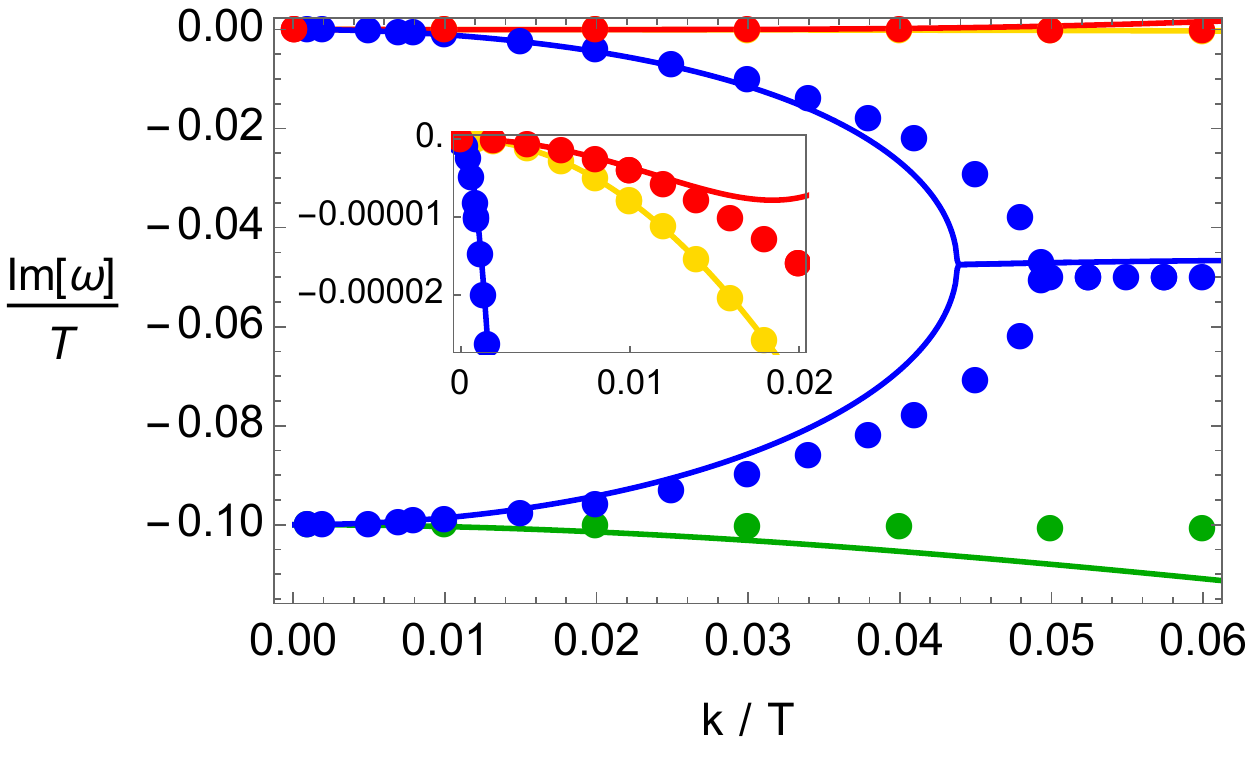} \label{zeroden1b}}
 \caption{Dispersion relations of the lowest QNMs at zero density ($\mu/T=0$) and $B/T^2=0$. }\label{zeroden1}
\end{figure}

\begin{figure}[]
\centering
     {\includegraphics[width=7.3cm]{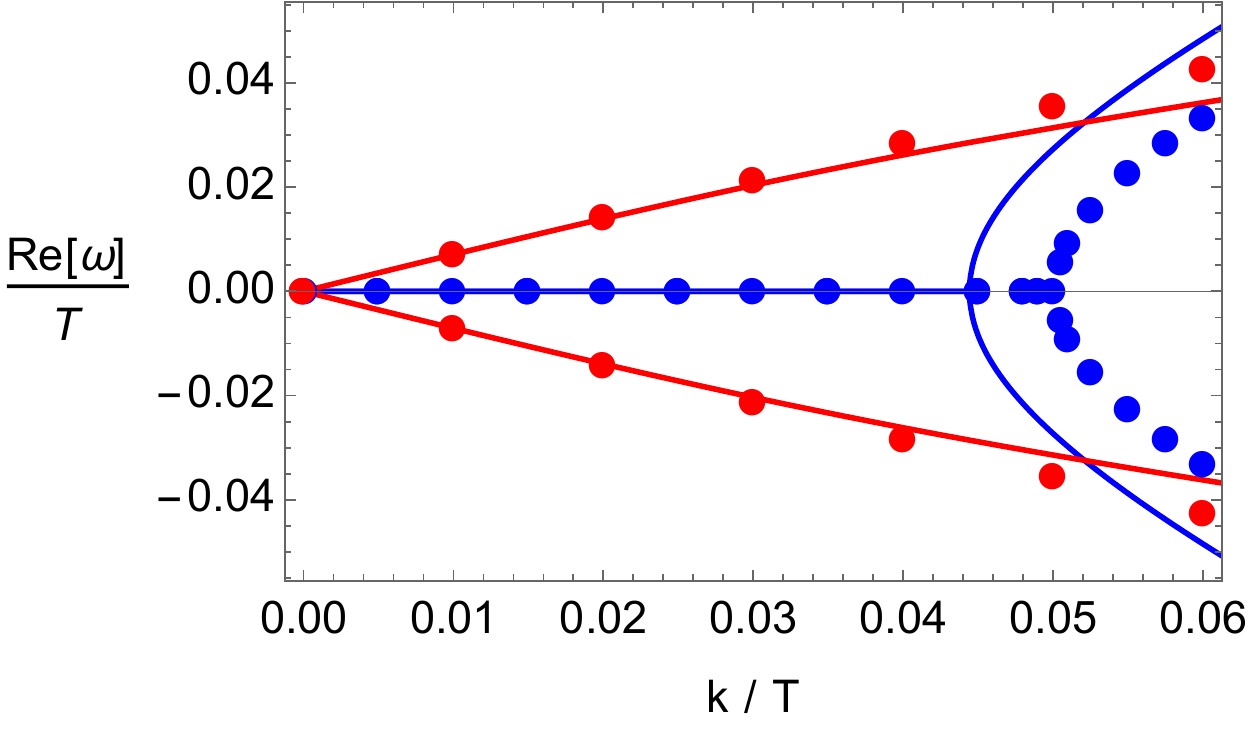} \label{}}
     {\includegraphics[width=7.3cm]{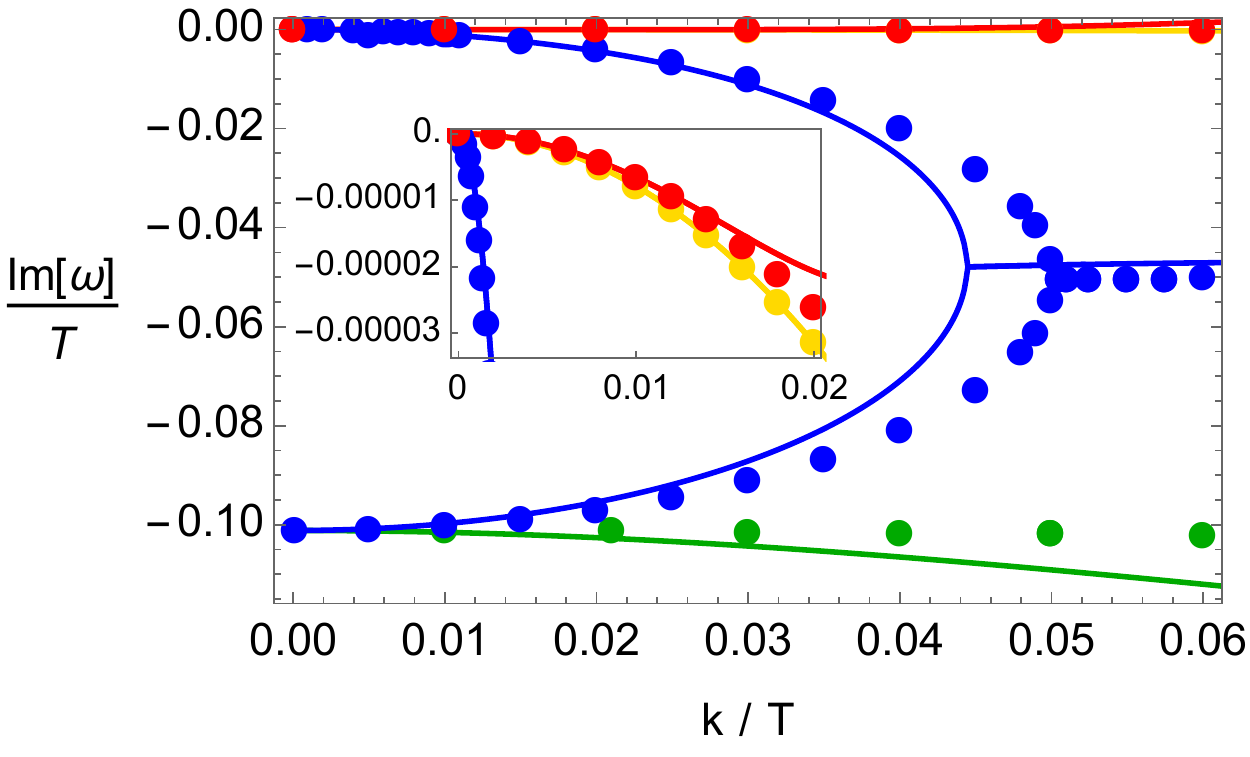} \label{}}

\vspace{0.2cm}

     {\includegraphics[width=7.3cm]{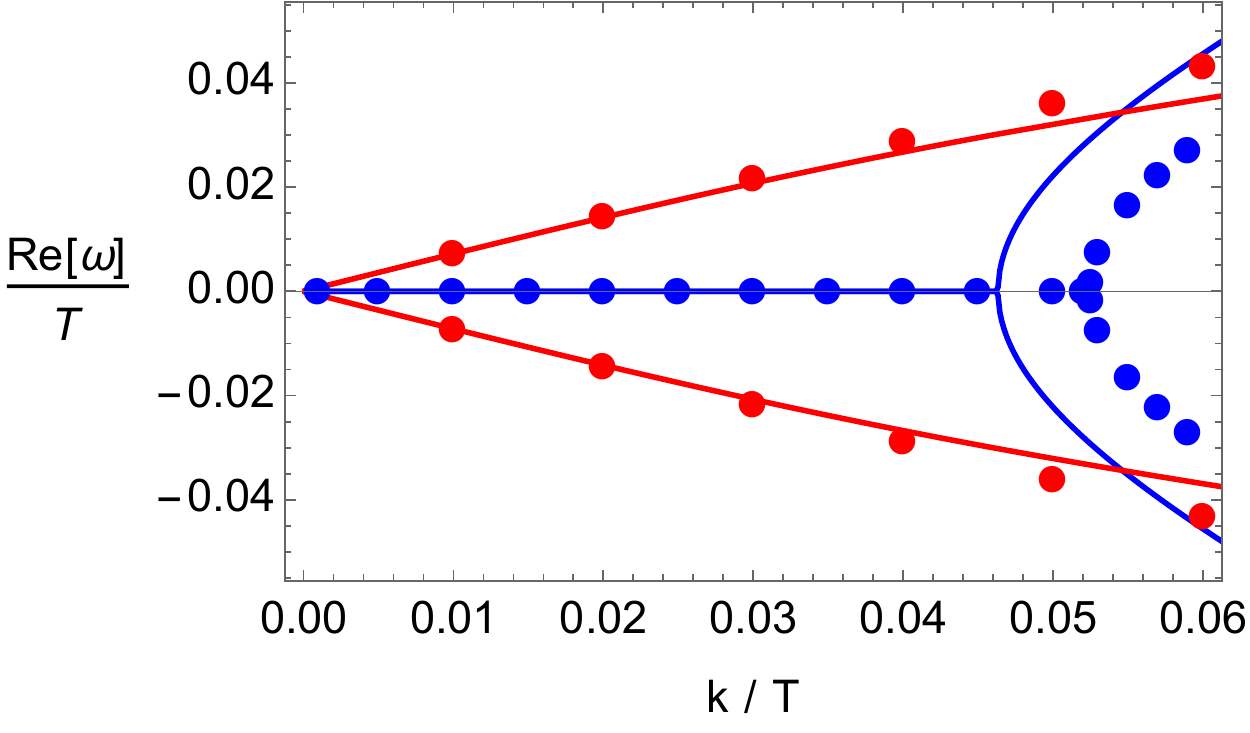} \label{}}
     {\includegraphics[width=7.3cm]{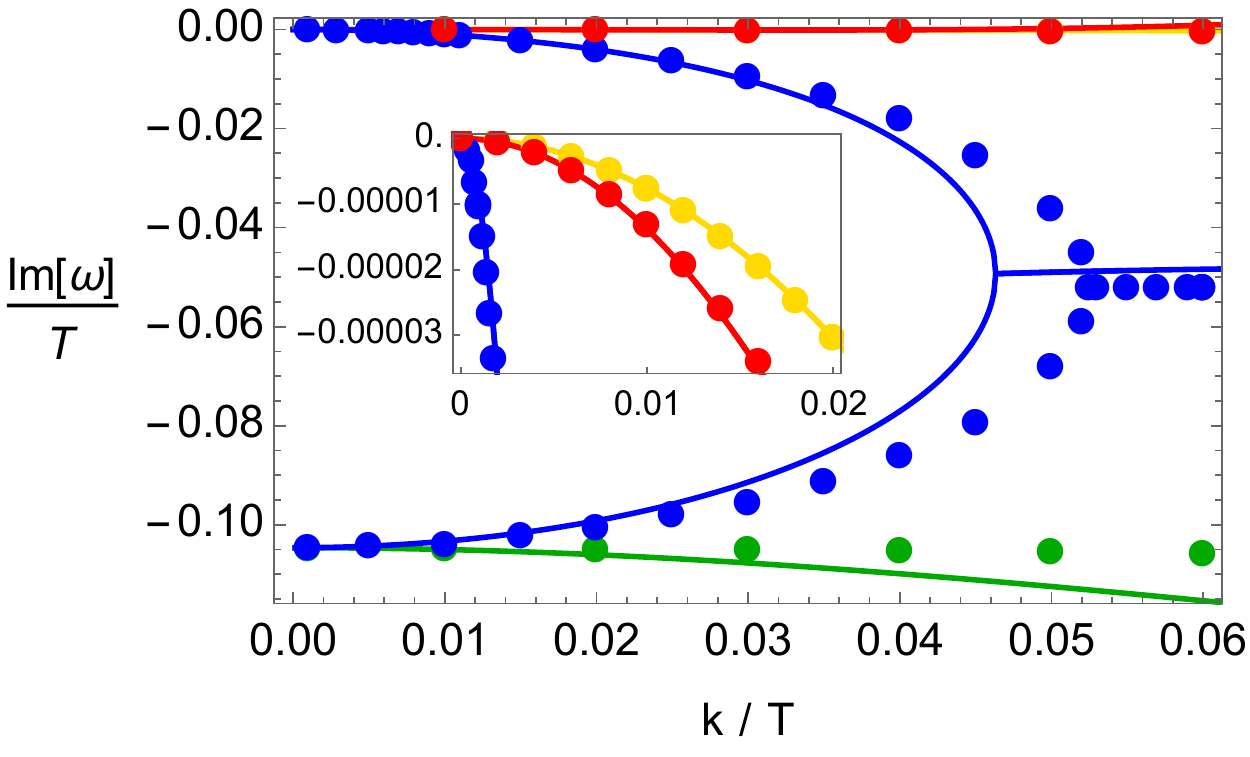} \label{zeroden3d}}
 \caption{Dispersion relations of the lowest QNMs at zero density ($\mu/T=0$) and $B/T^2 \neq0$. Top and bottom panels are respectively for $B/T^2=0.5,1$.}\label{zeroden3}
\end{figure}

Let us start from the case of zero density $\rho=\mu=0$.
We display the dispersion relation of the quasi-normal modes at zero magnetic field $B/T^2=0$ and small EM coupling $\lambda/T=0.1$ in Fig. \ref{zeroden1}. We find that the numerical results are well matched with the dispersion relations from hydrodynamics in the expected range, $k/T \ll 1$. In particular, Fig. \ref{zeroden1} presents:
\begin{itemize}
    \item the sound waves with dispersion as in Eq.\eqref{SWSHW} (red);
    \item the shear diffusion mode with dispersion as in Eq.\eqref{SWSHW} (yellow; see the inset);
    \item the EM waves as in Eq.\eqref{EMWAVE} (or \eqref{EMWAVE2}) (blue);
    \item the damped charge diffusion mode as in Eq.\eqref{DCDD} (green).
\end{itemize}
In Fig.~\ref{zeroden3}, we show the low energy modes in the case of finite magnetic field. In particular, we have:
\begin{itemize}
    \item magnetosonic waves with dispersion as in Eq.\eqref{SWSHW2} (red);
    \item shear diffusion as in Eq.\eqref{SWSHW2} (yellow);
    \item magnetic diffusion mode as in Eq.\eqref{EMWAVE3} (blue);
    \item gapped mode as in \eqref{DCDD2} (green).
\end{itemize}

The numerical results are still well fitted by the hydrodynamic formulae. The validity of the hydrodynamic framework and the match to the numerical data upon dialing the value of the magnetic field will be discussed in more detail in Section \ref{largeB}.

\subsection{Finite density}\label{FD}

From the hydrodynamic analysis of Section \ref{SECMHDDIS222}, at finite density, we do expect two gapless modes, Eq.\eqref{FINITEDEN1}, and four gapped modes, Eq.\eqref{FINITEDEN2}. In particular, depending on how large the density is, the gapped modes can exhibit distinct behaviors given by:
\begin{enumerate}
    \item[(I)] Eq.\eqref{SMDG} for small density ($\sigma^2/\epsilon_\text{e}^2 \,\gg\, 4 \, \Omega_p^2$);
    \item[(II)] Eq.\eqref{LAGDG} for large density ($\sigma^2/\epsilon_\text{e}^2 \,\ll\, 4 \, \Omega_p^2$).
\end{enumerate}
In this section, as representative examples for each case, we choose $\mu/T=0.5$ for the small density case and $\mu/T=5$ for the large density case. A more detailed discussion about the role of the chemical potential and the transition between the two regimes will be presented in Section \ref{appenb}. 

\begin{figure}[]
\centering
     {\includegraphics[width=7.3cm]{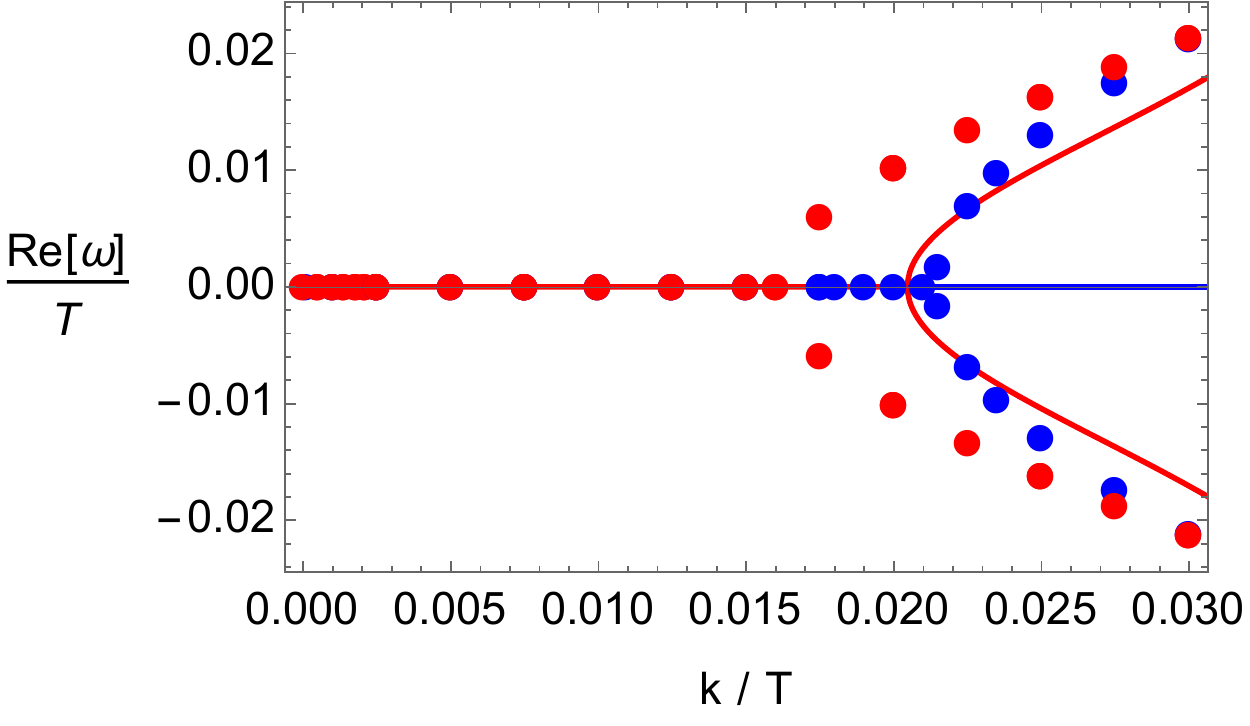} \label{finden1a}}
     {\includegraphics[width=7.3cm]{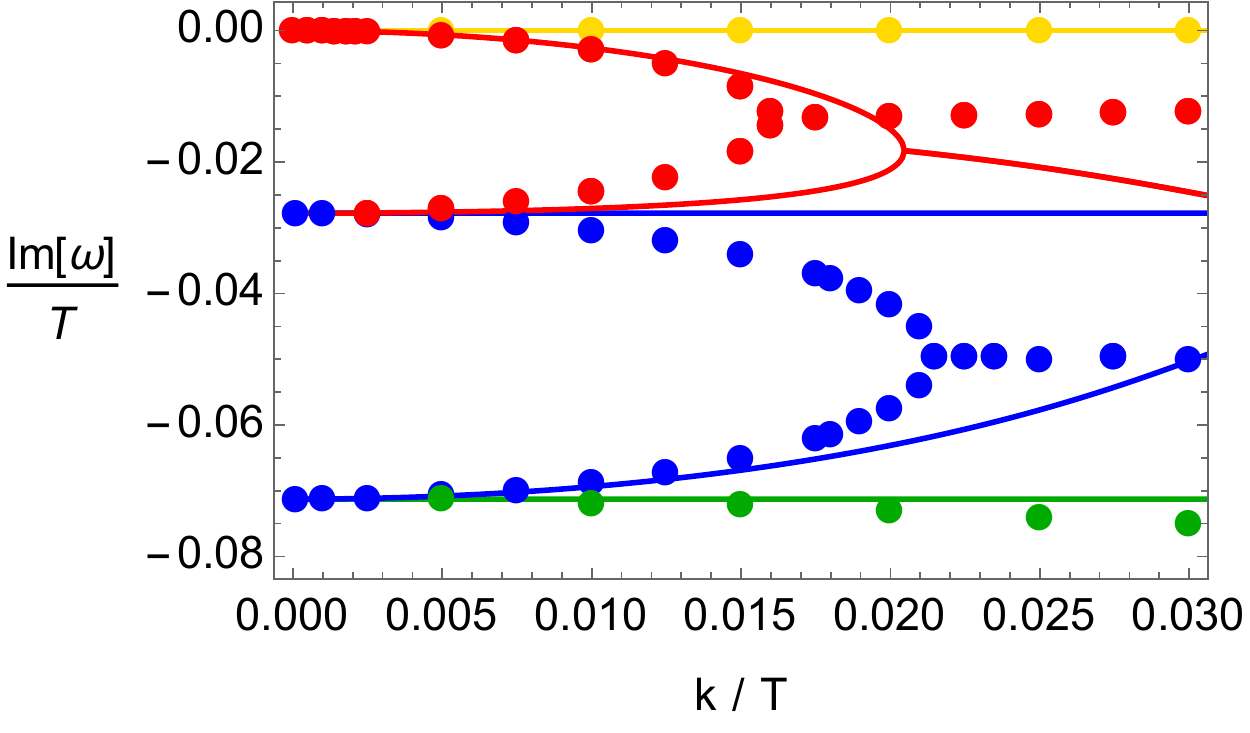} \label{finden1b}}

     \vspace{0.2cm}
     
     {\includegraphics[width=7.3cm]{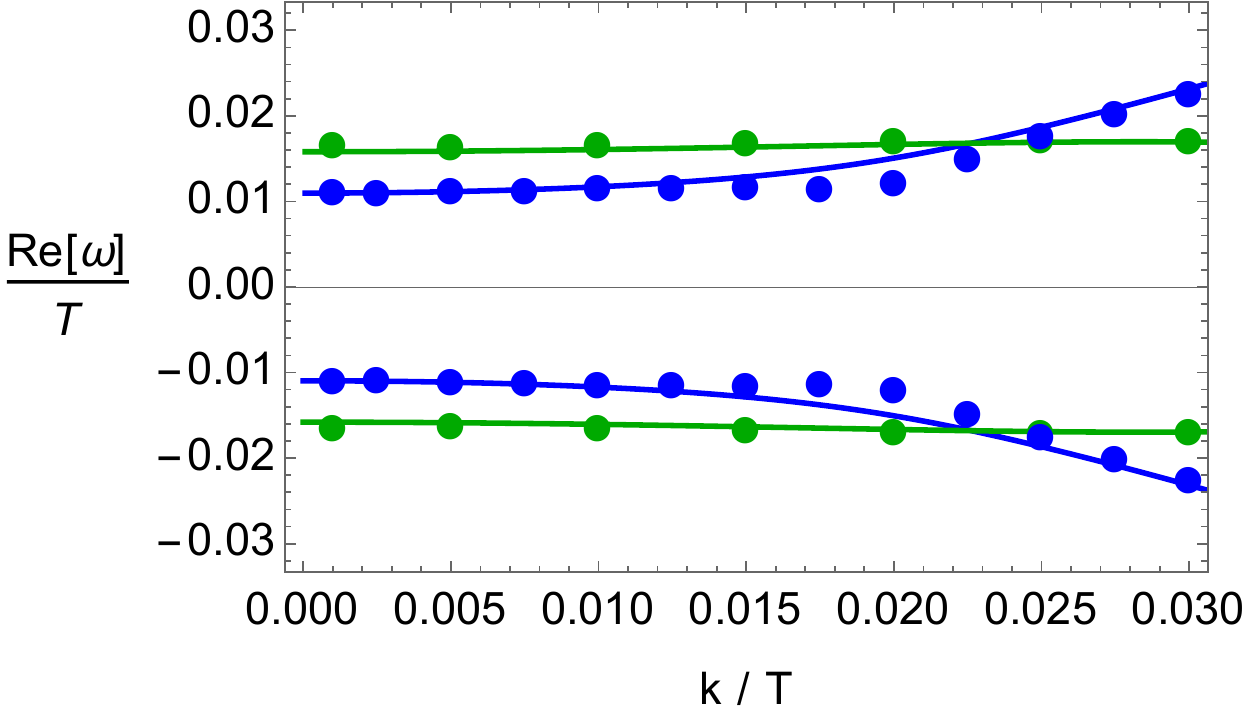} \label{finden1c}}
     {\includegraphics[width=7.3cm]{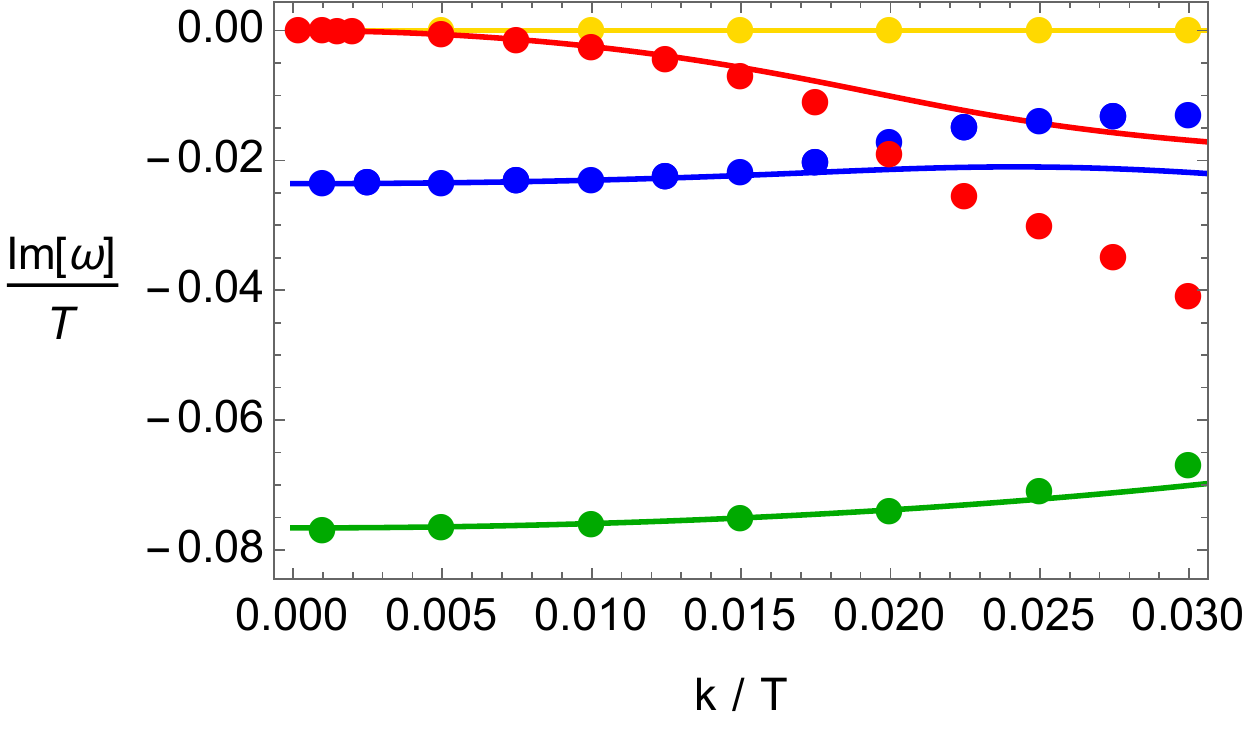} \label{finden1d}}
 \caption{Dispersion relations of the lowest QNMs at finite density ($\mu/T=0.5$). Top and bottom panels refer respectively to $B/T^2=0,0.5$.}\label{finden1}
\end{figure}

In Fig. \ref{finden1}, we display the quasi-normal modes at $\mu/T=0.5$ both for zero (top panel) and finite, but small, magnetic field (bottom panel).
In both cases, the red data correspond to the diffusive mode in Eq.\eqref{FINITEDEN1}, the yellow data to the subdiffusive mode Eq.\eqref{FINITEDEN1} and the green/blue data to the gapped modes Eq.\eqref{SMDG}. The strongest effect of the finite magnetic field appears in the gapped mode, Eq.\eqref{SMDG}. There, $B$ produces a real gap which is absent for $B=0$ (see the difference between top and bottom panels). In the case of small charge (top panel of Fig.\ref{finden1}), first order hydrodynamics is able to generally predict the existence of the k-gap but not its location accurately if that is large (in units of $k/T$). To be more precise, the k-gap curve appears well fitted by hydrodynamics only when the momentum gap is small, i.e. in the so-called quasihydrodynamic regime \cite{Grozdanov:2018fic} (see Fig. \ref{zeroden2} later in the discussion). 

\begin{figure}[]
\centering
     \includegraphics[width=7.2cm]{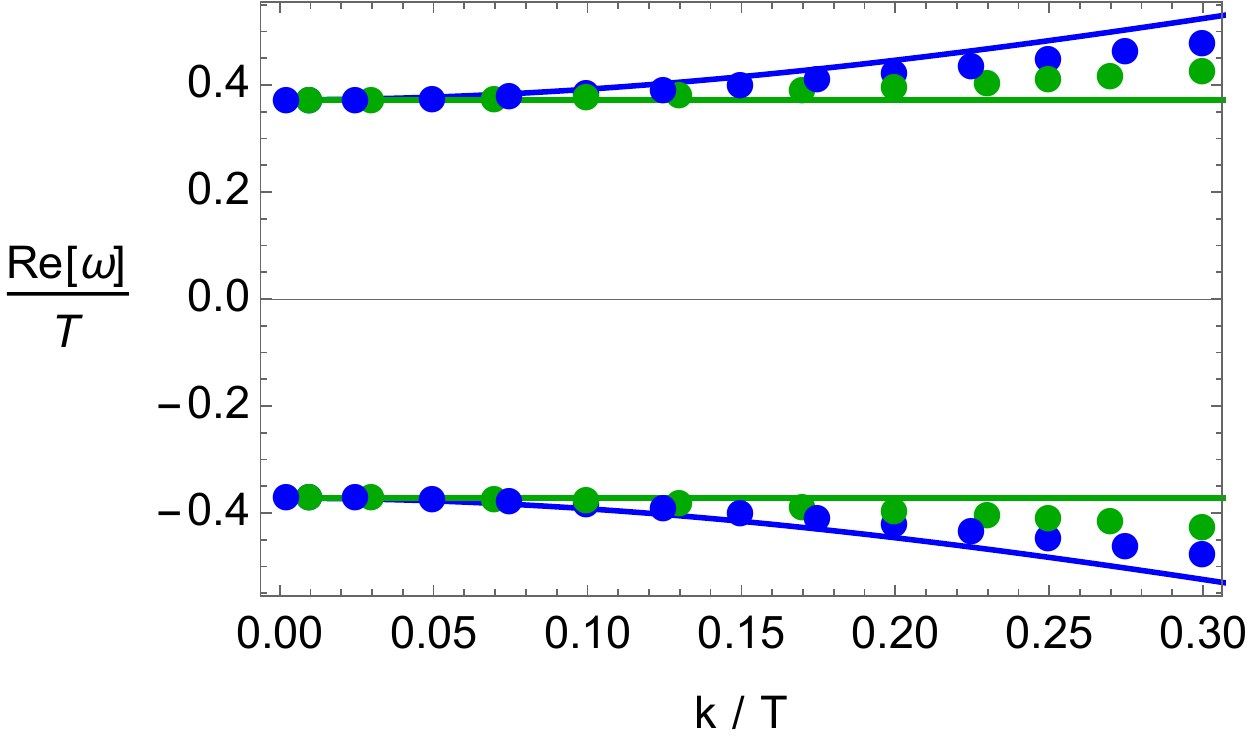}
     \includegraphics[width=7.4cm]{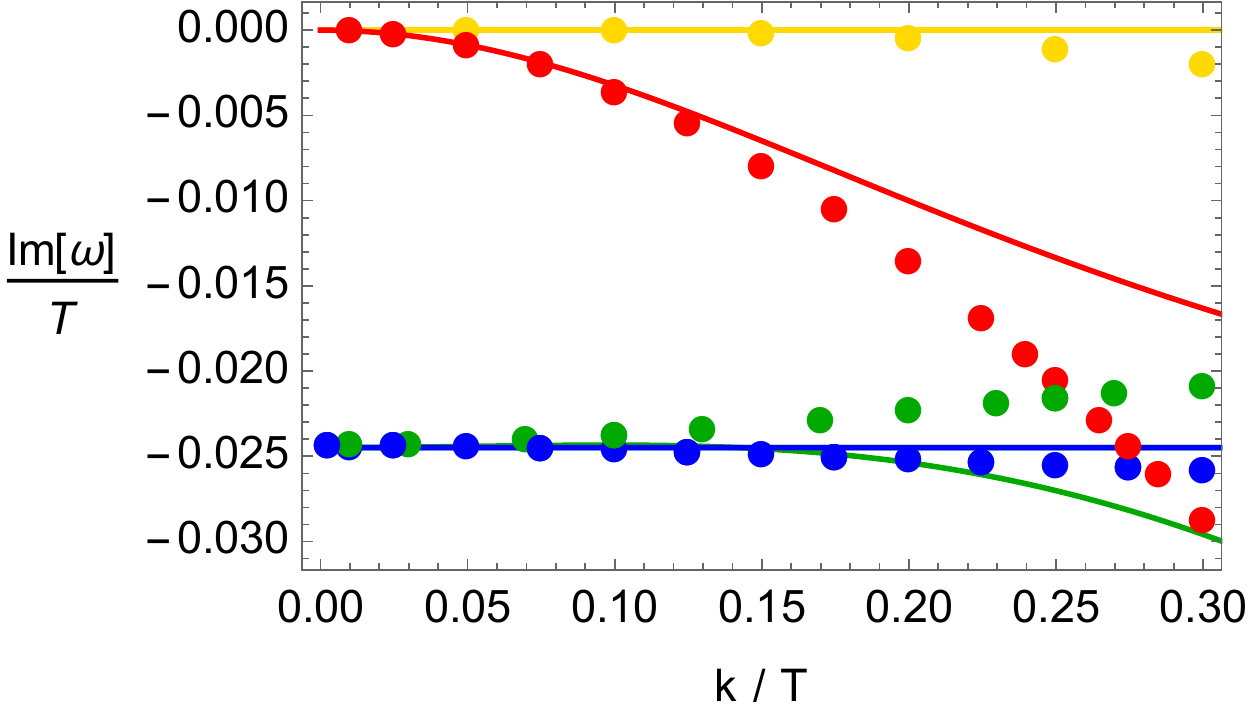}
     
     \includegraphics[width=7.2cm]{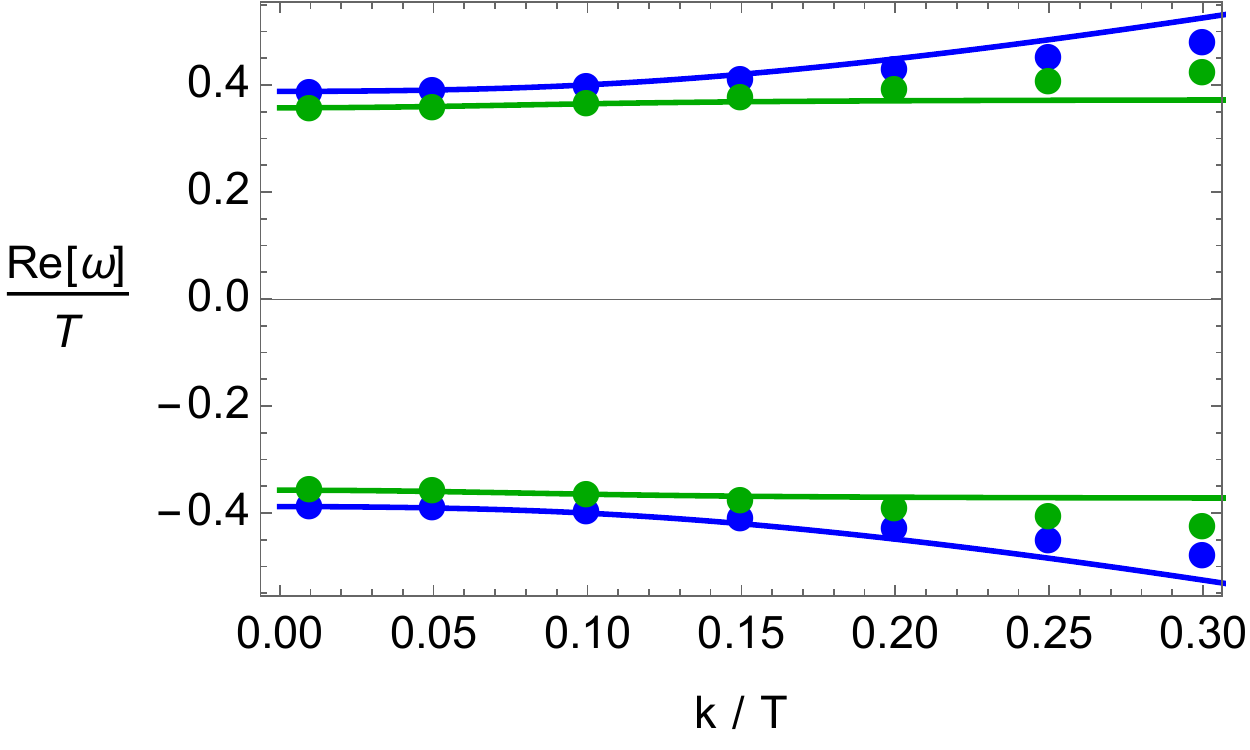} 
     \includegraphics[width=7.4cm]{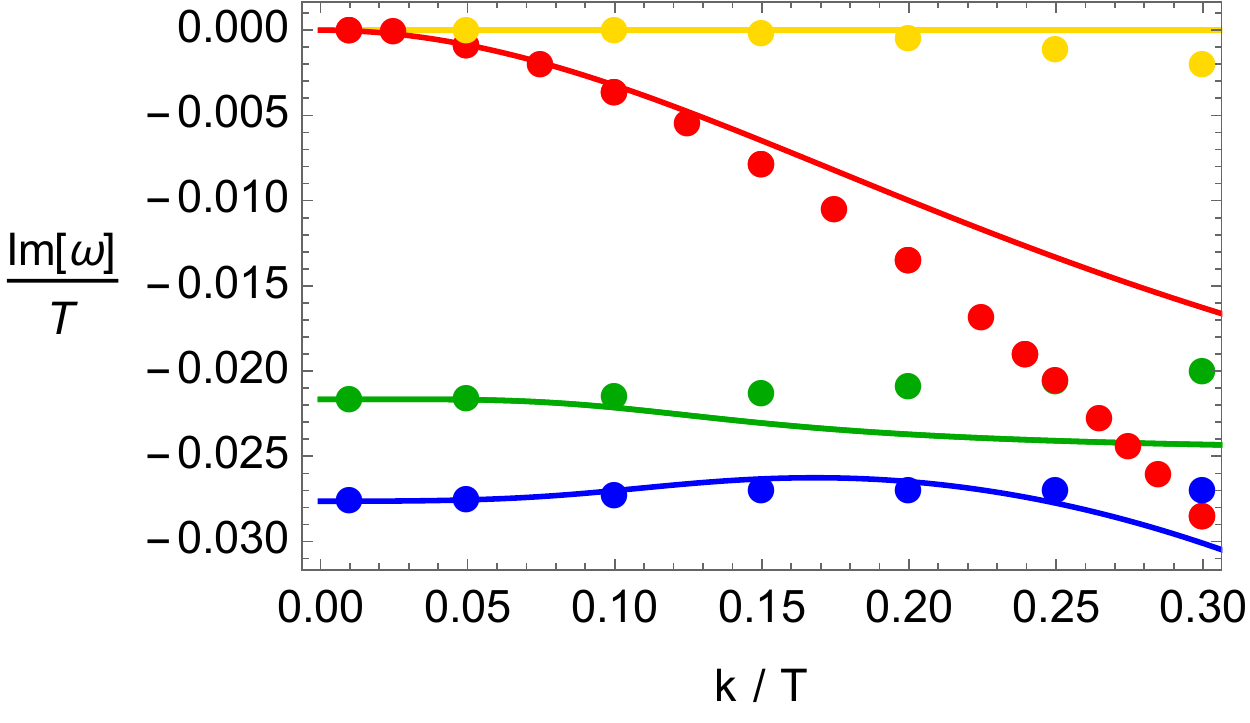} 
 \caption{Lowest QNMs at finite density ($\mu/T=5$). Top and bottom panels refer respectively to $B/T^2=0,0.5$.}\label{finden2}
\end{figure}

In Fig. \ref{finden2}, we show the quasi-normal modes at large chemical potential for both zero and finite magnetic field (respectively top and bottom panels therein). In all figures, the red data correspond to the diffusive mode in \eqref{FINITEDEN1}, the yellow data to the subdiffusive mode \eqref{FINITEDEN1} and finally the green/blue data to the gapped modes \eqref{LAGDG}. Note that the dispersion of the gapped modes is now given by Eq.\eqref{LAGDG}.

At $B=0$, the mode in Eq.\eqref{LAGDG} exhibits a real gap which was absent in the case of zero charge density (Eq.\eqref{SMDG}). The value of the gap corresponds to the plasma frequency $\text{Re}(\omega) = \pm \Omega_p$.
Additionally, at finite charge density, the magnetic field $B$ contributes to both the real and imaginary parts in the limit of zero wave-vector.

In summary, using mixed boundary conditions as explained in the previous sections, all the quasi-normal modes and their dispersion relations are consistent with the expectations from magnetohydrodynamics at low wave-vector. This is the concrete proof that the modified boundary conditions are indeed rendering the gauge field at the boundary dynamical and that the boundary physics is accurately described by relativistic magnetohydrodynamics.

%
\subsection{The effects of a finite chemical potential}\label{appenb}

In the previous sections, we did not explore in detail the role of the chemical potential on the dispersion relations of the low-energy modes.
Here, we present additional results about the $\mu$ dependence at fixed magnetic field. For this purpose, we fix $\lambda/T=0.1$ and $B/T^2=0$.

In order to proceed with this analysis, we first distinguish the two regimes of small and large chemical potential as
\begin{align}\label{ABOVEAPPB}
\begin{split}
\text{(I)} \quad \sigma^2/\epsilon_\text{e}^2  \,>\,  4 \, \Omega_p^2 \,, \qquad \text{(II)} \quad\sigma^2/\epsilon_\text{e}^2  \,<\,  4 \, \Omega_p^2 \,.
\end{split}
\end{align} 
For $\lambda/T=0.1$ and $B/T^2=0$, the same inequalities can be directly expressed as follows
\begin{align}\label{MURAG}
\begin{split}
\text{(I)} \quad 0 \,<\, \mu/T \,\lesssim\, 0.56\,, \qquad \text{(II)} \quad \mu/T  \,\gtrsim\,  0.56 \,,
\end{split}
\end{align} 
where $\mu/T\sim0.56$ corresponds to $\sigma^2/\epsilon_\text{e}^2  \,\sim\,  4 \, \Omega_p^2$.

\begin{figure}[]
\centering
     \includegraphics[width=4.8cm]{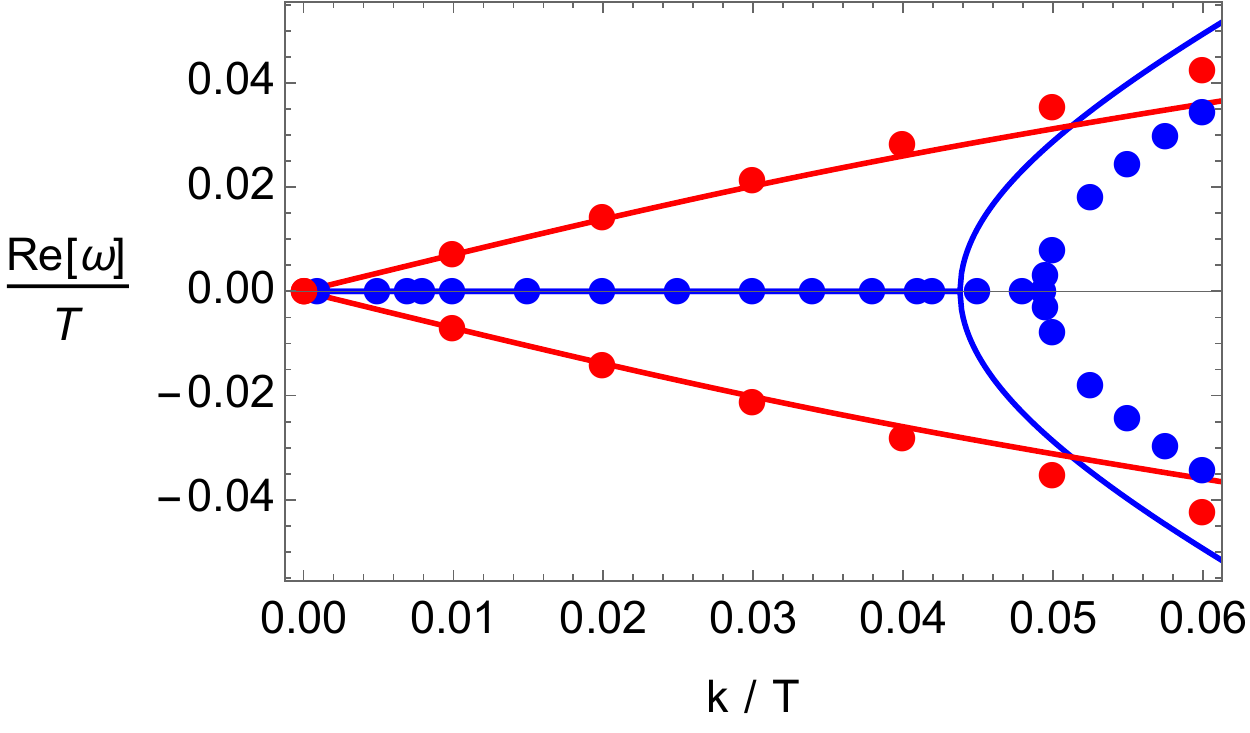} 
     \includegraphics[width=4.8cm]{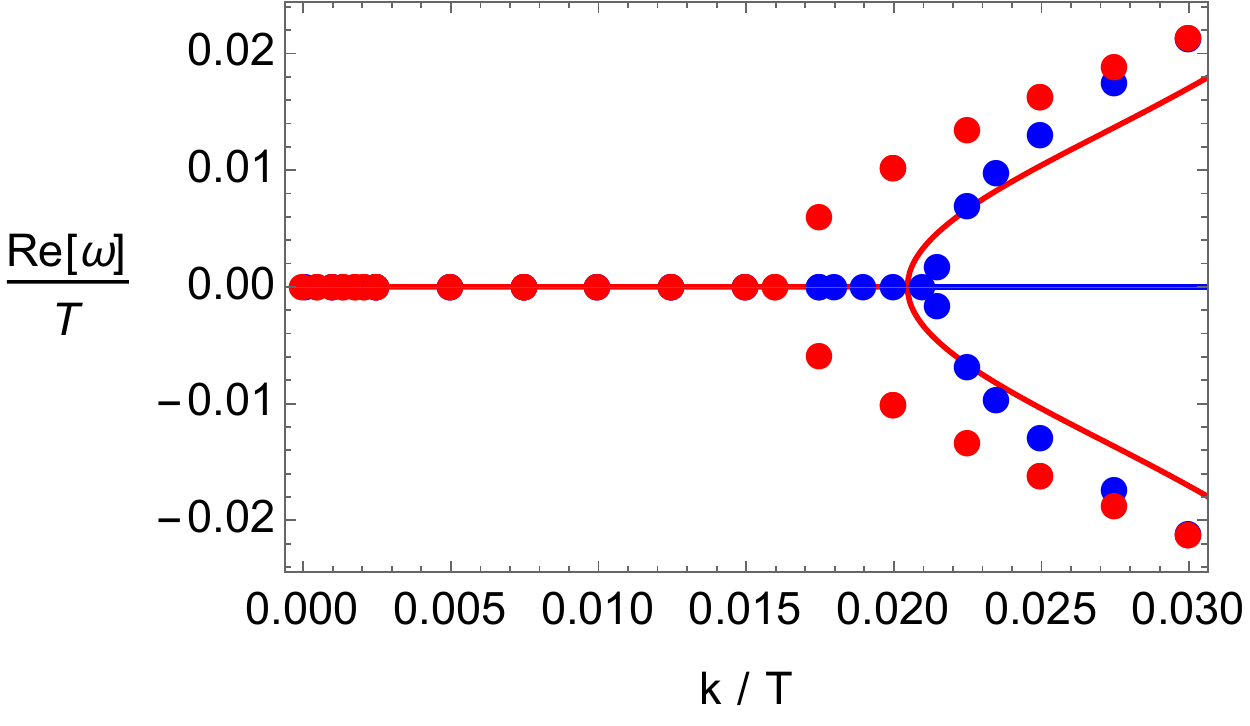} 
     \includegraphics[width=4.8cm]{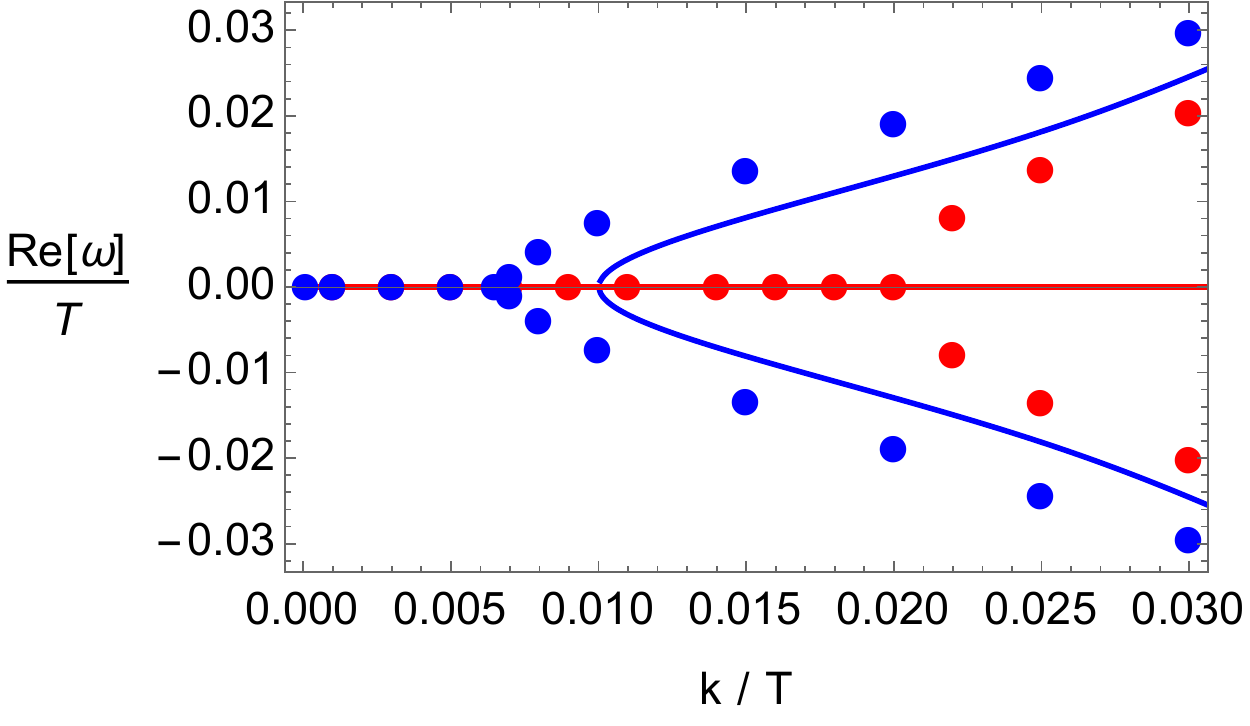} 
     
     \includegraphics[width=4.8cm]{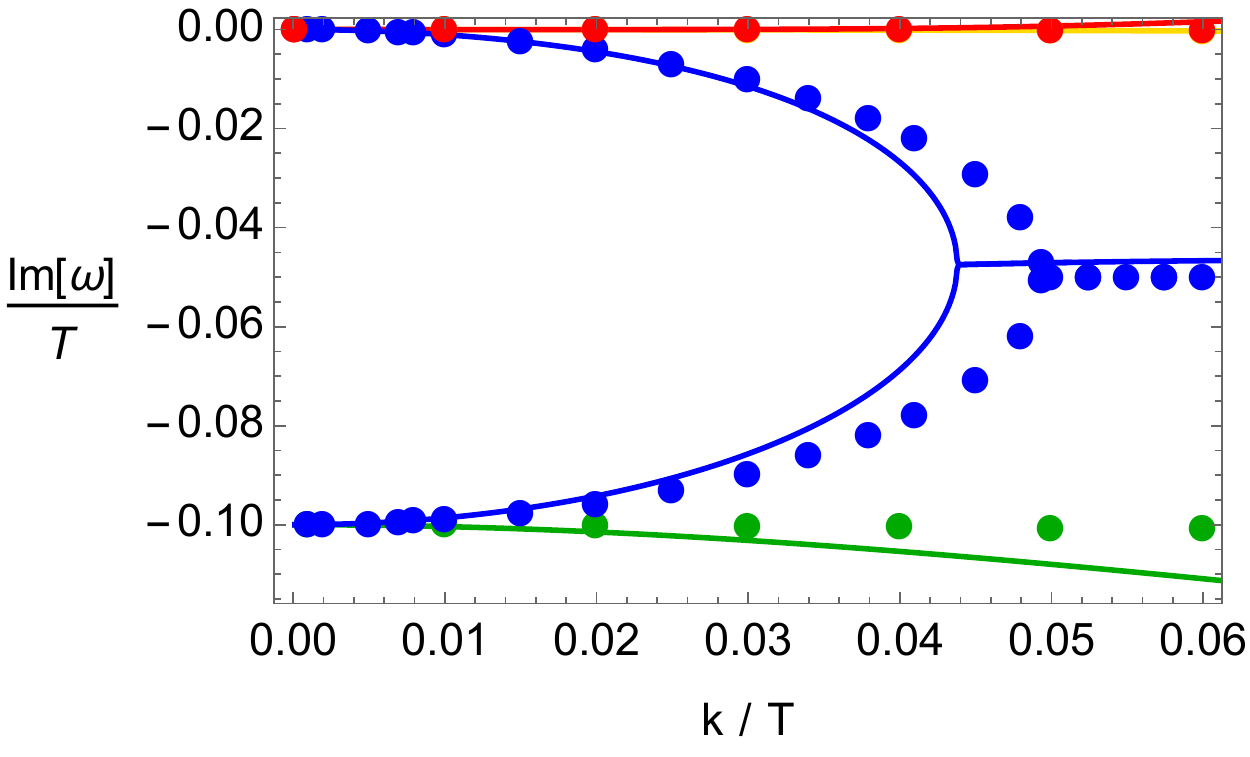} 
     \includegraphics[width=4.8cm]{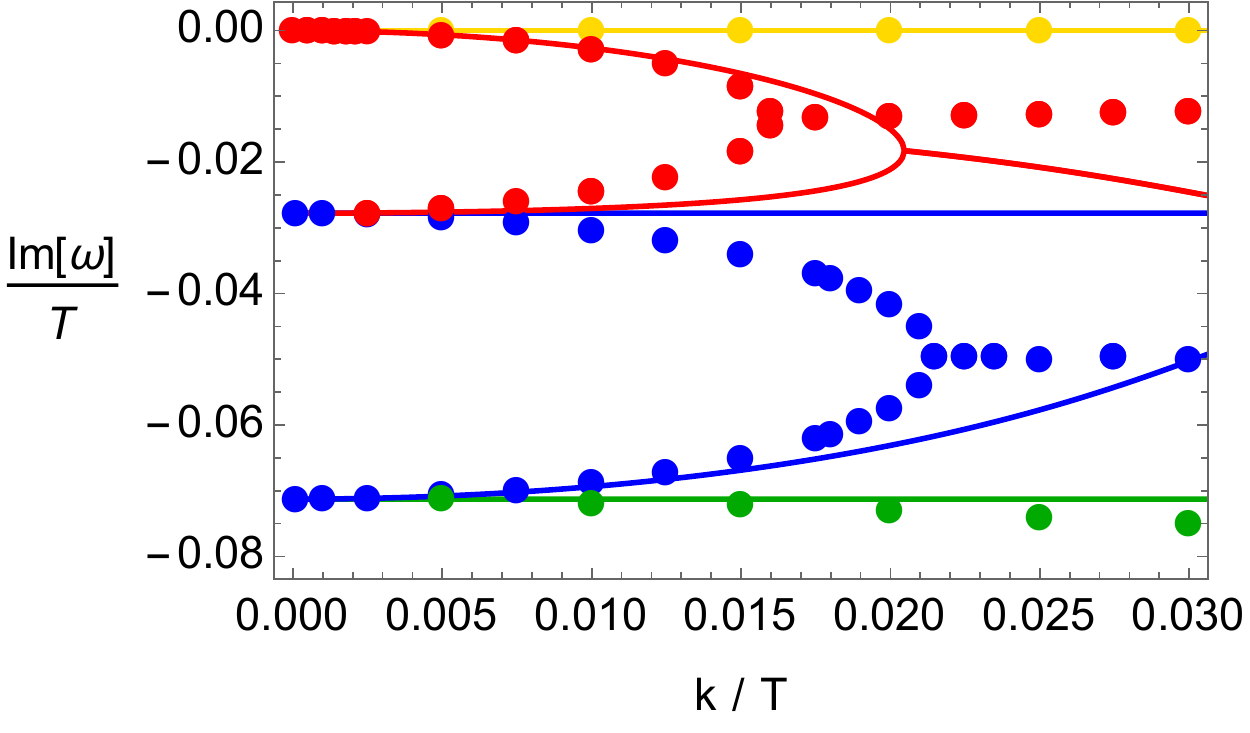} 
     \includegraphics[width=4.8cm]{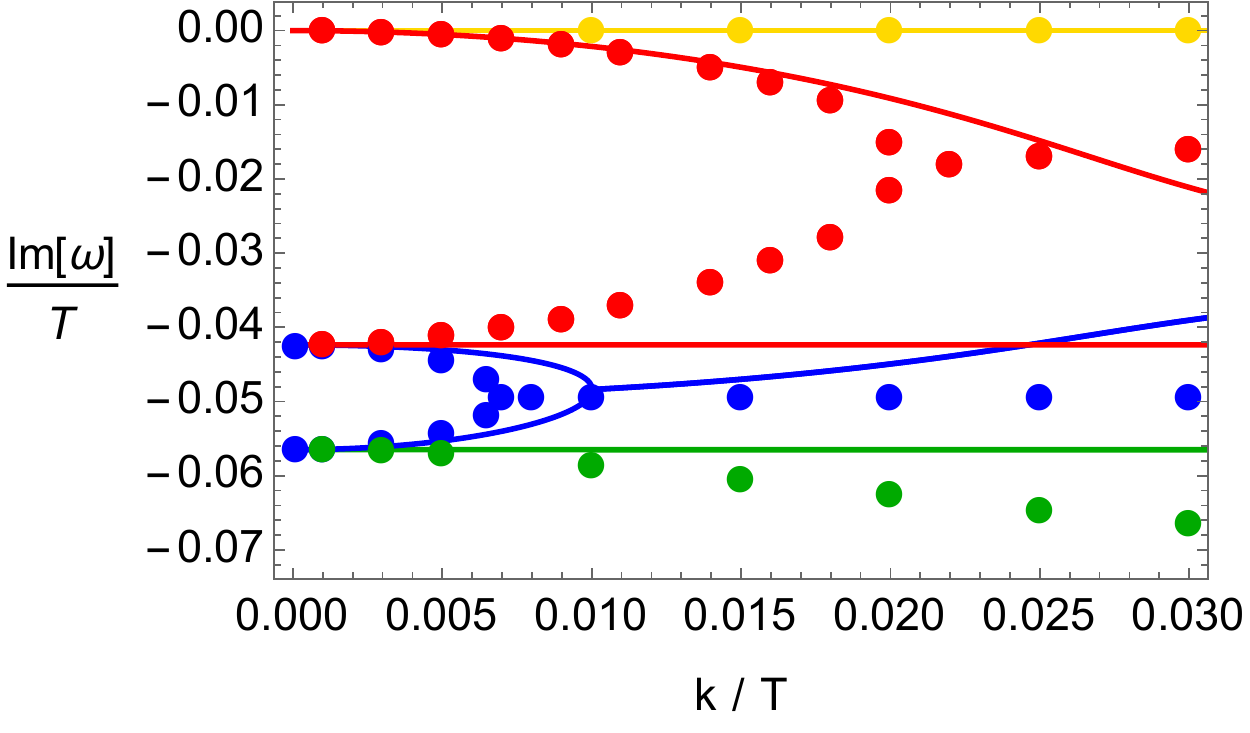}

 \caption{Dispersion relation of the low-energy modes. Left, center and right panels correspond respectively to $\mu/T=0,0.5,0.55$.}\label{APPBF1}
\end{figure}

Let us start with the first regime of small chemical potential. In Fig. \ref{APPBF1}, we display the dispersion relations of the low-energy excitations at finite wave-vector.

\begin{figure}[]
\centering
     
     \includegraphics[width=4.8cm]{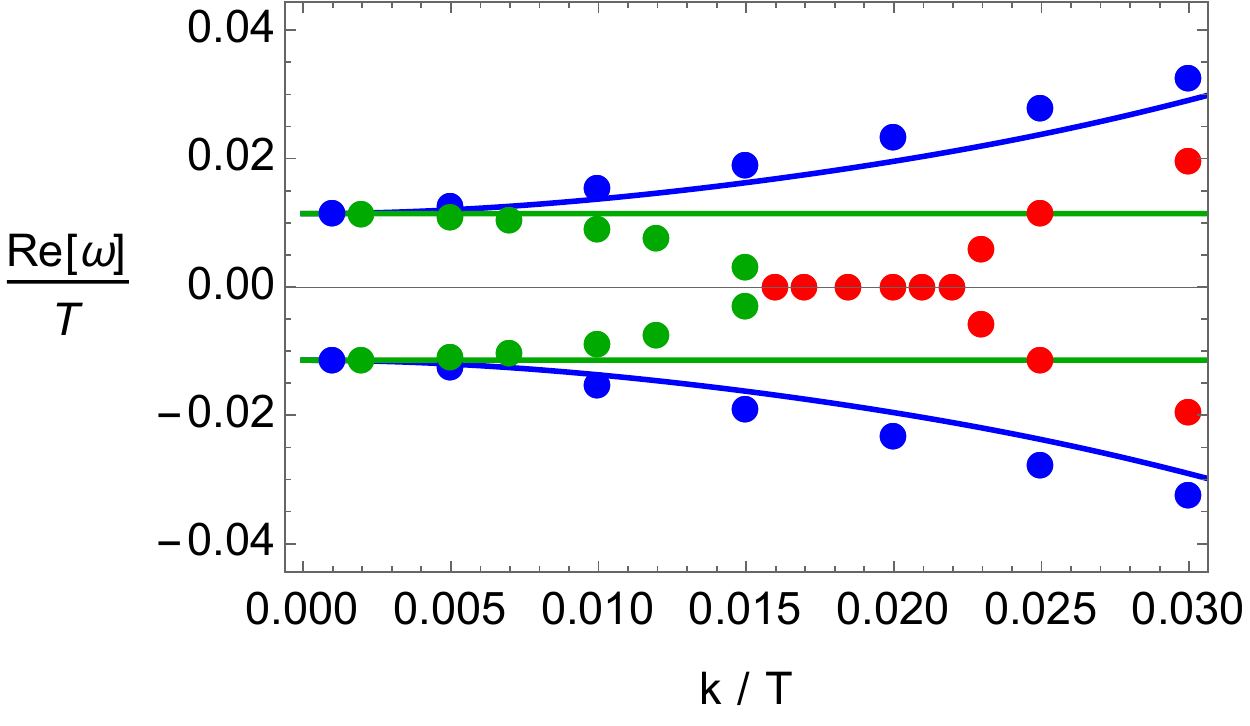} 
    \includegraphics[width=4.8cm]{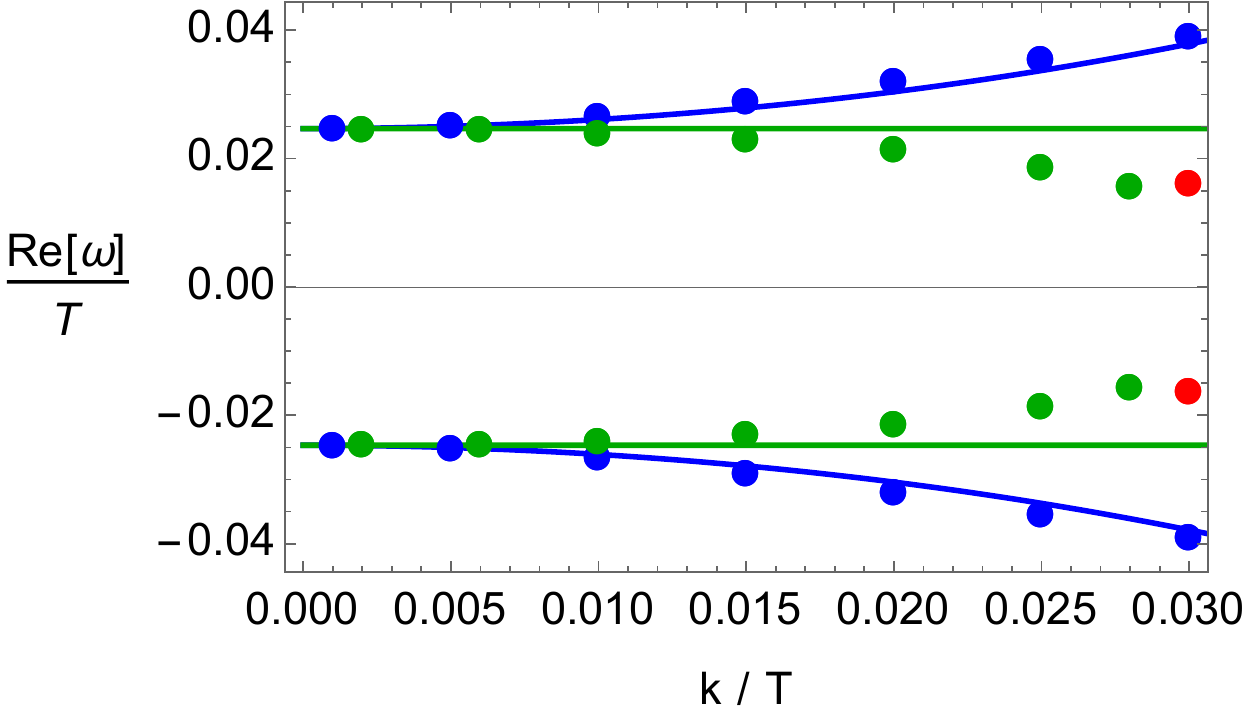} 
    \includegraphics[width=4.8cm]{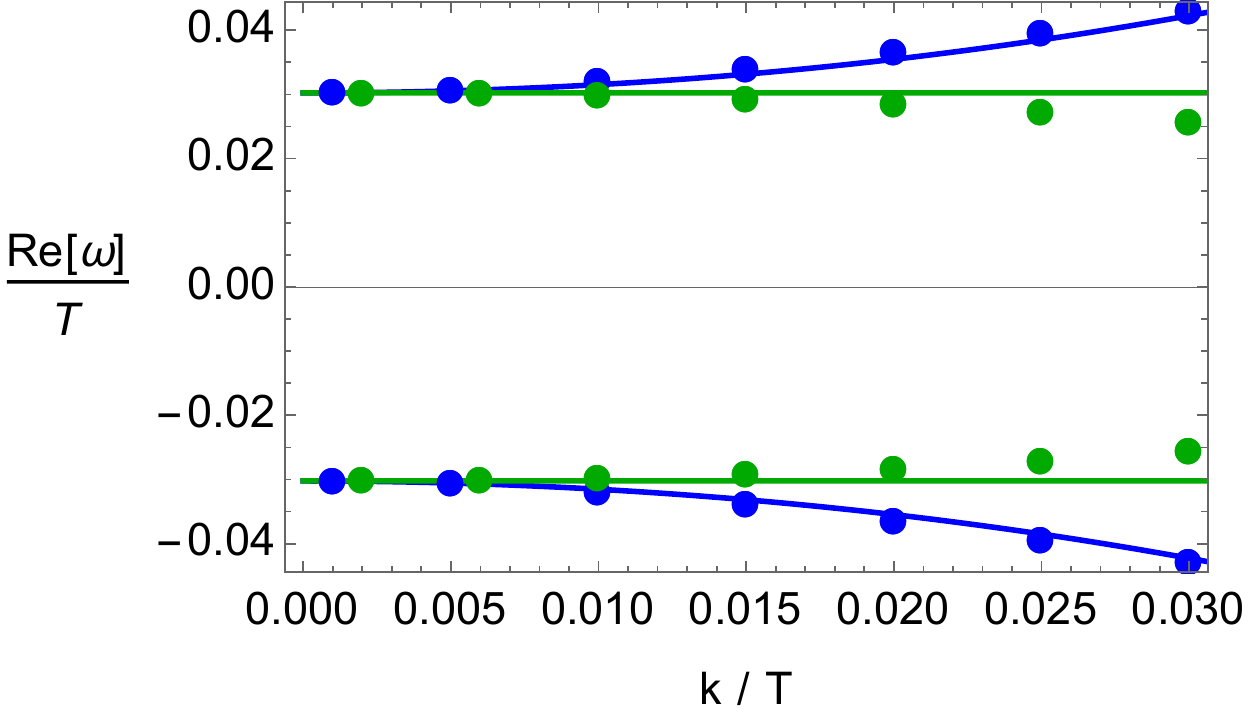}

     \includegraphics[width=4.8cm]{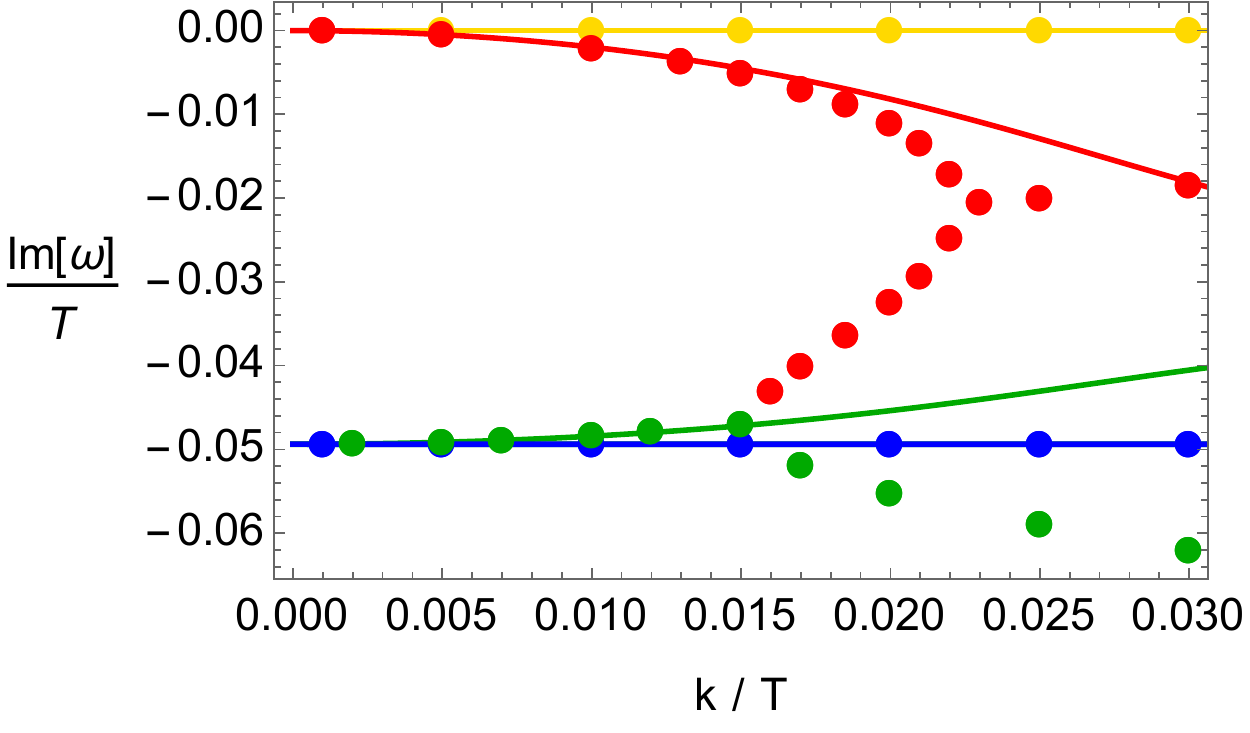} 
    \includegraphics[width=4.8cm]{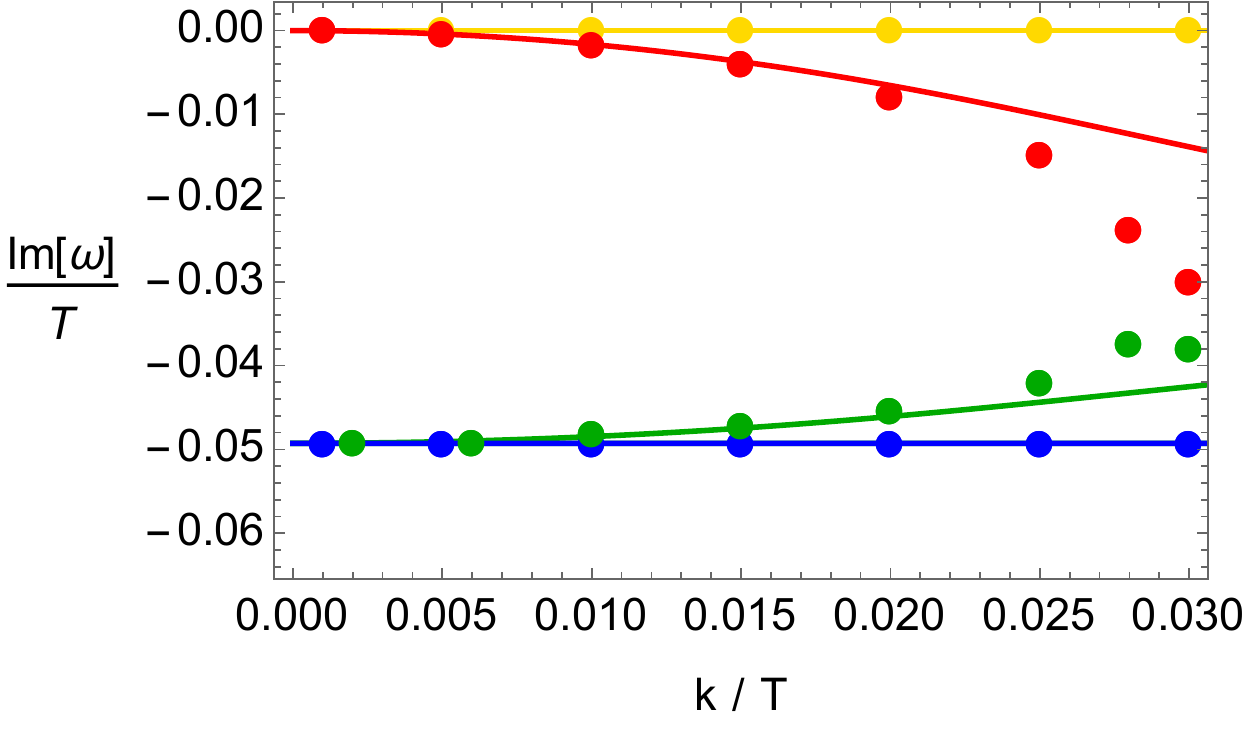} 
     \includegraphics[width=4.8cm]{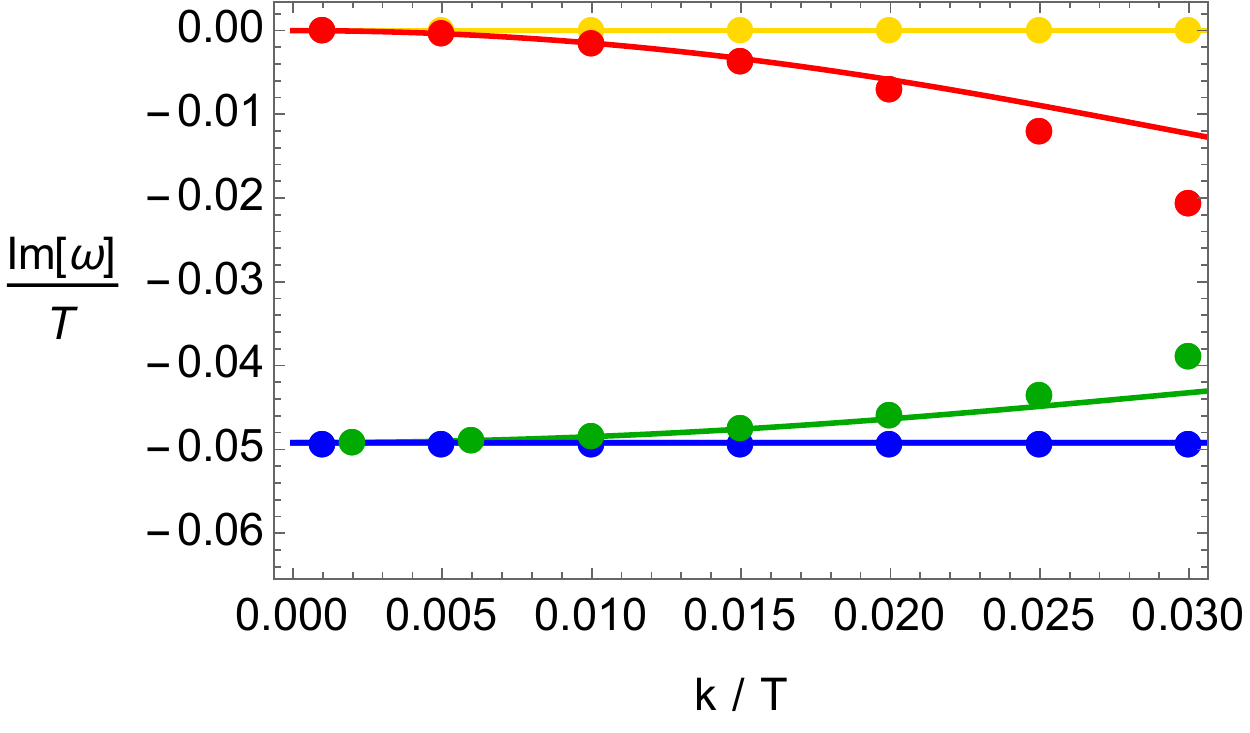} 
 \caption{Dispersion relation of the lowest QNMs. Left, center and right panels correspond respectively to $\mu/T=0.57,0.62,0.65$.}\label{bb}
\end{figure}

At zero chemical potential (left panel of Fig. \ref{APPBF1}), we observe a gapless sound mode together with the diffusive mode which substitutes the propagating EM wave. The left panel of Fig. \ref{APPBF1} can also be found in Fig. \ref{zeroden3}. By making the chemical potential finite, also the sound mode acquires a wave-vector gap and stops propagating at small $k$. The non-hydrodynamic modes (see for example the central panel in Fig. \ref{APPBF1}) are located, in the limit of small density, at:
\begin{align}\label{GAPSDENSI22}
\begin{split}
 \omega &\,=\, -\, i \, \frac{\epsilon_{\text{e}}}{\sigma} \, \Omega_p^2 \,, \qquad\, \omega  \,=\,  - i  \, \frac{\sigma}{\epsilon_{\text{e}}} \,+\, i \, \frac{\epsilon_{\text{e}}}{\sigma} \, \Omega_p^2   \,, 
\end{split}
\end{align} 
as derived in Eq.\eqref{GAPSDENSI}. As we increase $\mu$ further, the two values in Eq.\eqref{GAPSDENSI22} gets closer and the different non-hydrodynamic modes approach each other on the negative imaginary frequency axes (see right panel in Fig. \ref{APPBF1}). Exactly at the critical value, $(\mu/T)^* \sim 0.56$ , all the imaginary parts of the non-hydrodynamic modes are equal and given by $i \omega \,=\, \, \sigma/(2 \epsilon_{\text{e}})$. Moreover, increasing the charge density the k-gap of EM waves becomes smaller, while its imaginary part at large wave-vector remains constant. At the critical value, the k-gap becomes exactly zero and after that a real gap appears. The dispersion relations of the modes above the critical value $(\mu/T)^*\sim 0.56$ are shown in Fig. \ref{bb}.

Indeed, as we increase $\mu$ further, beyond the critical value $\mu/T \sim 0.56$, all the lowest non-hydrodynamic modes discussed before acquire a finite real gap and show the same dispersion
\begin{align}\label{GAPSDENSI33}
\begin{split}
\omega &\,=\,  \pm\,\, \Omega_p -\, i \, \frac{\sigma}{2 \epsilon_{\text{e}}}   \,,
\end{split}
\end{align} 
at zero wavevector, $k=0$, as already discussed in Eq.\eqref{GAPSDENSI}. This behavior is confirmed by the numerical data in Fig. \ref{bb}.
%

\subsection{The strong magnetic field limit}\label{largeB}
Now, we extend our analysis to the regime of strong magnetic field, $B/T^2\gg1$. 
Our main scope is to better understand the regime of validity of the magnetohydrodynamic description beyond the small $B$ limit. 
Here, we present the results at fixed $\lambda/T=0.1$.

For this purpose, we first investigate the thermodynamic parameters ($\epsilon, p, s$) and the transport coefficients ($\sigma, \eta$, diffusion constant, etc) appearing in the magnetohydrodynamic description. We recall that $p$ is the thermodynamic pressure and does not correspond to the mechanical pressure.



\subsubsection{Thermodynamics and transport coefficients}\label{SECEOS}

\begin{figure}[]
\centering
     \includegraphics[width=0.48 \linewidth]{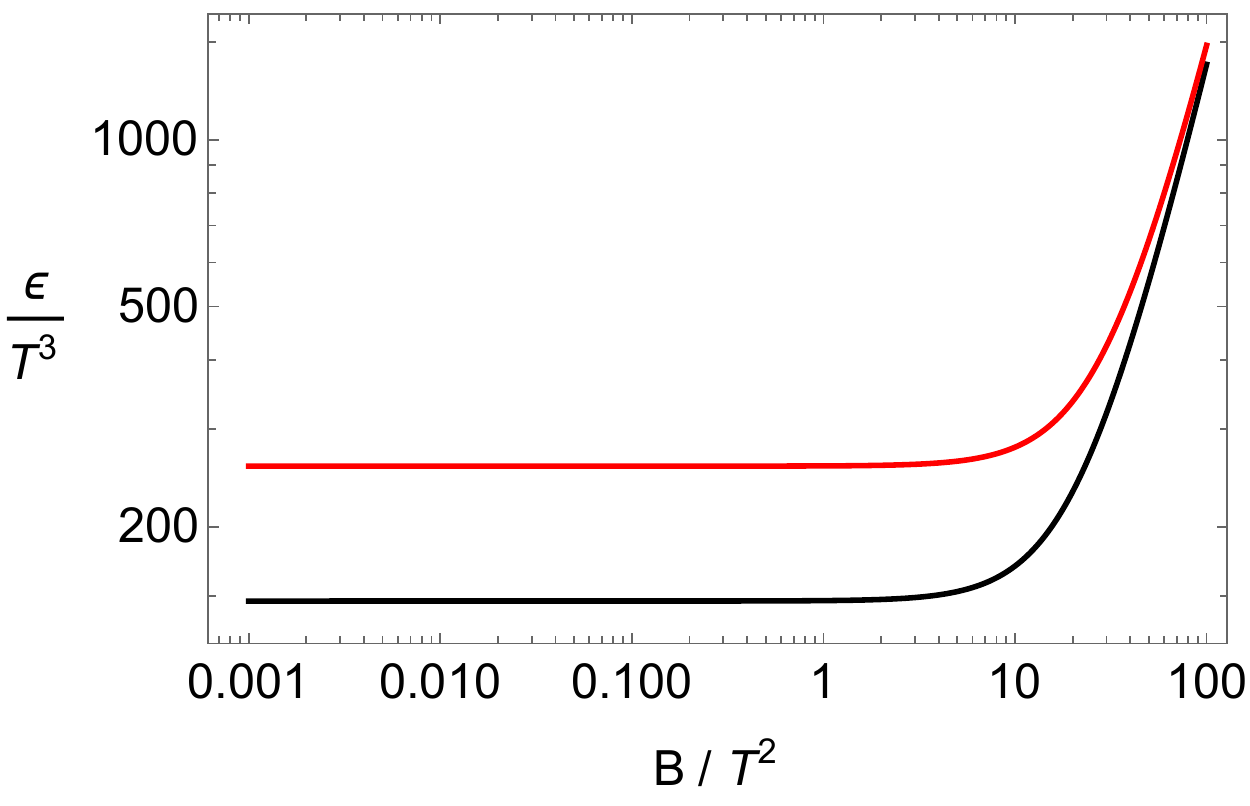} \includegraphics[width=0.47 \linewidth]{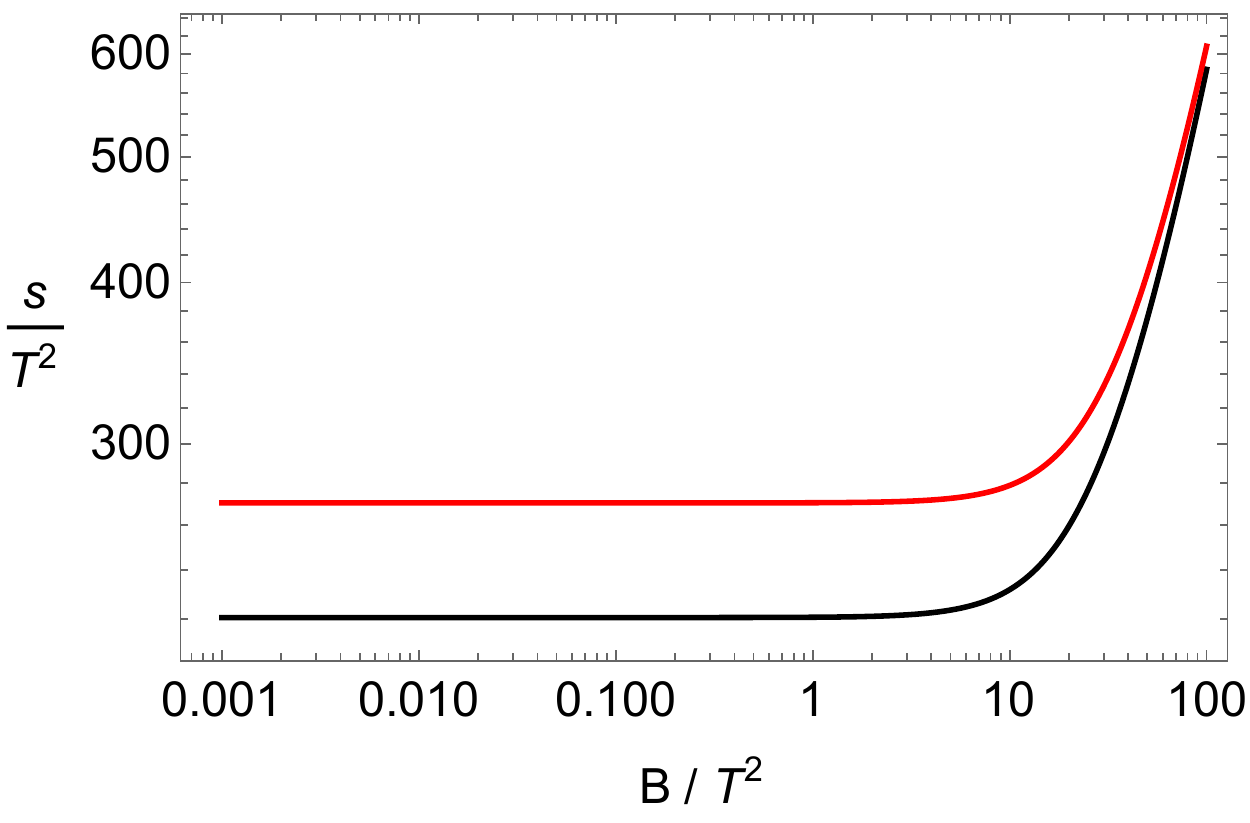} 
     
     \vspace{0.3cm}
     
     \includegraphics[width=0.48 \linewidth]{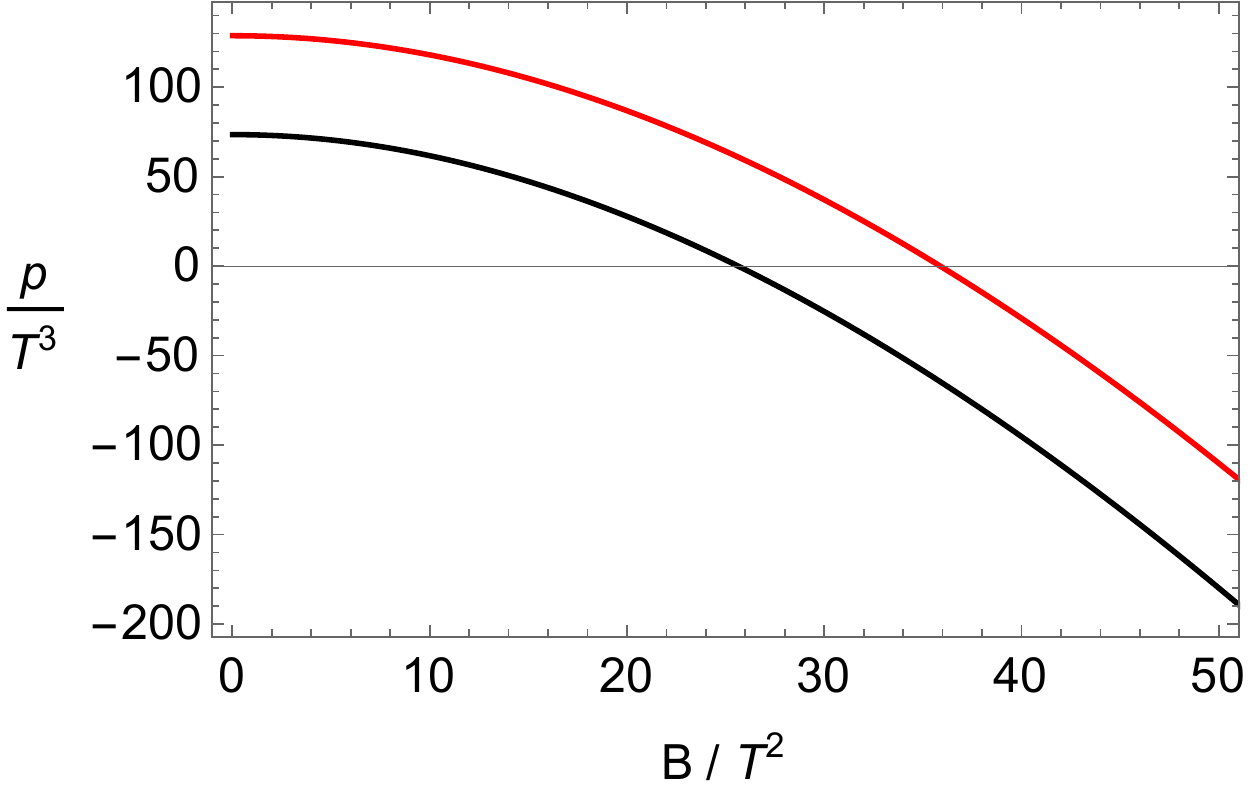} 
     \includegraphics[width=0.45 \linewidth]{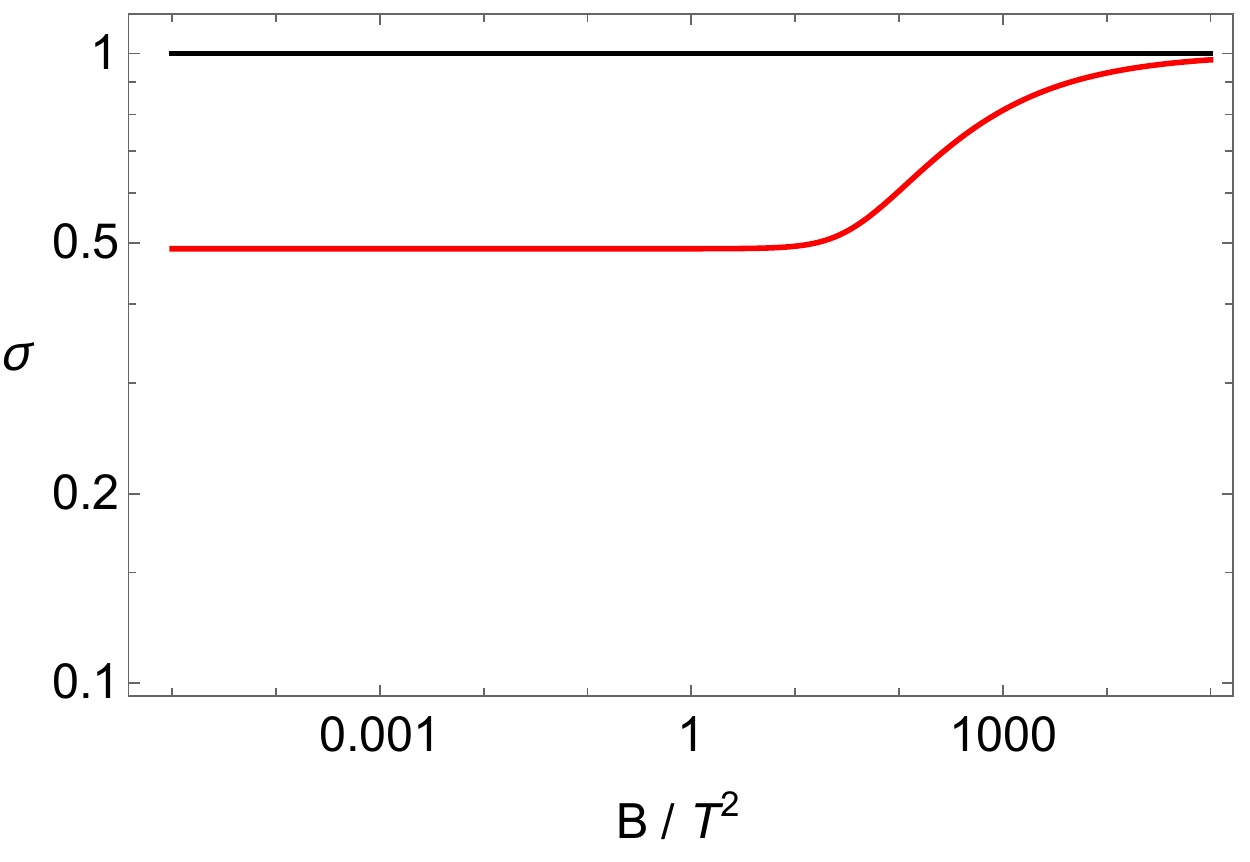} \label{EOSC4}
 \caption{Thermodynamic parameters ($\epsilon, p, s$) and conductivity $\sigma$ from weak to strong magnetic field at $\mu/T=0, 5$ (black, red).}\label{EOSC1234}
\end{figure}

\begin{figure}[]
\centering
     {\includegraphics[width=7.2cm]{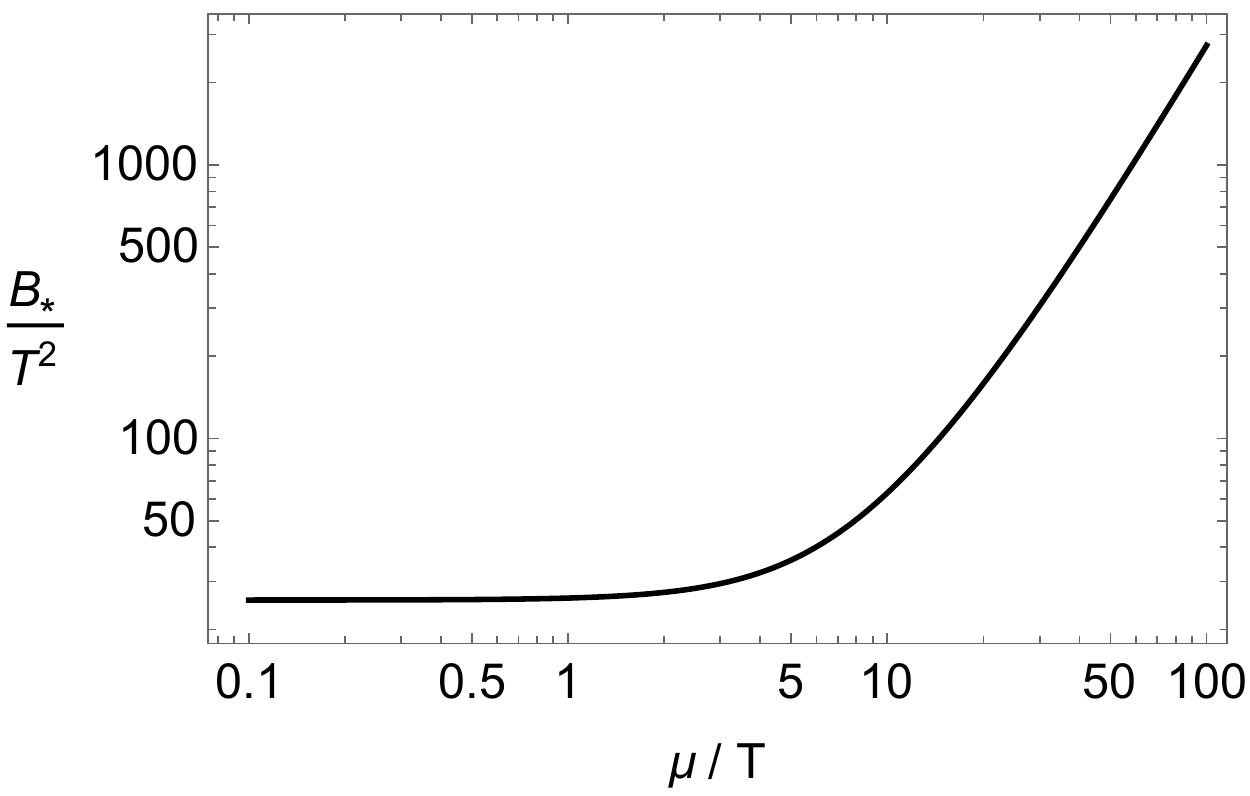} }
 \caption{Critical magnetic field $B_*$ vs. $\mu$.}\label{CRIBFIG}
\end{figure}

In Fig. \ref{EOSC1234}, we display the thermodynamic parameters ($\epsilon, p, s$) together with the electric conductivity $\sigma$ at fixed density as a function of the magnetic field. The shear viscosity $\eta$ is trivially related to the entropy $s$ via the KSS relation \eqref{etasigma} and it is therefore not shown.\\ 
In the weak $B$ field regime ($B/T^2 \ll1$), all the observables are $\mu$-dependent constants. On the other hand, in the strong $B$ field regime ($B/T^2 \gg1$), we find the following  asymptotic behaviors:
\begin{align}
    & \frac{\epsilon}{T^3}\sim  \frac{B^{3/2}}{T^3}\,,\qquad \frac{s}{T^2}\sim \frac{B}{T^2}\,,\qquad \frac{p}{T^3}\sim \frac{B^{3/2}}{T^3}\,,\qquad \sigma\sim const \,,
\end{align}
where all the coefficients are independent of the ratio $\mu/T$.\\
Interestingly, increasing the value of $B$, the thermodynamic pressure $p$ becomes negative at a critical value $B_{*}$. This critical value can be obtained analytically from Eq.\eqref{HAWKINGT}. Its behavior as a function of $\mu/T$ is showed in Fig. \ref{CRIBFIG}.

Using the Smarr relation ($ p= - \epsilon +  s  T + \mu \rho$), one can also understand the behavior of both $p$ and $\sigma$ in the strong $B$ regime. 
In that limit, $\epsilon\sim{B}^{3/2}$ scales faster than the entropy, $s\sim{B}$. As a result, at large $B$ one has $p\sim - \epsilon$ and therefore a negatively divergent thermodynamic pressure.
In addition, the electric conductivity can also be rewritten  as $\sigma = \left(\frac{s T}{s T+\mu \rho}\right)^2$ which approaches unity in the strong ${B}$ limit. Similar results have been obtained in AdS$_5$ using a higher-form language ~\cite{Grozdanov:2017kyl}.

\subsubsection{Transport coefficients and magnetohydrodynamics}\label{SECEOS2}

We now turn to the analysis of the dispersion relations of the low-energy modes (lowest QNMs) and in particular of the coefficients appearing up to order $k^2$. Our task is to verify the validity of the hydrodynamic description at large magnetic field.

Let us start with the simplest case of a neutral plasma. Here we do expect four gapless 
modes: two magnetosonic waves together with the two (shear/magnetic) diffusive modes
\begin{align}\label{DISP0D1}
    \omega\,=\,\pm\,v_\text{ms}\,k \,-\,i\frac{\Gamma_\text{ms}}{2}\,k^2 \,,\qquad \omega = -i\,D_\text{shear}\,k^2 \,, \qquad  \omega=-i\,D_\text{mag}\,k^2 \,.
\end{align}
The specific expression for the coefficients above is lengthy and is provided in the GitHub repository available \href{https://github.com/sicobysico/MHD_HOLO}{here}.

\begin{figure}[]
\centering
     \includegraphics[width=7.1cm]{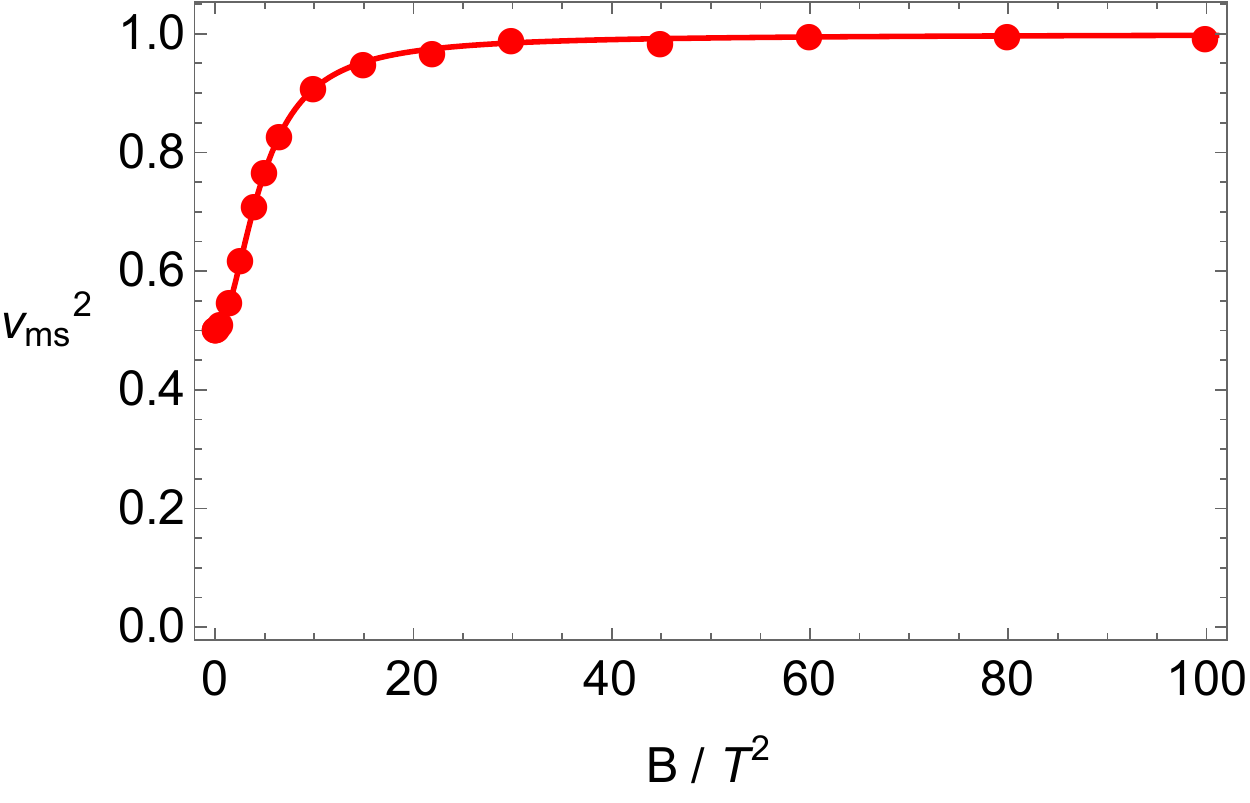} 
     \includegraphics[width=7.5cm]{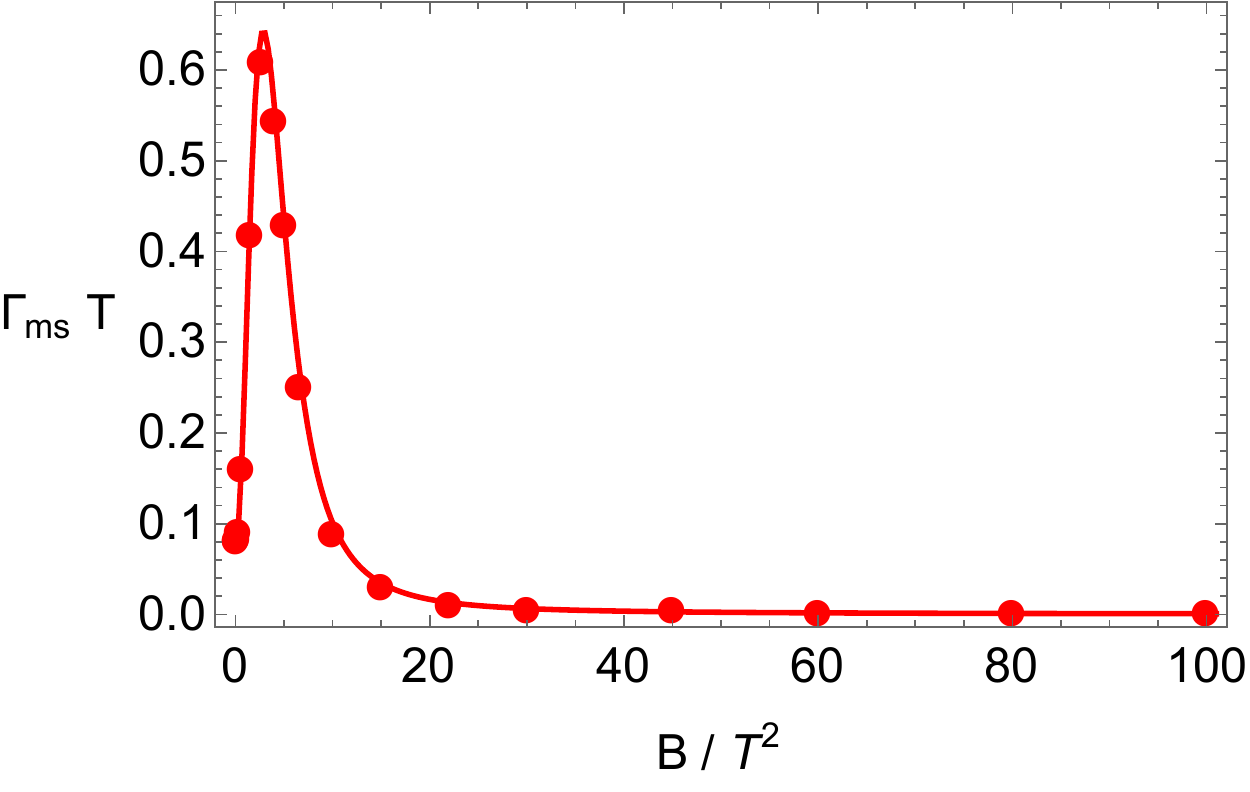} 
 \caption{Speed and attenuation constant of magnetosonic waves ($v_{\text{ms}}, \Gamma_{\text{ms}}$) at $\mu=0$.}\label{TC12}
\end{figure}

In Fig. \ref{TC12}, we display ($v_{\text{ms}}, \Gamma_{\text{ms}}$) as a function of $B$.
We find that the speed of magnetosonic waves interpolates between the conformal sound speed $v_{\text{ms}}^2 = 1/2$ at weak $B$ field  and the speed of light $v_{\text{ms}}^2 = 1$ at strong $B$ field. The normalized attenuation constant $\Gamma_{\text{ms}} T$ displays a non-monotonic behavior. It vanishes at large $B$ and it asymptotes a constant at zero $B$.
\begin{table}[]
\begin{center}
\begin{tabular}{| C{2.2cm} | C{4.6cm} | C{6.9cm}  |}
    \hline
        &  \cellcolor{green!08} \textbf{ Weak $B$ field}   &   \cellcolor{blue!08} \textbf{Strong $B$ field }    \\ 
    \hline
    \hline
      $v_\text{ms}^2$  &   $\frac{1}{2}\,+\, \mathcal{O}\left(B/T^2\right)^2$   &  $1-\frac{\sqrt{3} \pi}{2} \,{\lambda}/{T}\,\left(B/T^2\right)^{-1} \,+\, \mathcal{O}\left(B/T^2\right)^{-3/2}$   \\ 
    \hline
      $\Gamma_\text{ms}T$  &   $\frac{1}{4\pi} \,+\, \mathcal{O} \left(B/T^2\right)^{2}$   &   $\frac{{\lambda}/{T}}{2\sqrt{3}} \, \left(B/T^2\right)^{-1} \,+\, \mathcal{O}\left(B/T^2\right)^{-3/2}$  \\ 
    \hline
    \hline
      $D_\text{shear}T$  &   $\frac{1}{4\pi} \,-\, \mathcal{O}\left(B/T^2\right)^{2}$    &   $\frac{\lambda/T}{2\sqrt{3}} \left(B/T^2\right)^{-1}   \,+\, \mathcal{O}\left(B/T^2\right)^{-3/2}$   \\ 
    \hline
      $D_\text{mag}T$  &    $\frac{4\pi+3\lambda/T}{4 \pi \lambda/T} \,-\, \mathcal{O}\left(B/T^2\right)^{2}$  &    $\,\,\,\frac{3^{1/4}}{\sqrt{2}} \, \left(B/T^2\right)^{-1/2}\,+\, \mathcal{O}\left(B/T^2\right)^{-1} $ \\
    \hline
\end{tabular}
\end{center}
\caption{Approximate asymptotic behavior for the transport coefficients appearing in the dispersion relations of the hydrodynamic modes in weak- and strong-field limits at zero density.}\label{0Dtable}
\end{table}
Notice that, using the asymptotic forms presented in table \ref{0Dtable}, a negative value of $\lambda$ would result in a superluminal magnetosonic waves at strong magnetic field. This is not surprising since $\lambda$ must be taken as positive.

\begin{figure}[]
\centering
  
     \includegraphics[height=2.9cm]{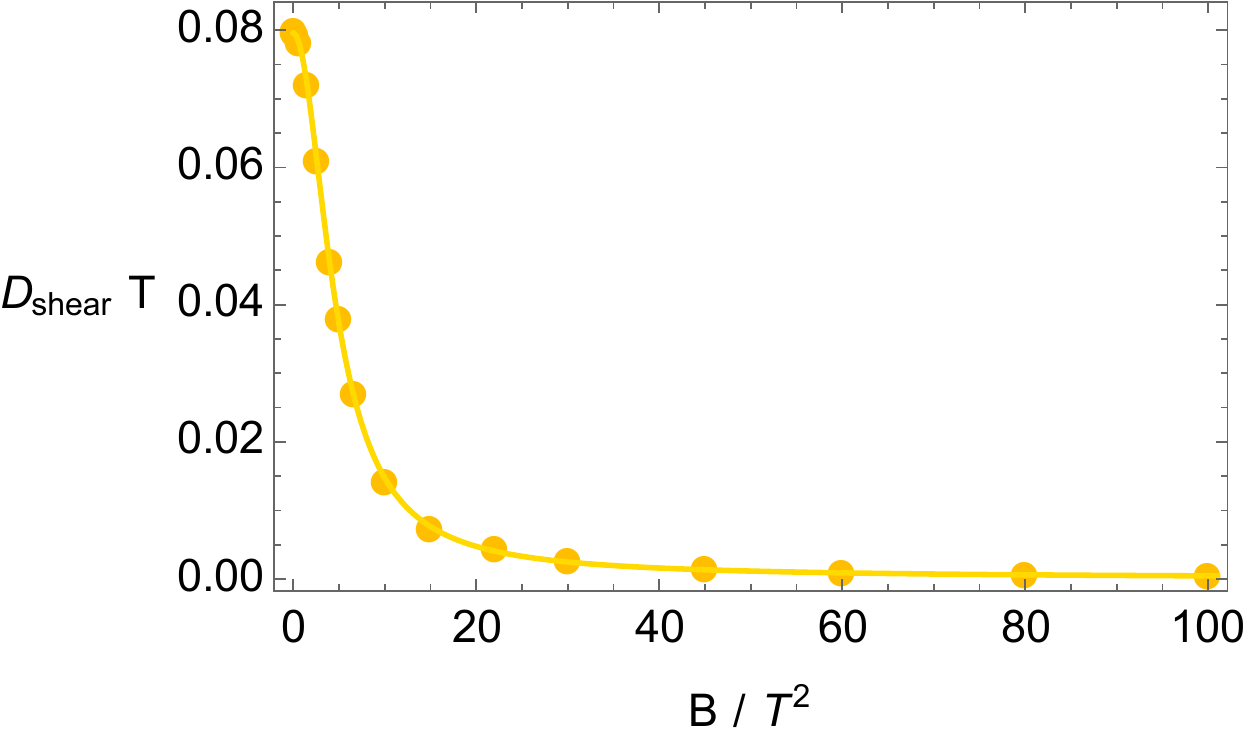} \includegraphics[height=2.9cm]{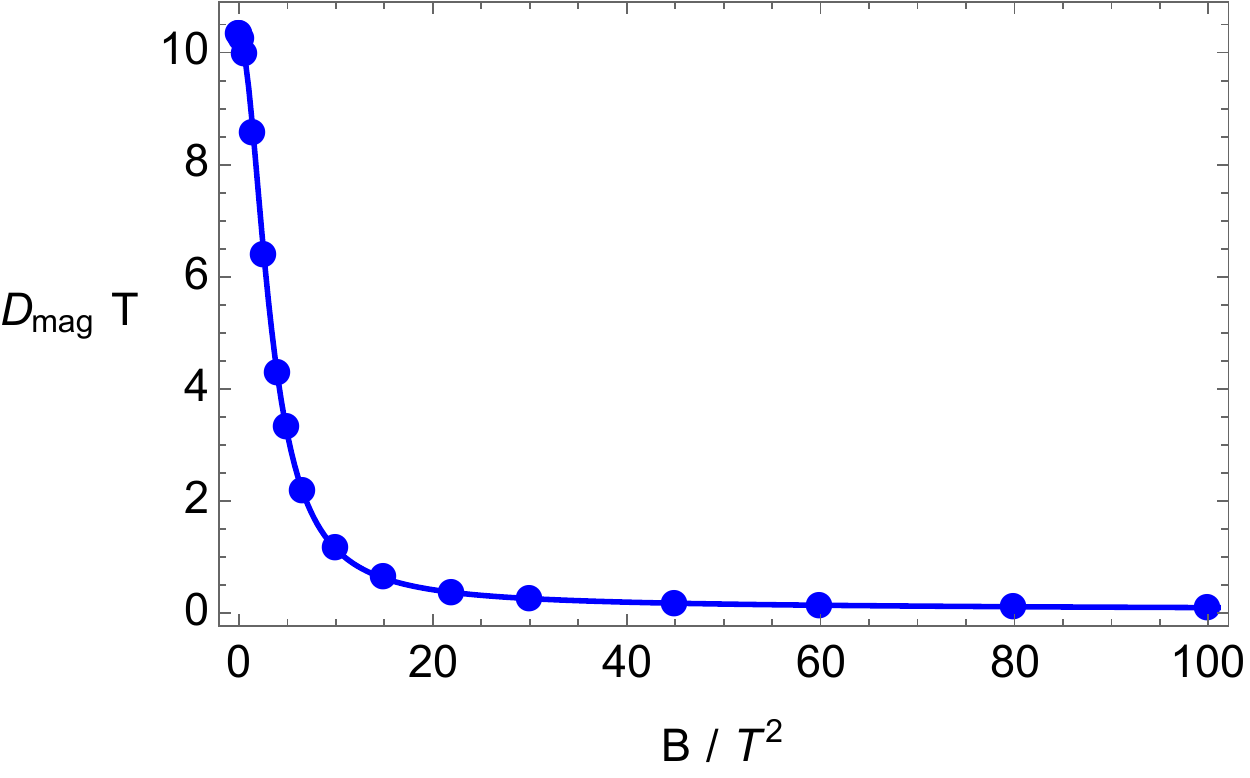} 
    \includegraphics[height=2.9cm]{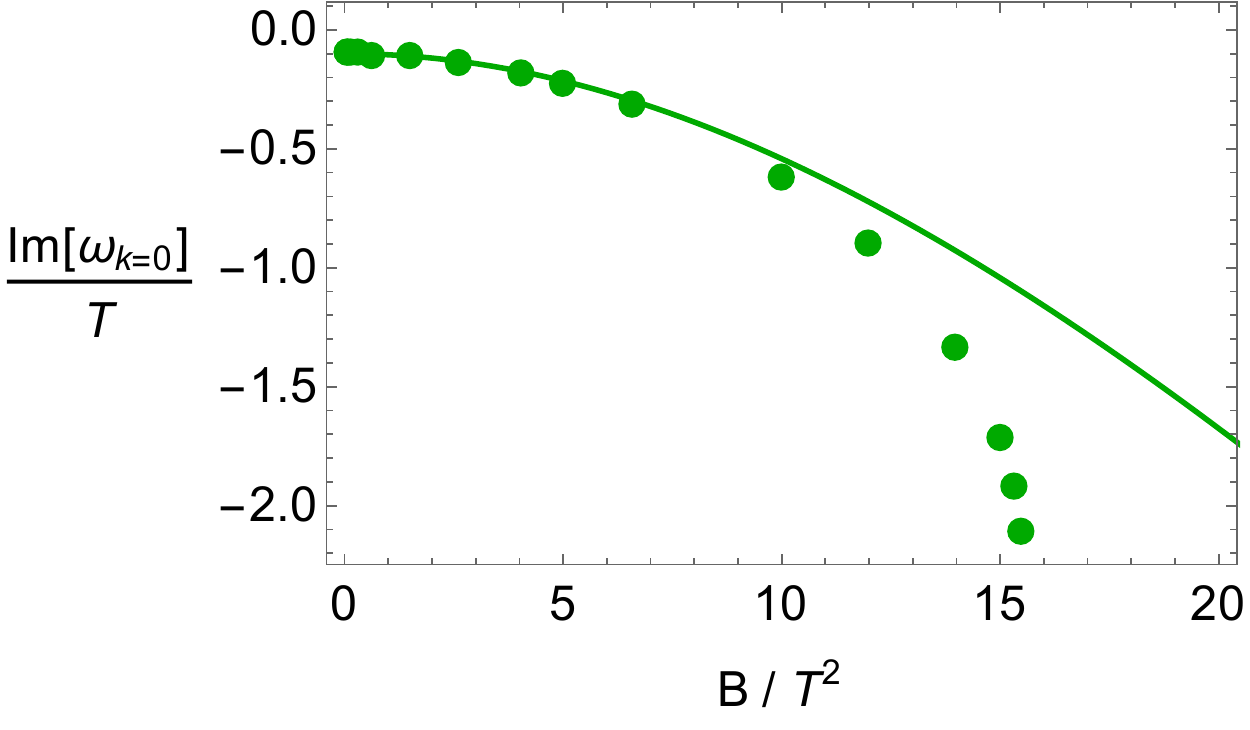} 
 \caption{Diffusion constants for shear and magnetic diffusive modes and damping of the first non-hydro mode as a function of $B/T^2$ at $\mu=0$.}\label{TC345}
\end{figure}

We also show the $B$-dependence of the diffusion constants for shear and magnetic diffusion in Fig. \ref{TC345}.
We find that all the diffusion constants vanish in the strong $B$ regime, which is consistent with the previous literature ~\cite{Grozdanov:2017kyl,Jeong:2021zhz,Jeong:2022luo}. 

We find that all the transport coefficients of the gapless modes, obtained by fitting the numerical dispersion relations, are in perfect agreement with the magnetohydrodynamic formulae even in the large $B$ limit. This is somehow surprising but, as we will see, strongly dependent on the value of the EM coupling $\lambda$. {We will comment more on this point in the next sections and in the conclusions.} This agreement implies that the formula for the conductivity $\sigma$ given in Eq.\eqref{etasigma} works well for all values of $B$, even beyond the small $B$ regime. To complete this section, we can also analytically derive the asymptotic behavior of all these coefficients. For the zero density case, these are given in table \ref{0Dtable}. 

In Fig. \ref{TC345} we also show the value of the damping of the first non-hydrodynamic mode in function of the dimensionless magnetic field. In this case, the prediction from magnetohydrodynamics are not in well agreement with the numerical data for $B/T^2 \gg 1$. This is not surprising. It is simply due to the fact that the imaginary part of this non-hydrodynamic mode becomes large, i.e., $\mathrm{Im}(\omega) \sim T$, and the mode moves away from the regime of validity of linearised hydrodynamics.

Next, we perform the similar analysis at finite density. In this case, as demonstrated in Section \ref{FDskl}, we have two gapless modes, the longitudinal diffusive mode and subdiffusive mode, whose dispersions are given by
\begin{align}\label{Eq:finiteD}
    \omega\,=\,-i\,D_\text{long}\,k^2 \,,\qquad
    \omega\,=\,-i\,D_\text{subdiff}\,k^4 \,.
\end{align}
The concrete form of the diffusive parameters is cumbersome but can be found in the GitHub repository available \href{https://github.com/sicobysico/MHD_HOLO}{here}. For the caveats related to the validity of the subdiffusive dispersion within first order hydrodynamics see the discussions in Section \ref{nn} and appendix \ref{lala}.

\begin{figure}[]
\centering
    \includegraphics[width=7.2cm]{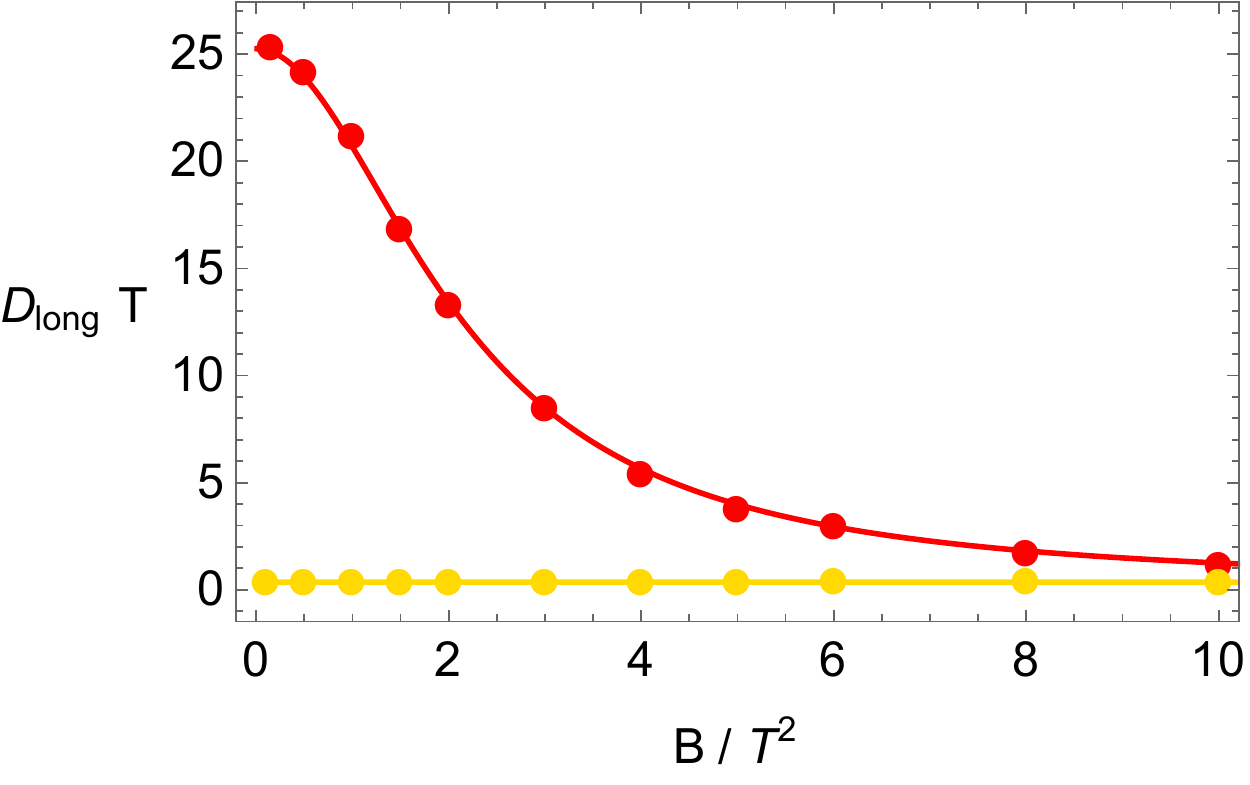} 
    \includegraphics[width=7.65cm]{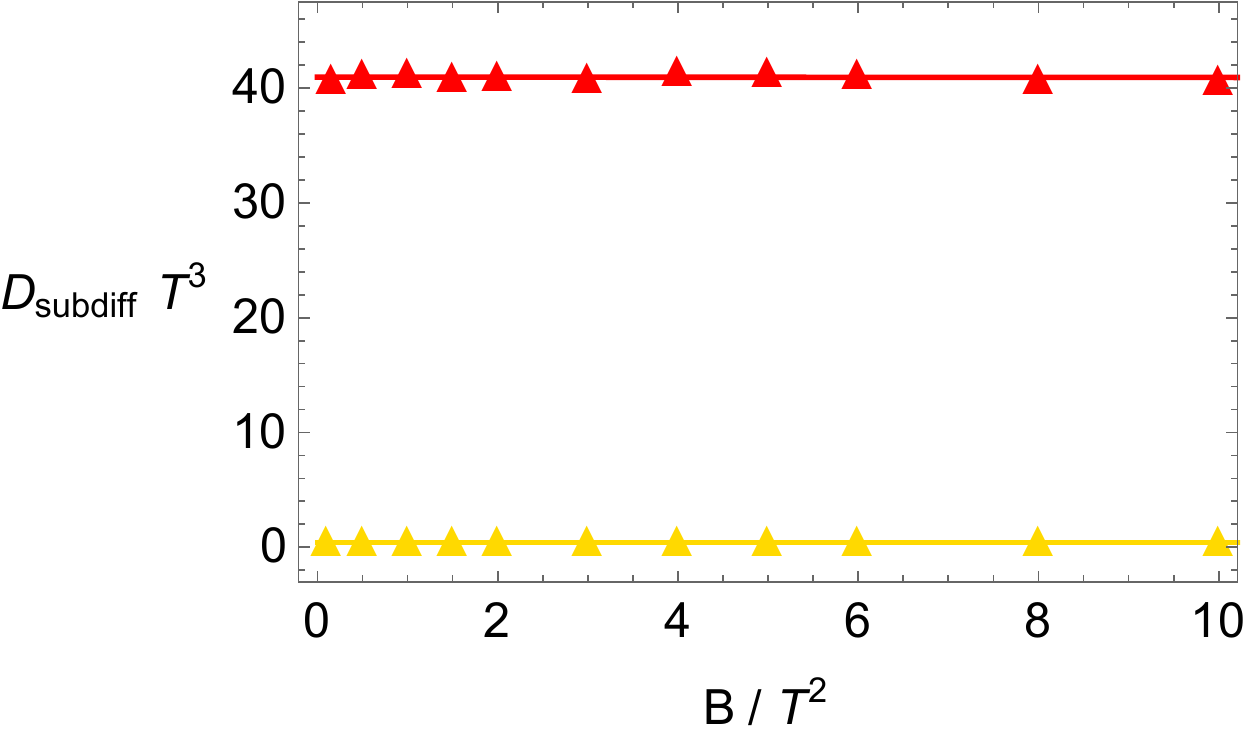} 
 \caption{The diffusive parameters for the longitudinal diffusive (circles) and subdiffusive (triangles) modes at finite charge density and magnetic field. Yellow lines are for $\mu/T=5$ while red lines for $\mu/T=0.5$. The fluctuations of the red triangles are due to numerical inaccuracy.}\label{TCmu0p5}
\end{figure}

In Fig. \ref{TCmu0p5}, we display the $B$-dependence of the two diffusive parameters. $D_\text{long}$ vanishes in the strong $B$ limit independently of the value of the charge density. On the contrary, $D_\text{subdiff}$ reaches a constant value at $B/T^2 \rightarrow \infty$. More precisely we find that in the strong $B$ limit:
\begin{equation}
    D_\text{long}T\sim \left(B/T^2\right)^{-1/2}+\dots\,,\qquad D_\text{subdiff}T^3\sim \left({\lambda}/{T}\right)^{-1} \left({\mu}/{T}\right)^{-2} +\dots \,.
\end{equation}
In addition, we also find that both transport coefficients are suppressed at larger density.
Once again, the results obtained by the fitting method are in good agreement with magnetohydrodynamic predictions at finite density even in the strong $B$ regime. The apparent fluctuations in the numerical data visible in Fig. \ref{TCmu0p5} are just due to numerical precision.

Finally, we discuss also the dynamics of the non-hydrodynamic modes at finite charge density by dialing the strength of the magnetic field $B$.

\begin{figure}[]
\centering
     \includegraphics[width=7.3cm]{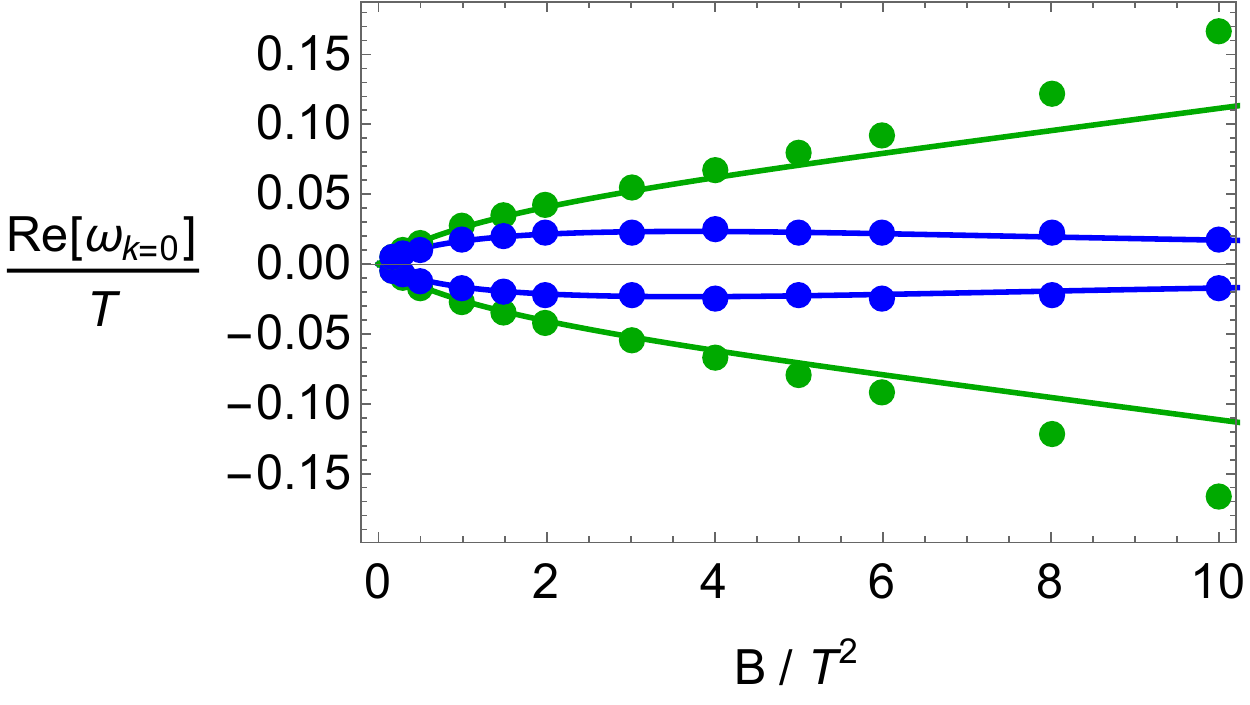} 
     \includegraphics[width=7.3cm]{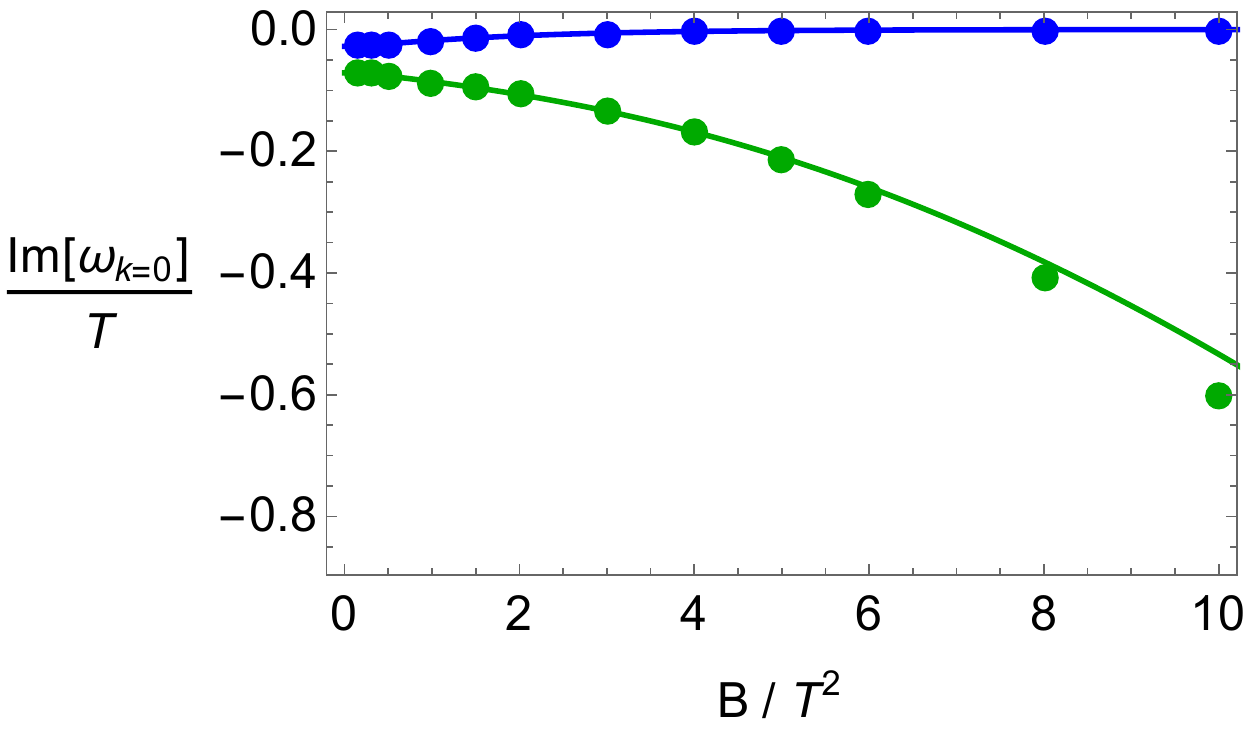} 
     \includegraphics[width=7.3cm]{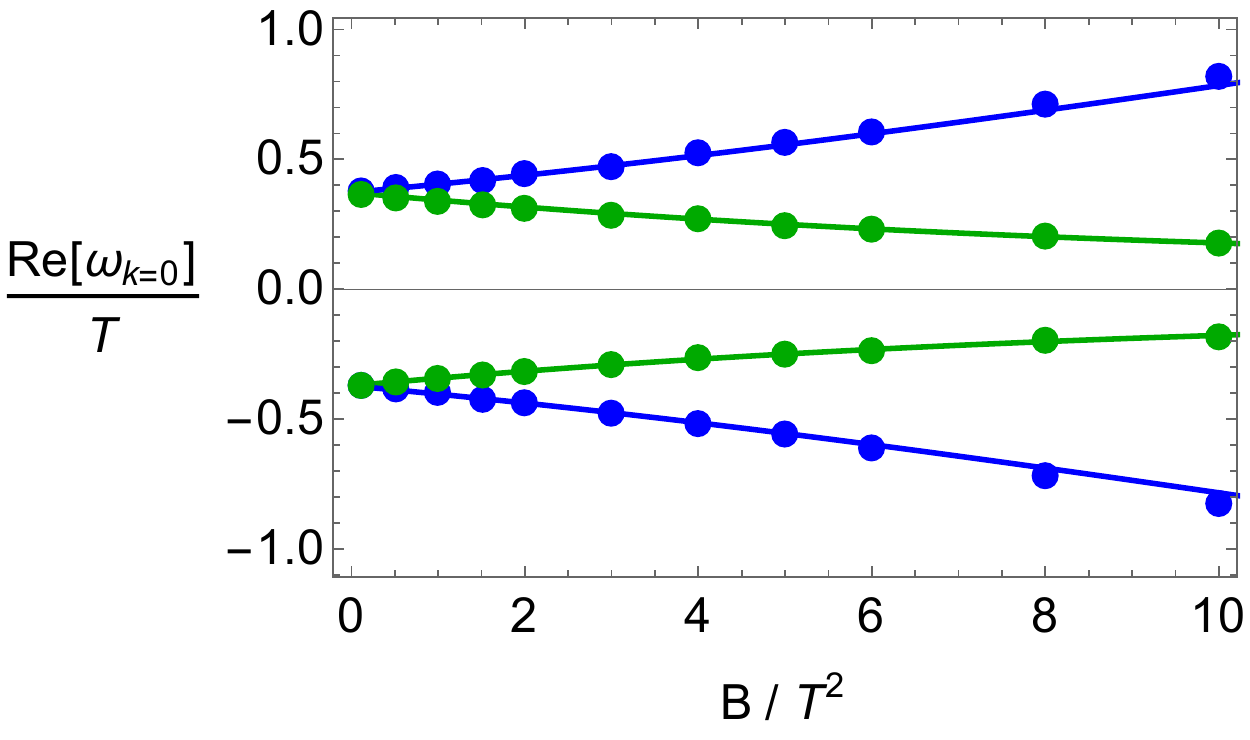} 
     \includegraphics[width=7.3cm]{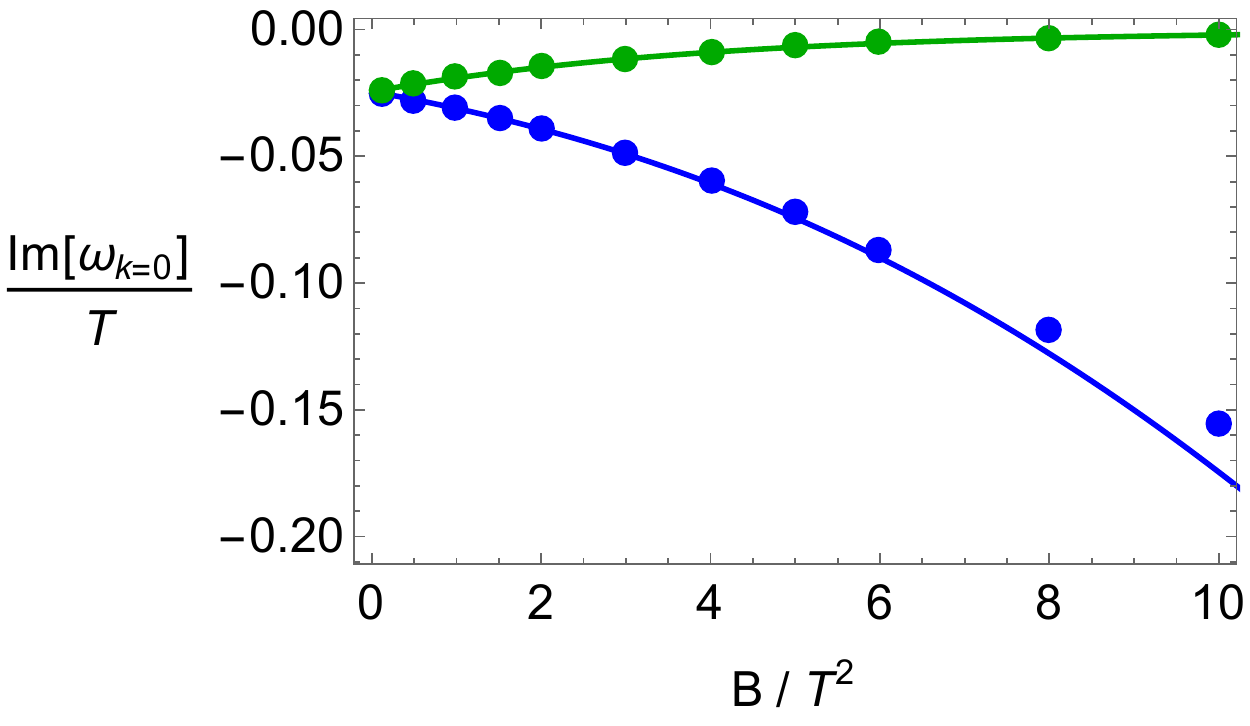} 
  \caption{The real and imaginary part of the first non-hydrodynamic modes at density $\mu/T = (0.5, 5)$ (upper panels, lower panels).}\label{TCmu5}
\end{figure}

In Fig. \ref{TCmu5}, we display the $B$-dependence of the imaginary and real gaps at zero wave-vector, $k=0$. In the weak $B$ regime, as demonstrated in Section \ref{FDskl}, the behavior of the gaps depend on the density. The precise expressions are provided in Eq.\eqref{SMDG} for small density and in Eq.\eqref{LAGDG} for large density. 

Interestingly, in the strong $B$ limit, we find that one of the two pair of modes approaches the origin of the complex frequency plane. More specifically, both its real and imaginary parts at zero wave-vector go to zero in the limit $B/T^2\rightarrow \infty$. This mode in the strong $B$ limit becomes an emergent propagating magnetosonic wave with speed $v_{\text{ms}}^2=1$ and vanishing attenuation constant. It is very tempting to describe this mode as an emergent photon in the strong $B$ limit. We are not aware of similar observations in the previous literature. This point needs further investigation in the future. On the other hand, the remaining pair is pushed away from the hydrodynamic limit since both its real and imaginary part diverge. Notice that whenever its imaginary part becomes too large, magnetohydrodynamics breaks down and its predictions are not anymore in good agreement with the numerical data.

The asymptotic behavior for these non-hydrodynamic modes in strong $B$ regime can be obtained analytically and it reads
\begin{align}
\frac{\omega_{k=0}}{T}&\,=\, \pm\, \frac{\left(\lambda/T\right)\left(\mu/T\right)}{\sqrt{2}\,3^{1/4}}\left(\frac{B}{T^2}\right)^{-1/2} \,-\, i\,\, \frac{\sqrt{2}\, \pi (\lambda/T)^3 (\mu/T)^2}{3^{3/4}}\left(\frac{B}{T^2}\right)^{-5/2} +\dots\,, \label{SFAB1} \\
\frac{\omega_{k=0}}{T}&\,=\, \pm\,\frac{3^{1/4} (\mu/T)}{2\,\sqrt{2}\, \pi}\left(\frac{B}{T^2}\right)^{1/2} \,-\, i\,\, \frac{\sqrt{3}}{2\pi}\left(\frac{B}{T^2}\right) +\dots \,, \label{SFAB2}
\end{align} 
where \eqref{SFAB1} corresponds to the pair approaching the hydrodynamic limit in Fig. \ref{TCmu5}, while \eqref{SFAB2} to the pair which is gapped away in the large $B$ limit.

In summary, we find that as long as the EM coupling $\lambda$ is small, the magnetohydrodynamic predictions for the hydrodynamic modes are in perfect agreement with the numerical results even in the strong $B$ regime. {At this point, we are not able to provide a solid derivation of why this is the case and how general this is. We will comment further on this point in the conclusion section.} In the next section, we discuss the effect of $\lambda$.

%
\subsection{The role of the electromagnetic coupling}\label{appena}
In the previous sections, we have fixed $\lambda/T=0.1$ and considered only the limit of small electromagnetic coupling. We now investigate the role of the EM coupling $\lambda$ and the interpolation between the two limits $\lambda \rightarrow 0$ and $\lambda \rightarrow \infty$. Similar analyses in the context of plasmons can be found in \cite{Baggioli:2019aqf,Baggioli:2019sio,Baggioli:2021ujk}.\\

\begin{figure}[]
\centering
     \includegraphics[width=6.3cm]{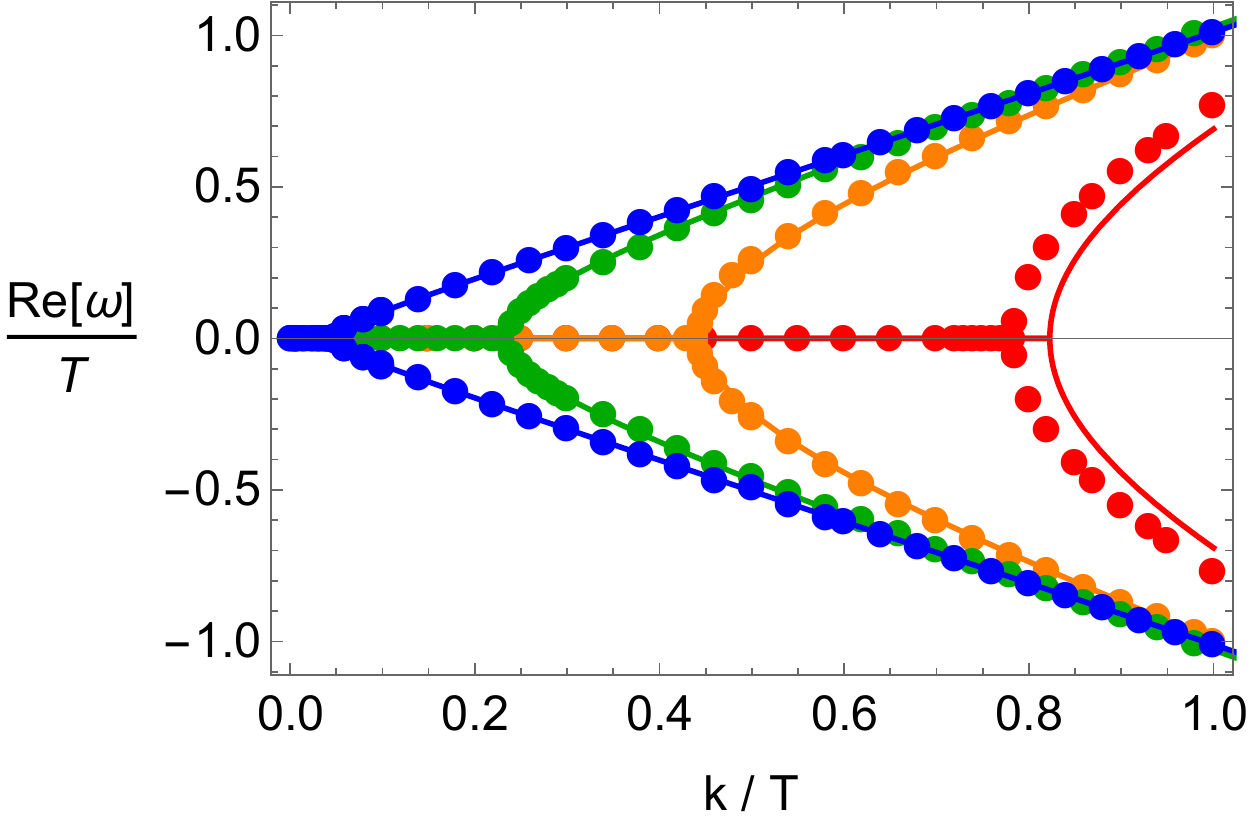} 
     \includegraphics[width=6.3cm]{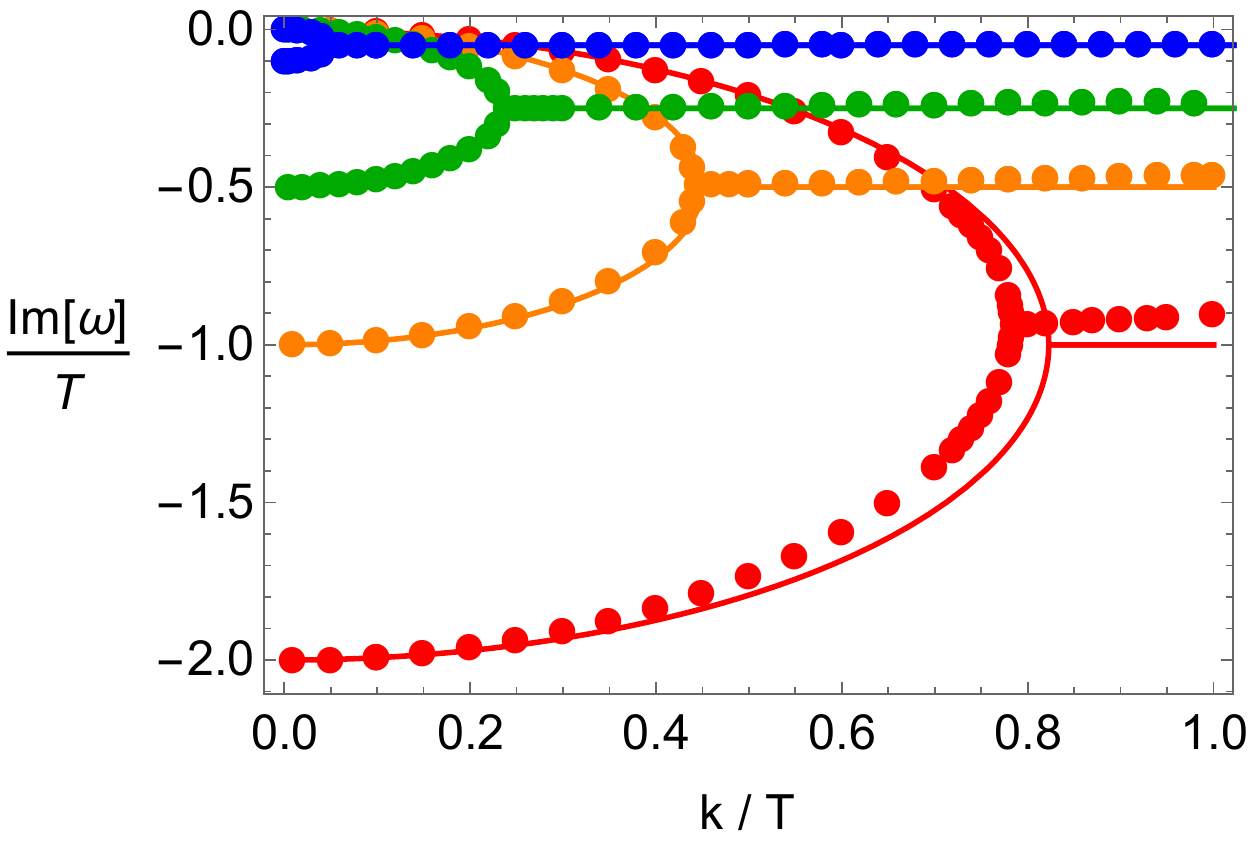} 
     \includegraphics[width=6.3cm]{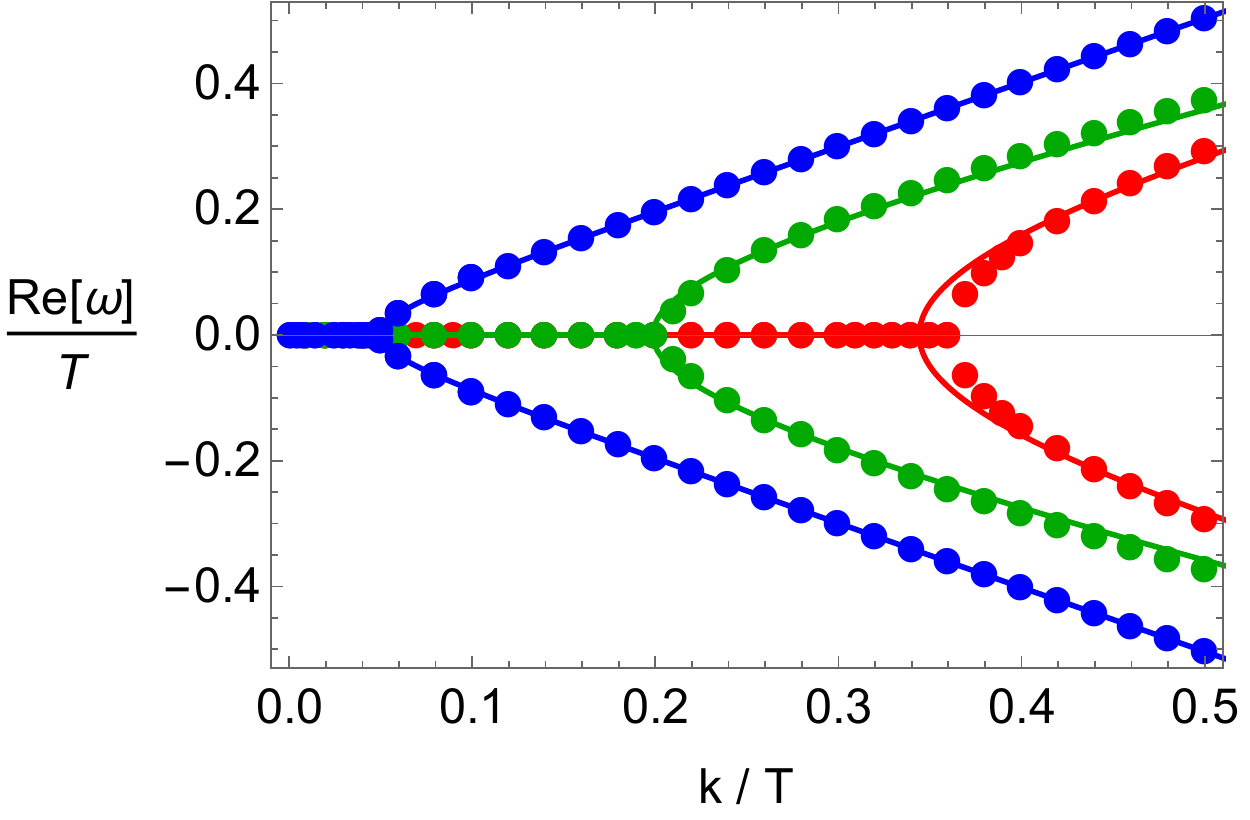} 
     \includegraphics[width=6.3cm]{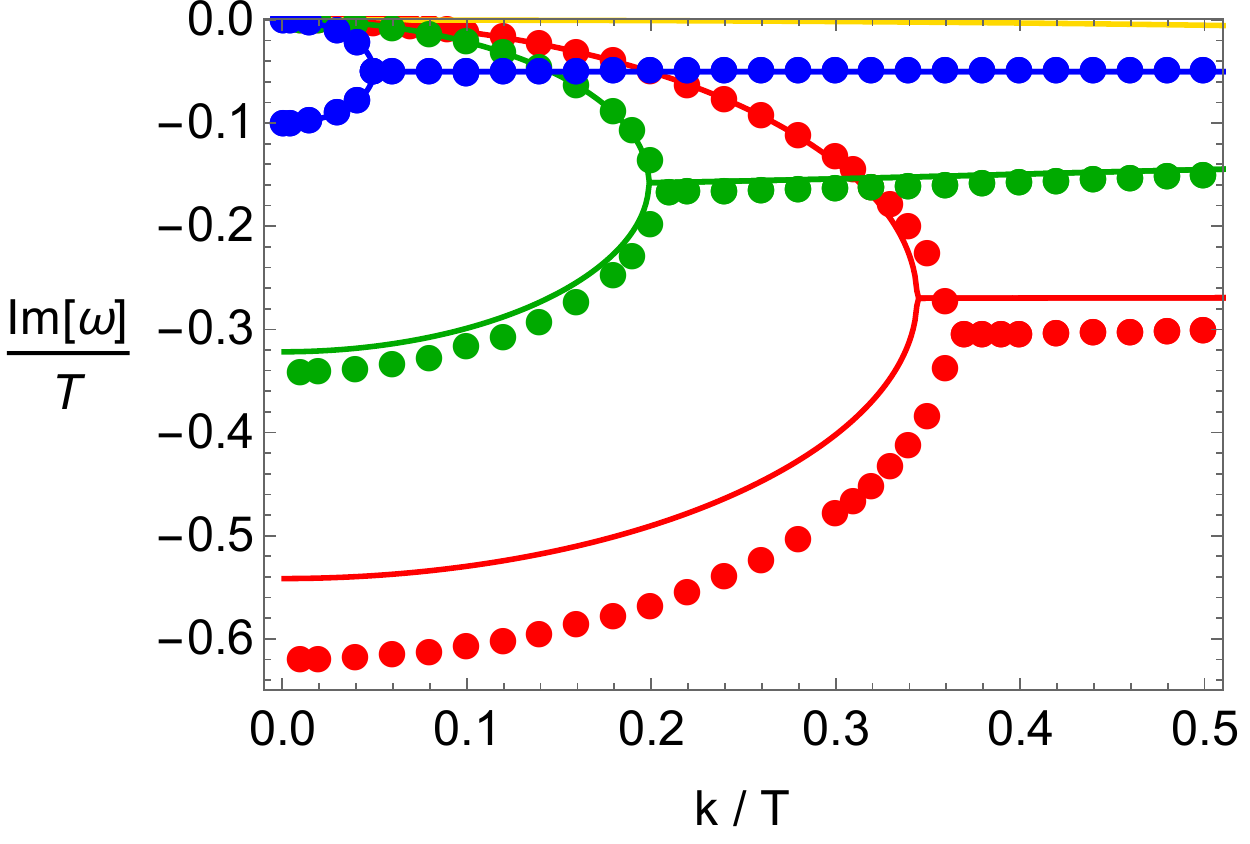} 
 \caption{\textbf{Top: }The dynamics of the EM waves at zero density and zero magnetic field.  Different colors, from blue to red, correspond to $\lambda/T=(0.1,0.5,1,2)$. The imaginary part of the non-hydrodynamic mode at $k=0$ is given analytically in Eq.\eqref{GAPGAP22}. \textbf{Bottom: }The dynamics of the EM waves at zero density and $\lambda/T=0.1$.  Different colors, from blue to red, correspond to $B/T^2=(0,7,10)$.
 }\label{zeroden2}
\end{figure}

For this purpose, we first consider the dynamics of EM waves which is given within the magnetohydrodynamic framework by Eq.\eqref{EMWAVE}. In Fig. \ref{zeroden2}, we compare the numerical QNMs data from the holographic model with the magnetohydrodynamic preditions. As expected, EM waves are screened by Coulomb interactions and they start propagating only above a certain cutoff wave-vector $k^\star$, the k-gap. The magnetohydrodynamic framework, in the first-order approximation, gives:
\begin{equation}
    k^\star\,=\,\frac{\sigma}{2\,c\,\epsilon_\text{e}}\,\qquad \text{with}\qquad c^2 := (\epsilon_\text{e}\,\mu_\text{m})^{-1}\,.
\end{equation}
From Fig. \ref{zeroden2}, we can indeed observe that at low wave-vector the EM waves are not propagating but they rather split into a diffusive mode and a non-hydrodynamic one, as predicted by magnetohydrodynamics. Moreover, for small values of the EM coupling, $\lambda/T \ll 1$, the magnetohydrodynamic formula, Eq.\eqref{EMWAVE}, is in very good agreement with the numerical data and accurately predicts the value of the cutoff wave-vector $k^\star$. The onset of propagation moves to larger wave-vectors by increasing the EM coupling $\lambda$ or equivalently the magnetic field $B$ (see for example the red data in Fig. \ref{zeroden2}) and the accuracy of the magnetohydrodynamic description decreases. This is simply the sign that higher order corrections in Eq.\eqref{EMWAVE} become important at $k/T \sim \mathcal{O}(1)$. Interestingly, the imaginary gap of the non-hydrodynamic mode can be derived analytically (see Eq.\eqref{GAPGAP22} below) and it is in perfect agreement with the numerical data even when $\mathrm{Im}[\omega]/T \gg 1$.

In order to derive the inverse relaxation time of the non-hydrodynamic mode analytically, we notice that the equations of motion for the gauge field fluctuations ($\delta a_{i=x,y}$) decouple in the limit of $k=0$ and $\mu=B=0$. Then, the equations can be solved analytically and give the following leading/subleading coefficients near the AdS boundary 
\begin{align}\label{BC33}
\begin{split}
\delta a_{i}^{(L)} = 1 \,,\quad  \delta a_{i}^{(S)} = i \,\omega \,.
\end{split}
\end{align} 
Continuing, we use the boundary conditions described in Section \ref{DTMF} which at $k=0$ are given by
\begin{equation}
    \omega^2 \, \delta a_{i}^{(L)} + \lambda \,\delta a_{i}^{(S)}=0\,.
\end{equation}
By combining the two last equations, we finally arrive at two independent solutions
\begin{align}\label{GAPGAP22}
\begin{split}
\omega=0\,,\qquad \omega \,=\, -i \lambda \,.
\end{split}
\end{align} 
The first mode corresponds to the hydrodynamic diffusive mode at $k=0$ while the second mode is the non-hydrodynamic one visible in Fig. \ref{zeroden2} (top panel). Taking into account that in the limit $\mu=B=0$ we have $\sigma=1$, this analytically confirms the identification of the parameter $\lambda$ in the b.c.s. as $\lambda=1/\epsilon_e$, at least in the regime of small $\lambda$ coupling.\footnote{Whenever $\lambda$ becomes large, the non-hydrodynamic mode acquires a very large imaginary gap. At that point, the prediction from first-order hydrodynamics are totally unreliable and the identification $\lambda=1/\epsilon_e$ does not hold anymore. This is confirmed numerically in Fig. \ref{PFSLDEP}.} These results are in agreement with those found in \cite{Baggioli:2021ujk}. It is interesting to notice that the inverse relaxation time of the non-hydrodynamic mode can be analytically derived for arbitrary values of the EM coupling $\lambda$, in perfect agreement with the numerical data.

\begin{figure}[]
\centering
     \includegraphics[width=6.3cm]{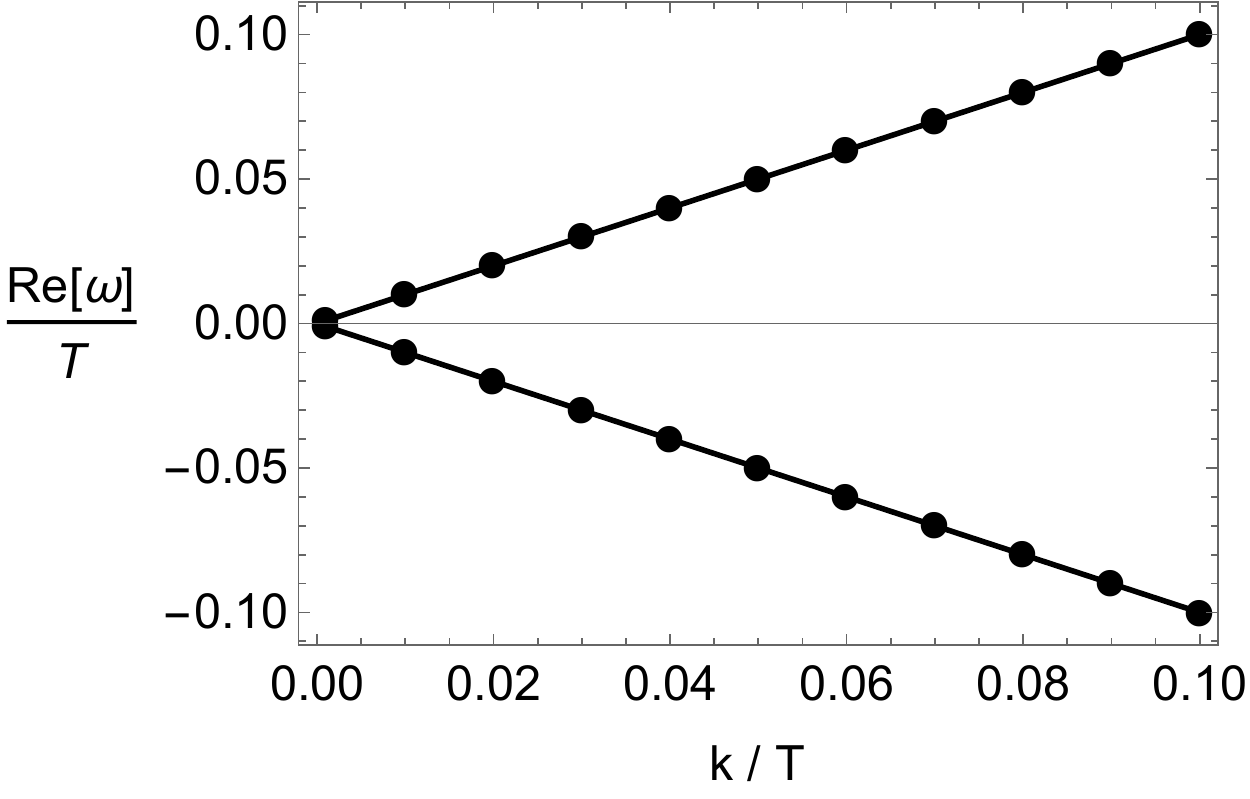} 
     \includegraphics[width=6.3cm]{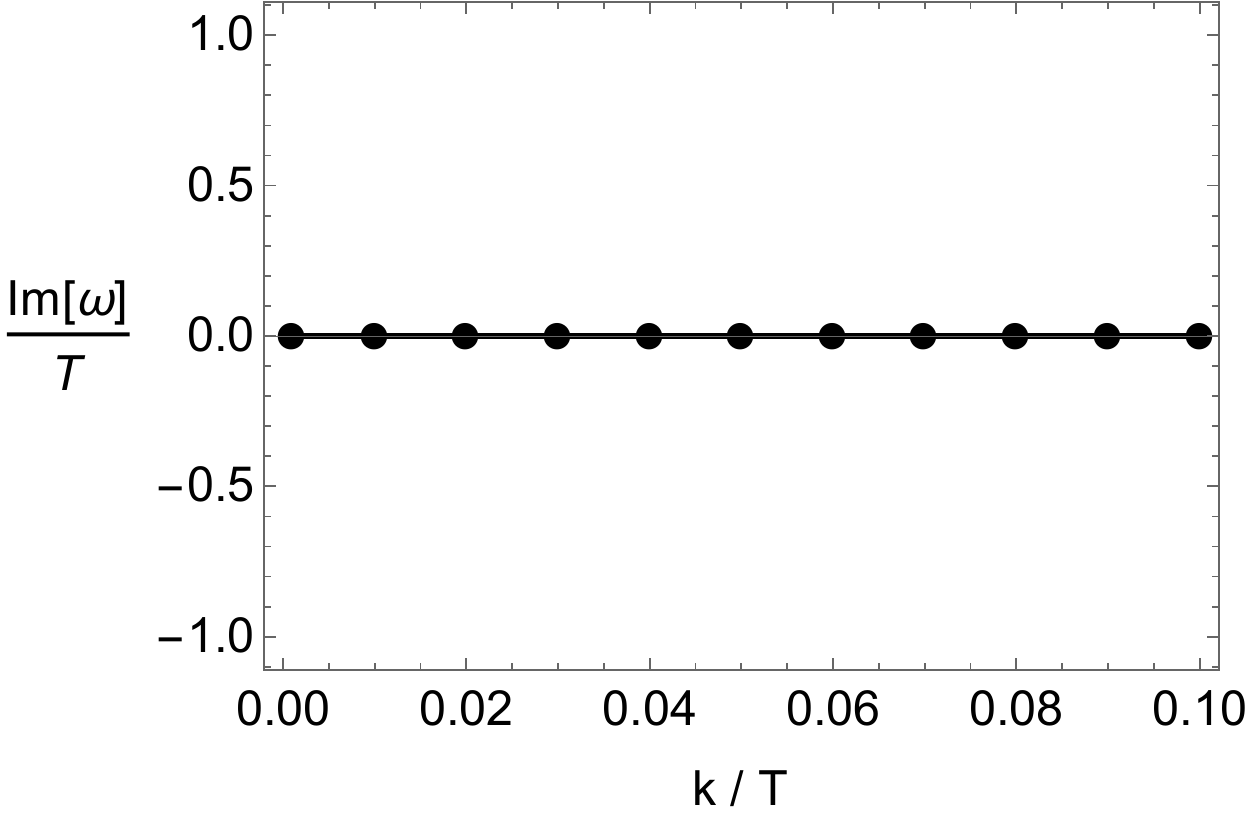} 
 \caption{The emergent propagating photon at zero density, zero magnetic field and zero EM coupling $\lambda/T=0$.}\label{zeroden3dd}
\end{figure}

Following the trend in Fig. \ref{zeroden2}, one could anticipate the appearance of a propagating free photon for $\lambda\rightarrow0$. This is indeed the case as explicitly shown in Fig. \ref{zeroden3dd} for $\lambda=0$.
This outcome is maybe not surprising since the mixed b.c.s. imposed reduce in the limit of $\lambda \rightarrow 0$ to the decoupled boundary Maxwell equations in vacuum:
\begin{equation}
    \partial_\mu F^{\mu\nu}=0 \,,
\end{equation}
which clearly displays a freely propagating photon. In other words, we can understand this limit as the one in which the EM interactions are vanishing and therefore all the effects of polarization and screening disappear. The dynamics of the Maxwell field decouples from the current. From a technical perspective, this comes from the fact that the b.c.s. used do not reduce to the standard Dirichlet ones in the limit of $\lambda \rightarrow 0$. On the contrary, they reduce to the Dirichlet b.c.s. times an independent and external factor $(\omega^2-k^2)=0$. The spectrum of the theory in this limit is then the same as the one for a CFT with non-dynamical U(1) symmetry times a freely propagating (and infinitely living) photon. Our results are consistent with those found using higher-form symmetries in \cite{Hofman:2017vwr}.

\begin{figure}[]
\centering
      \includegraphics[width=7.1cm]{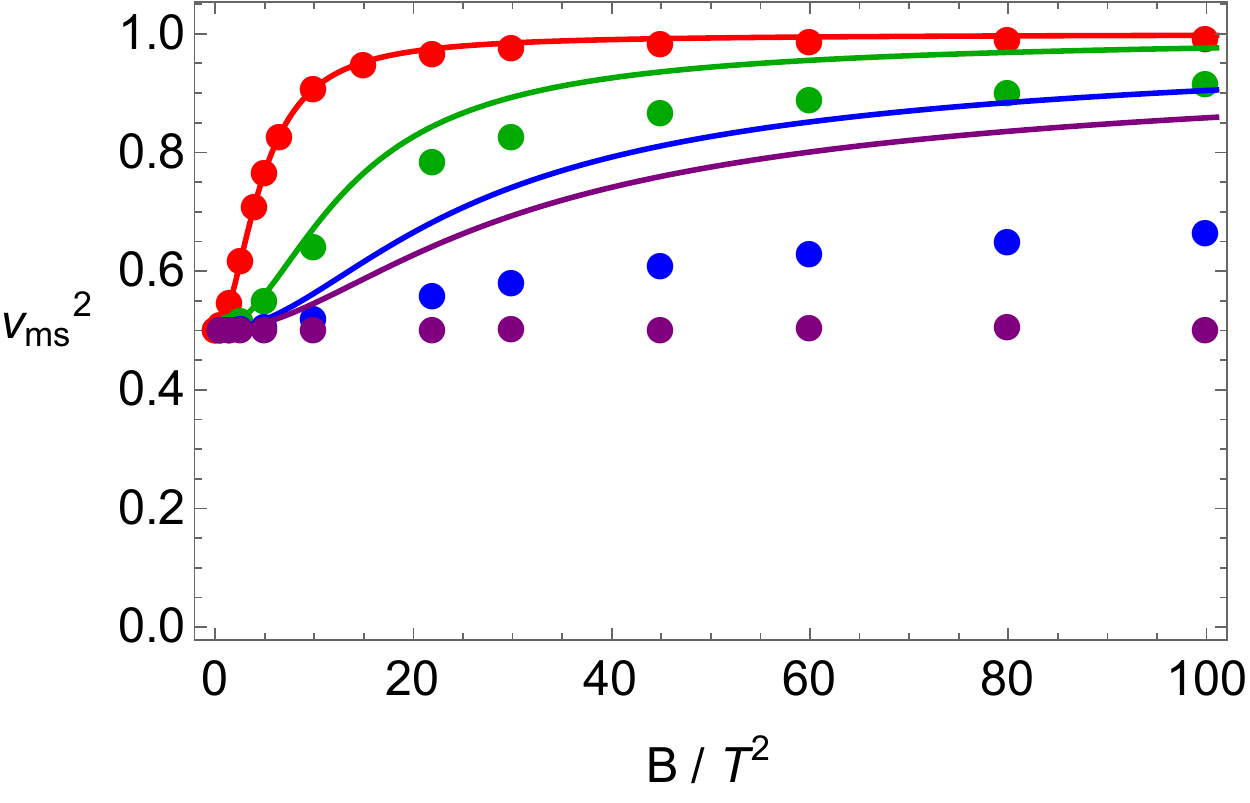}
     \includegraphics[width=7.1cm]{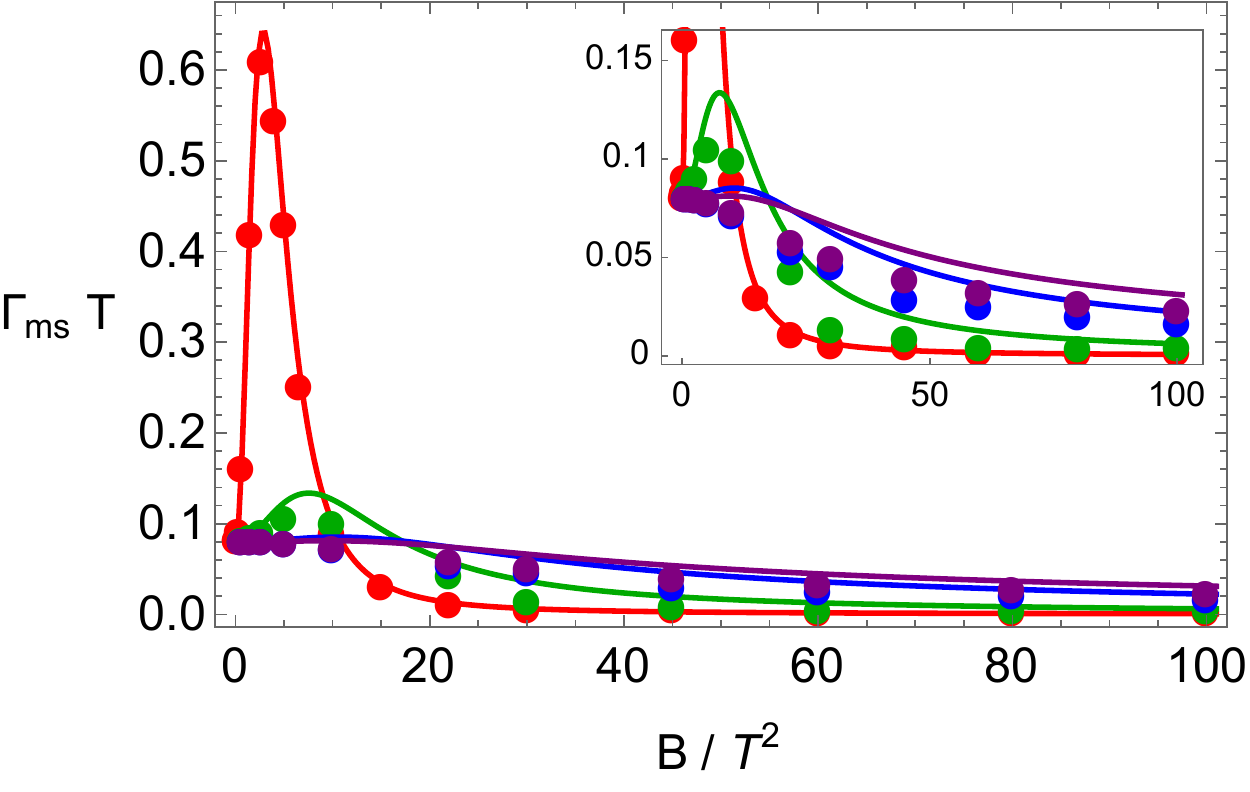}\\
      \includegraphics[width=7.3cm]{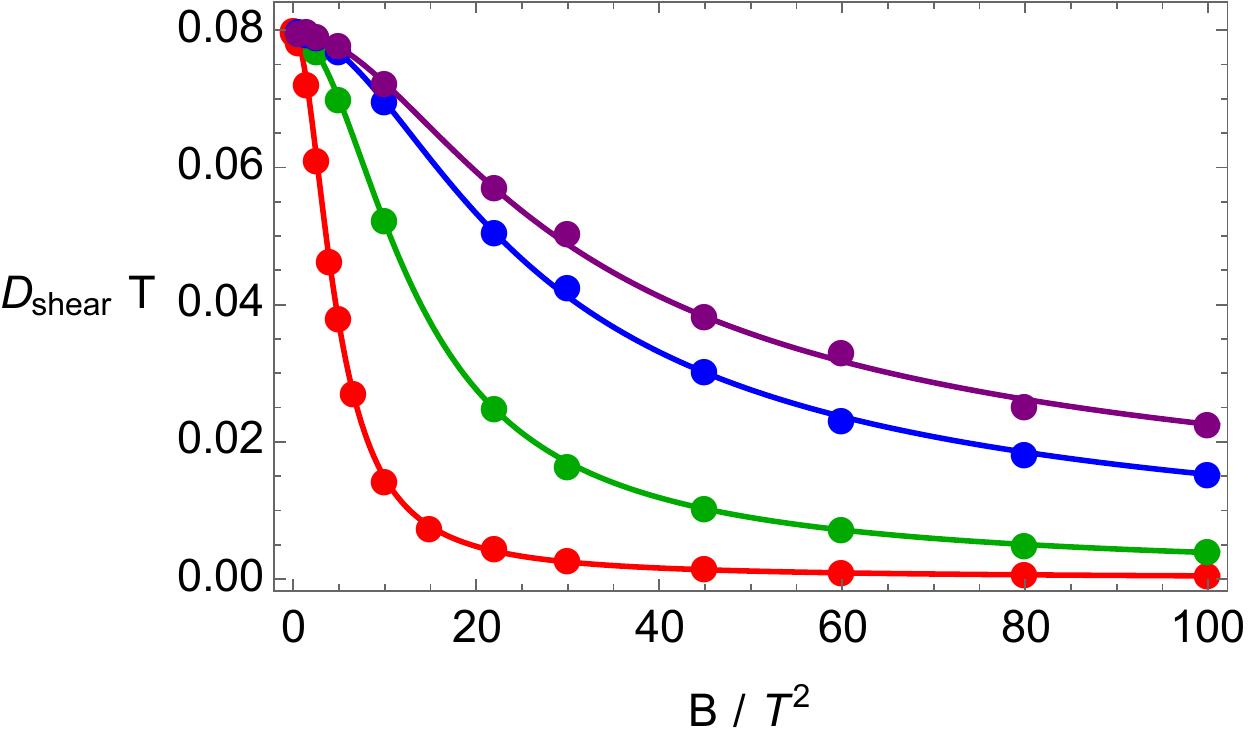}
     \includegraphics[width=7.2cm]{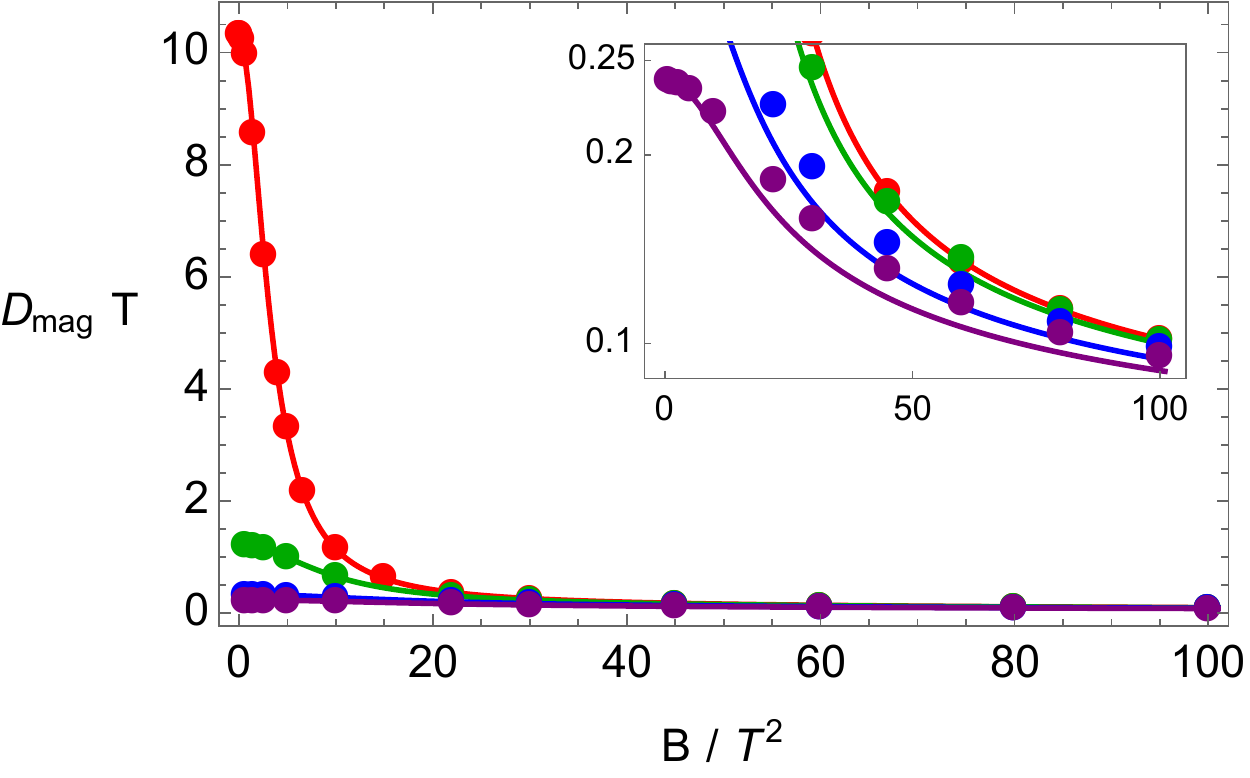}
\caption{\textbf{Top: }Speed and attenuation constant of magnetosonic waves. \textbf{Bottom: }The diffusion constants of shear and magnetic diffusion. The colors correspond to $\lambda/T$ = (0.1, 1, 10, 1000) (red, green, blue, purple). The insets are a zoom for the large $\lambda$ case (e.g., purple).}\label{LAMDEPFIG1}
\end{figure}

\begin{figure}[]
\centering
     \includegraphics[width=7.3cm]{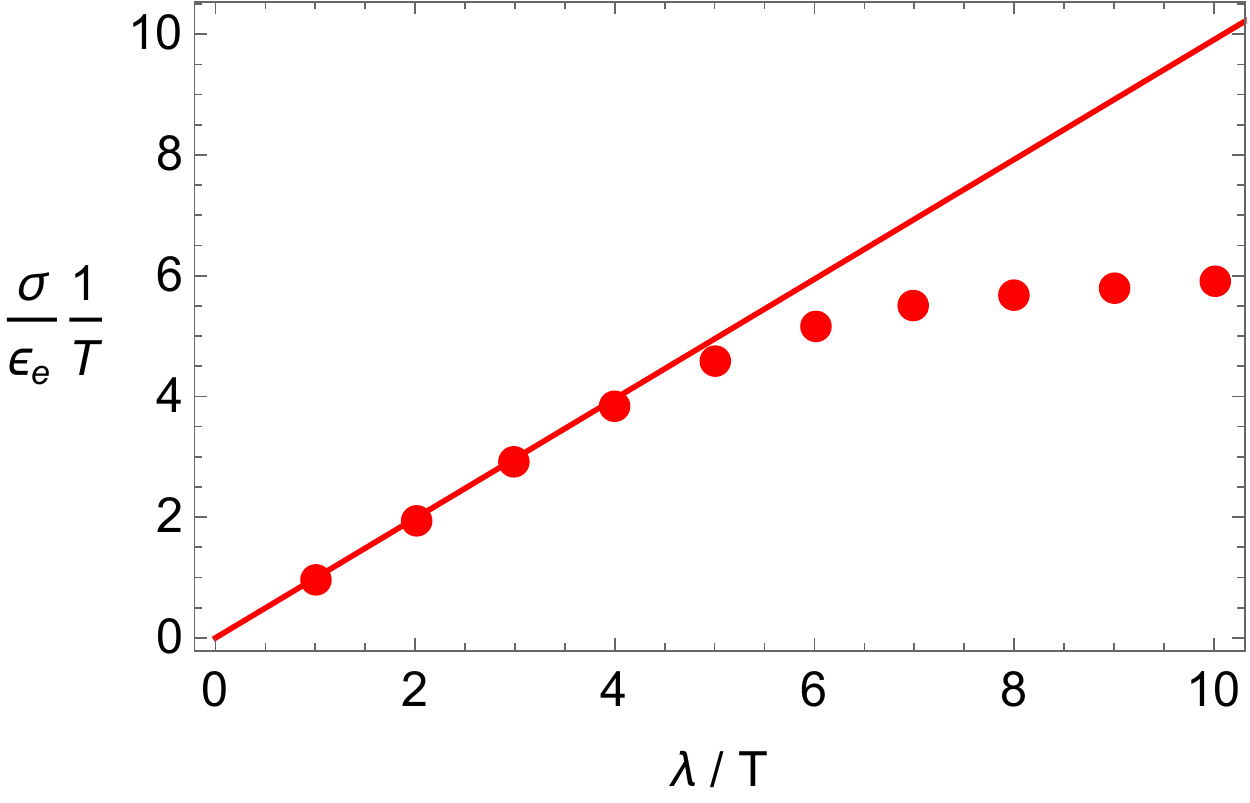}
     \includegraphics[width=7.3cm]{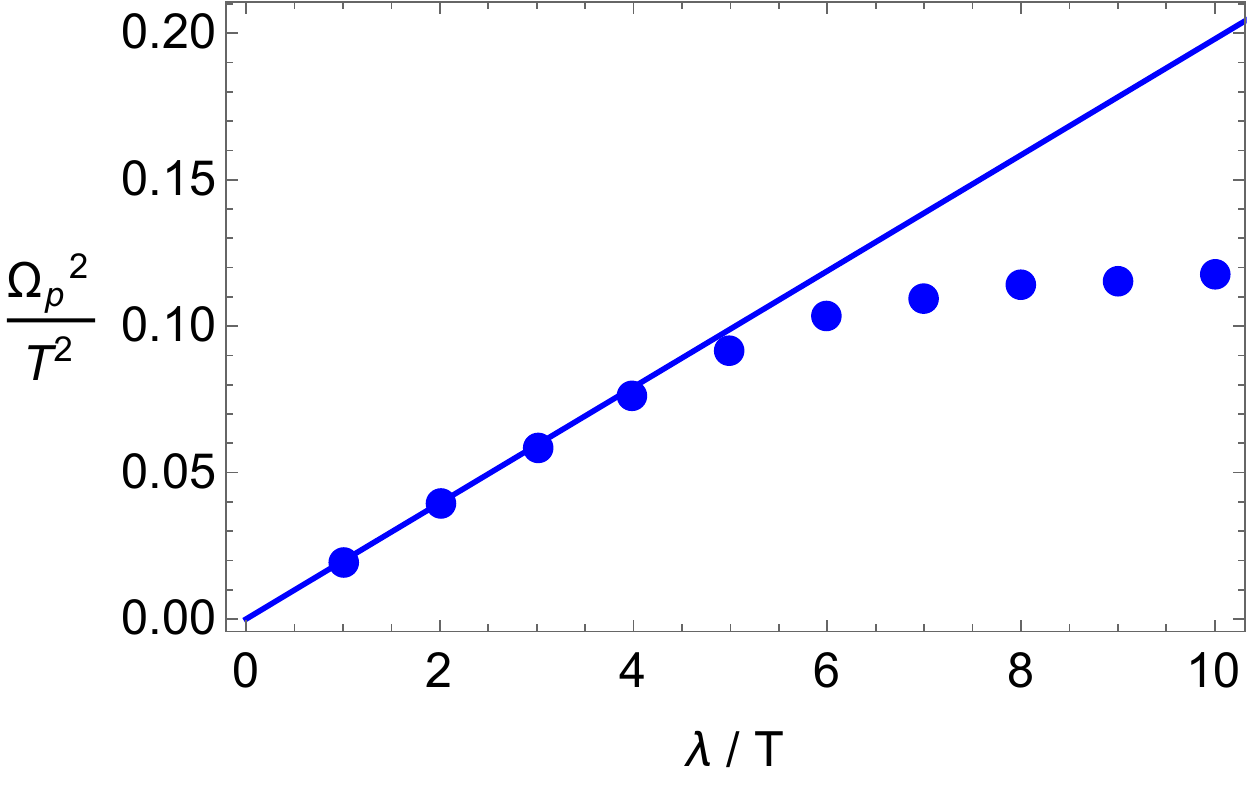}
\caption{Plasma frequency $\Omega_p$ and damping parameter ${\sigma}/{\epsilon_\text{e}}$  at finite density ($\mu/T=0.5$) and zero magnetic field.}\label{PFSLDEP}
\end{figure}

What about the opposite limit of $\lambda \rightarrow \infty$? As already hinted in Section \ref{DTMF}, our boundary conditions in this limit do not boil down to the ones usually defined as \textit{alternative quantization}. In particular, we do not fix directly the subleading term of the bulk gauge field, as for example done in \cite{Gao:2012yw}.
Before getting there, let us first ask a different question. Is the hydrodynamic framework of Section \ref{SECMHDDIS222} still reliable in the large $\lambda$ limit? From Fig. \ref{zeroden2}, one can already notice that, in the \textit{small} wave-vector regime, the gapless modes can be still well described by the hydrodynamic predictions even at large $\lambda$. In Fig. \ref{LAMDEPFIG1}, we show the speed and attenuation constant of the magnetosonic waves together with the diffusive parameters of shear and magnetic diffusion for different values of $B$ and $\lambda$. 
We do observe that the numerical data are not matching well the predictions from magnetohydrodynamics in the regime of large magnetic field and concomitant large EM coupling. Interestingly, the dynamics of shear diffusion is still perfectly described by hydrodynamics.

In addition to the hydrodynamic modes, one can further discuss the $\lambda$ dependence of the gap of the non-hydrodynamic mode. For this purpose, we focus on the behavior of Eq.\eqref{FINITEDEN2}. Fitting their dispersion relation, we obtain the value of the plasma frequency $\Omega_p^2$ and the damping parameter ${\sigma}/{\epsilon_\text{e}}$ numerically.

In Fig. \ref{PFSLDEP}, we find that at small EM coupling their values are in good agreement with our expectations from magnetohydrodynamics:
\begin{equation}
    \Omega_{p}^2\sim\lambda\,,\quad \tau_e^{-1}=\sigma/\epsilon_{\text{e}}\sim\lambda\,.
\end{equation}
Away from the small $\lambda$ limit, both the plasma frequency and the inverse relaxation time approaches a constant which is not anymore well approximated by the hydrodynamic predictions. {Two comments regarding this discrepancy are in order. First, this might imply that, for large values of $\lambda$, the non-hydrodynamic mode are already too far away from the hydrodynamic window and, not surprising, the predictions from hydrodynamics, Eq.\eqref{FINITEDEN2}, are not reliable anymore. Second, note that for the solid lines in Fig. \ref{PFSLDEP}, we set $\chi_{EE}$ to zero and assume that $\chi_{EE}$ does not depend on $\lambda$. This is probably not the case. In particular, we do not expect $\chi_{EE}$ to be generically zero in our holographic model. It would be interesting to find an independent way to calculate $\chi_{EE}$ holographically. The main difficulty is that, by switching on a background electric field, time dependence is unavoidable.}
%
%

%
\section{Alternative quantization and a bulk experiment with non-canonical kinetic term}\label{SECnew1}
In this section, we discuss the possibility of modifying the nature of the dual field theory not by using boundary conditions nor by a Hodge duality in the bulk but rather by substituting the original Maxwell term $F^2$ with a higher derivative action of the form \eqref{eq2}. Moreover, we compare the results of this \textit{experiment} with the results obtained in the $\lambda \rightarrow \infty$ limit.

\subsection{Higher derivative bulk action}
Let us consider a higher derivative bulk action as 
\begin{equation}\label{GENMODELXX11}
\begin{split}
S = \int \dd^4x \sqrt{-g} \left[ R + 6 - \frac{1}{4}\left(F^2\right)^{{N}/{2}} \right]   \,,
\end{split}
\end{equation}
where the AdS asymptotic behavior of the gauge field reads
\begin{align}\label{bcNgenb}
\begin{split}
A_{\mu}\left(r,t,\Vec{x}\right) \,\underset{r\rightarrow \infty}{\sim}\,   A_{\mu}^{(0)}\left(t,\Vec{x}\right) \,+\, A_{\mu}^{(1)}\left(t,\Vec{x}\right) \, r^{\frac{{N}-3}{{N}-1}}  \,.
\end{split}
\end{align}
Note that the higher derivative bulk action in Eq.\eqref{GENMODELXX11} reduces to the standard Maxwell action in Eq.\eqref{ACTIONH} when ${N}=2$.

\begin{figure}[]
\centering
     \includegraphics[width=10.0cm]{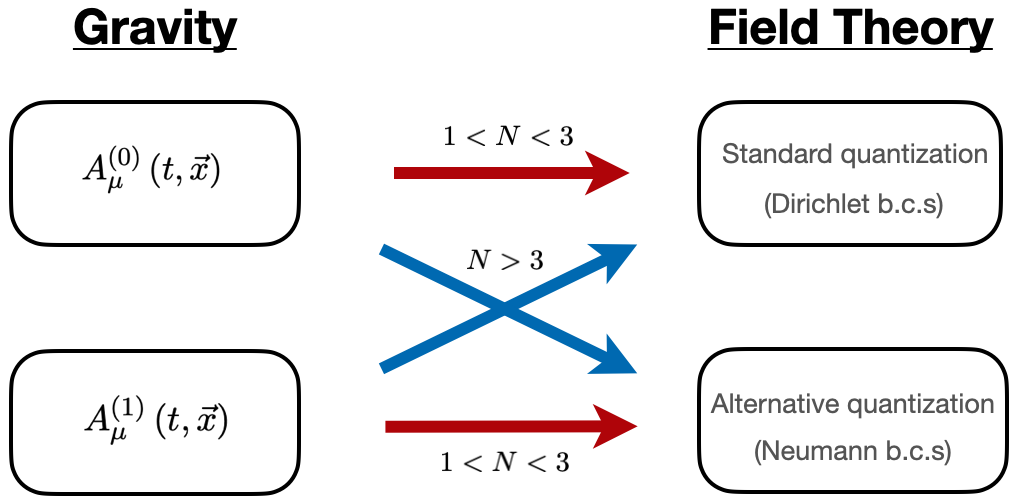}
 \caption{A representation of the role of the gauge field coefficients at the asymptotic boundary and the quantization scheme in the dual field theory for different $N$.}\label{SADN}
\end{figure}

Depending on the value of ${N}$, the coefficient $A^{(0)}_\mu\left(t,\Vec{x}\right)$ in the asymptotic expansion of the gauge field \eqref{bcNgenb} can be \textit{leading} or \textit{subleading}, i.e.,
\begin{align}\label{TABLEDFADSF11}
\begin{cases}
\quad{1} \,<\, {N} \,<\, {3}&:   \quad A_{\mu}^{(0)}\left(t,\Vec{x}\right) \,  \text{is \textit{leading}}\,, \,\,\,  \quad\,\, A_{\mu}^{(1)}\left(t,\Vec{x}\right) \,  \text{is \textit{subleading}}\,, \\
\quad\qquad\, {N} \,>\, {3}&:   \quad A_{\mu}^{(0)}\left(t,\Vec{x}\right) \,  \text{is \textit{subleading}}\,, \,\,\,  A_{\mu}^{(1)}\left(t,\Vec{x}\right) \,  \text{is \textit{leading}} \,.
\end{cases}
\end{align}
This implies that, for the bulk action \eqref{GENMODELXX11} with ${N}>3$, the coefficient $A^{(0)}_\mu\left(t,\Vec{x}\right)$ is not anymore a source for the external field $A_\mu$ but rather the expectation value of the conjugated current $J^\mu$. In other words, the standard quantization for the bulk theories with $N>3$ appears equivalent to the alternative scheme for those with ${N}<3$ (see Fig. \ref{SADN} for a graphic summary). Fixing the value of $A^{(0)}_\mu\left(t,\Vec{x}\right)$ in a theory with $N>3$ does not correspond to setting the value of a non-dynamical external gauge field (the \textit{source} in common jargon).\\
The scope of this section is to try to make sense of a bulk theory with a non-canonical kinetic term with $N>3$ and in particular to understand which is the nature of its dual field theory. Finally, we would like to ask whether this bulk theory is equivalent, and in which sense, to using the standard Maxwell kinetic term ($N=2$) but with alternative boundary conditions. For brevity, we will often refer to the $N=4$ theory with Dirichlet b.c.s. as the ``$F^4$ theory" and to the Maxwell action $N=2$ with alternative b.c.s. as the ``$F^2$ theory".

\subsection{Low-energy modes and magnetohydrodynamics}
As a concrete example, we will focus on the neutral state and compute the dispersion relation of the lowest quasi-normal modes in the  $F^2$ model (${N}=2$) with alternative quantization (Neumann b.c.s) and in a high-derivative $F^4$ model (${N}=4$) with standard quantization (Dirichlet b.c.s). Note that in both cases the boundary condition corresponds to fix the value of $A_{\mu}^{(1)}\left(t,\Vec{x}\right)$ at the boundary. Notice also that, for theories with $N\neq 2$, one cannot safely take the limit of zero charge and zero magnetic field since the bulk fluctuations would then suffer of a strong coupling problem. One can explicitly check this fact by looking at the generalized bulk Maxwell equation:
\begin{equation}\label{FLMEOM}
\begin{split}
 \nabla_{\mu} \left[ \left(F^2\right)^{\frac{N-2}{2}} F^{\mu\nu} \right] = 0 \,,
\end{split}
\end{equation}
in which the effective EM coupling in the bulk would be given by:
\begin{equation}
    \frac{1}{g_{\text{eff}}^2}\sim\left(F^2\right)^{\frac{N-2}{2}} \,, 
\end{equation}
such that $g_{\text{eff}} \longrightarrow \infty$ when $F^2\rightarrow 0$. This is totally analogous to the case in which the bulk action is a higher-derivative theory for massless scalar, see discussion in \cite{Alberte:2017oqx}.

Let us now consider the neutral state with a finite magnetic field for which the background  bulk solution is given by
\begin{equation}\label{BGFN}
\begin{split}
 f(r) &\,= r^2 - \frac{m_{0}}{r} \,+ \, \frac{2^{\frac{N}{2}-3} B^{{N}}}{2{N}-3} r^{2(1-{N})}  \,, \quad m_{0} = r_{h}^3\left( 1 +  \frac{2^{\frac{N}{2}-3} B^{{N}}}{(2{N}-3)\, r_{h}^{2{N}}} \right) \,, 
\end{split}
\end{equation}
with the corresponding thermodynamic variables
\begin{align}\label{THFN}
\begin{split}
 T \,=\, \frac{1}{4\pi} \left( 3\,r_{h} \,-\, \frac{2^{\frac{N}{2}-3}B^{{N}}}{r_{h}^{2{N}-1}}  \right) \,,  \quad s \,=\, 4\pi \, r_{h}^2  \,, \quad \epsilon \,=\,  2 m_0  \,, \quad \bar{P} \,=\, m_0 \,. 
\end{split}
\end{align}
For convenience, we have defined $\bar{P} =: \langle T_{xx} \rangle$. Note that, $\bar{P}$ is not equal to the thermodynamic pressure in presence of a magnetic field~\cite{Hartnoll:2007ih,Jensen:2011xb,Kovtun:2016lfw,Jeong:2022luo,Hartnoll:2007ip}.\footnote{This is not so uncommon. Within the axion model~\cite{Baggioli:2021xuv}, one can also find that $\bar{P}$ is different from the thermodynamic pressure defined as minus the free energy.}
We then consider the fluctuations defined in Eq.\eqref{FLUCOURSETUP} to study numerically the quasi-normal modes of the system. We impose the Neumann/Dirichlet b.c.s for the gauge fields, while we keep the Dirichlet b.c.s for the metric fluctuations. In what follows, Neumann/Dirichlet b.c.s denote the boundary conditions for the gauge fields only.

\begin{figure}[]
\centering
     \includegraphics[width=6.3cm]{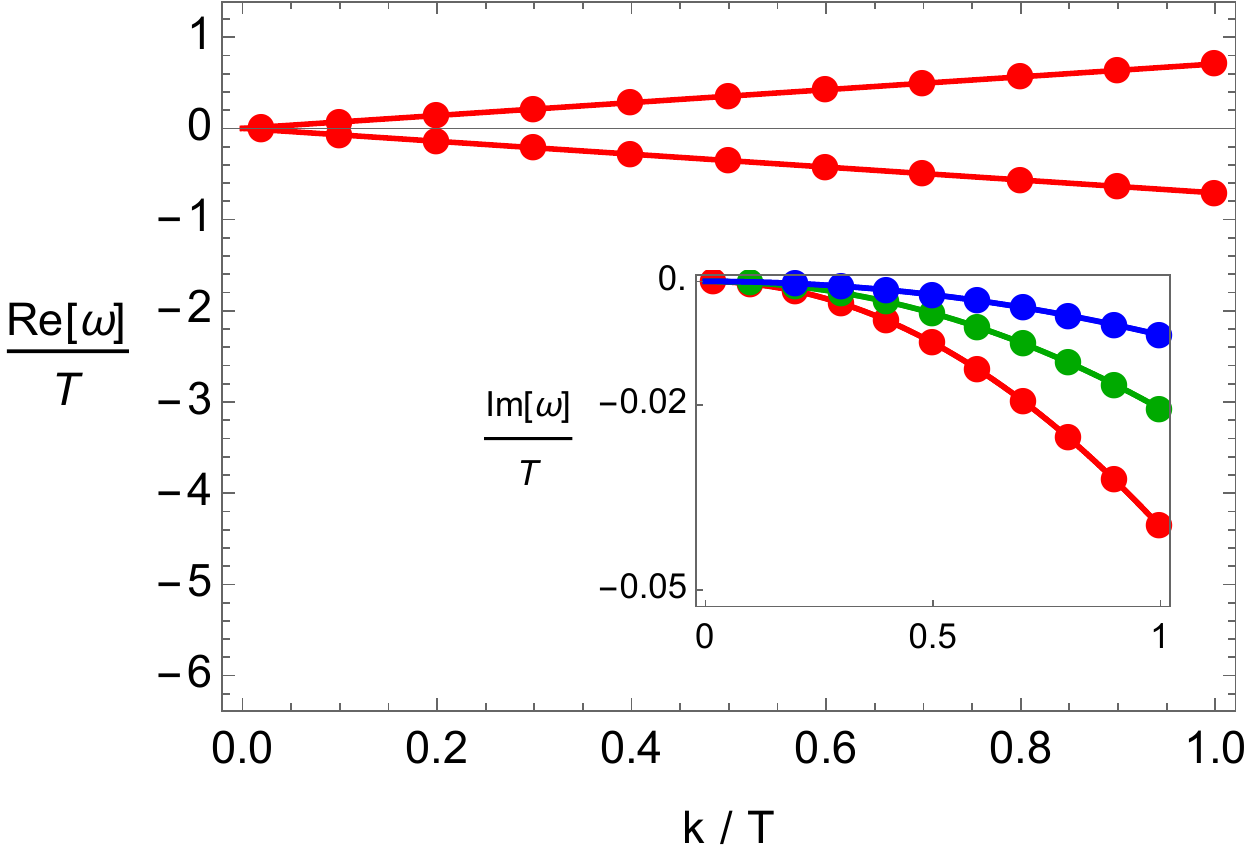}
     \includegraphics[width=6.3cm]{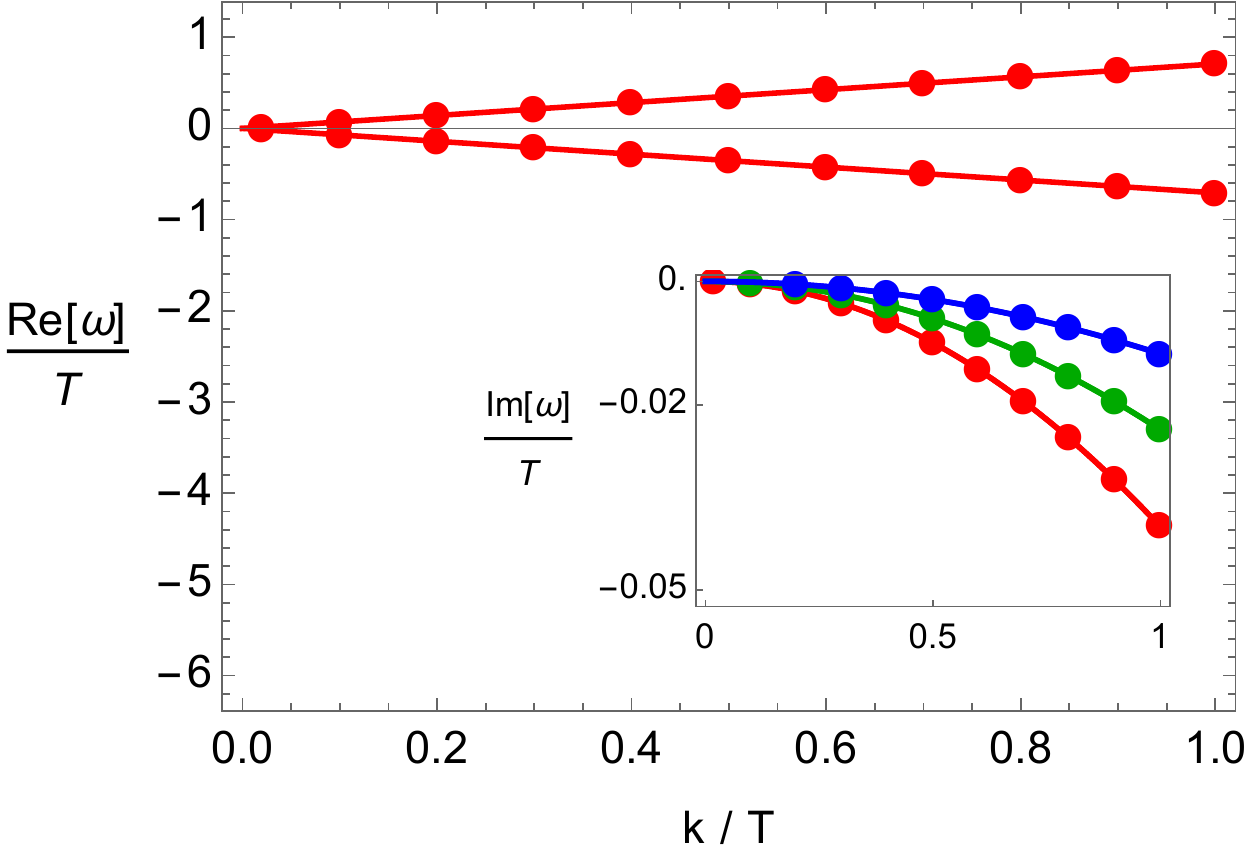}  
     \includegraphics[width=6.3cm]{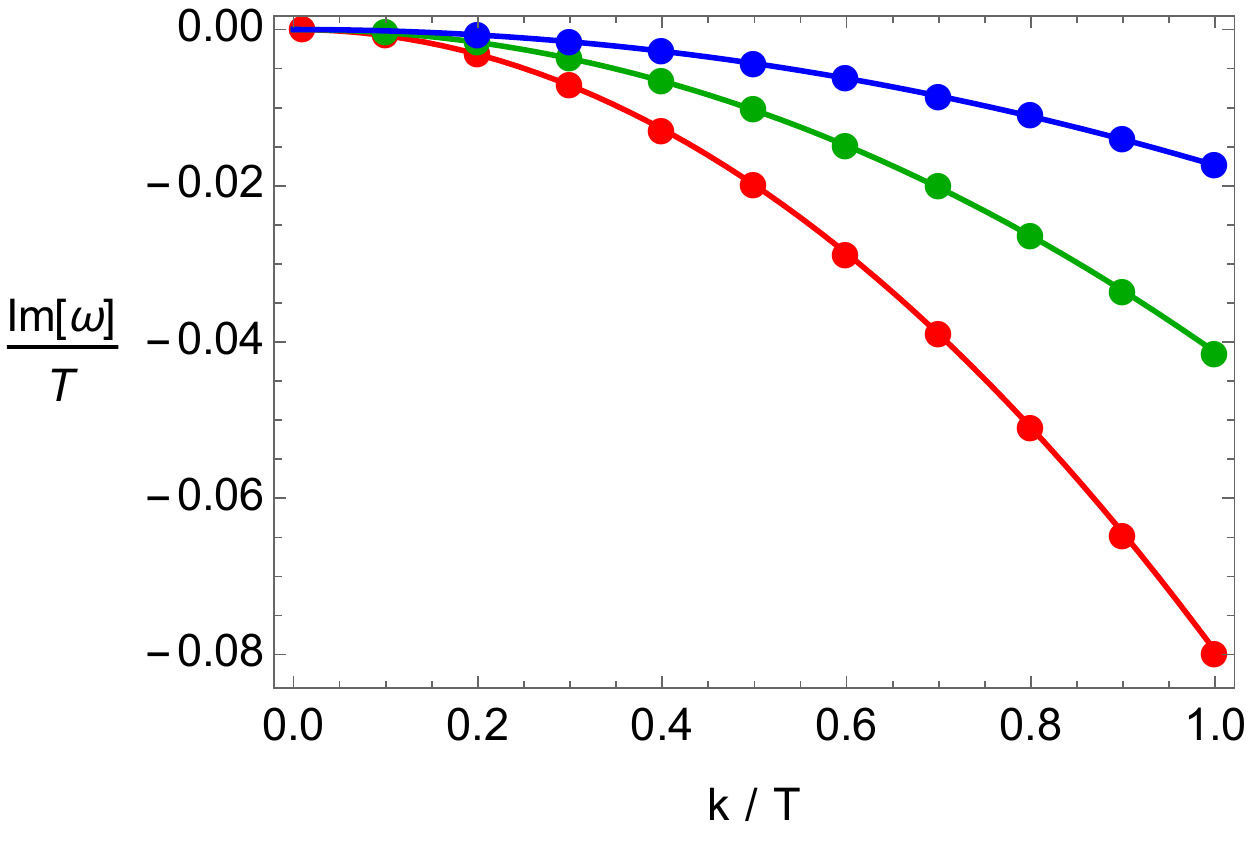} 
     \includegraphics[width=6.3cm]{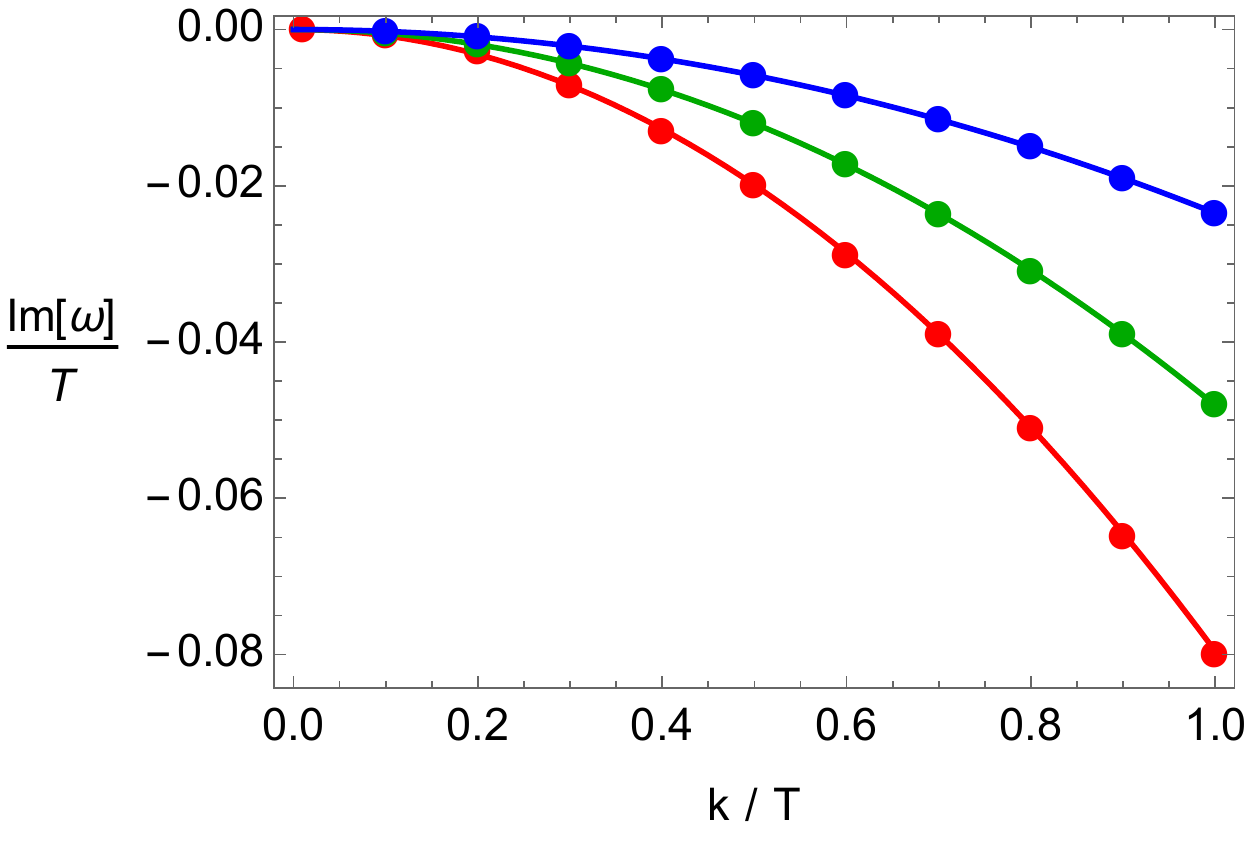} 
 \caption{\textbf{Left:} $F^2$ theory with Neumann boundary conditions; \textbf{right:} $F^4$ theory with Dirichlet boundary conditions. \textbf{Top:} Sound waves. \textbf{Bottom: } Shear diffusion. The colors correspond to $B/T^2$ = $10^{-6}$, 40, 150 (red, green, blue).}\label{F2F4THF1}
\end{figure}

We find that the quasi-normal modes for both the $F^2$ model (with Neumann b.c.s) and $F^4$ model (with Dirichlet b.c.s) exhibit four gapless modes: a pair of sound waves, a shear diffusion mode and a magnetic diffusion mode. Moreover, we empirically observe that their dispersion relations at finite magnetic field are well approximated by the following formulae: 
\begin{align}\label{HBEDIS}
\begin{split}
\omega &= \pm \sqrt{\frac{\partial \bar{P}}{\partial \epsilon}} \, k - i \frac{\eta}{2 \left( \epsilon + \bar{P} \right)} \, k^2\,, \qquad \omega = -i \frac{\eta}{\epsilon + \bar{P}} \, k^2 \,, \qquad   \omega = -i D \, k^2 \,.
\end{split}
\end{align} 
The meaning of the diffusion constant $D$ is associated with the magnetic diffusion constant $D_{\text{mag}}$ at small $B$ explained below.
%
The numerical results for the dispersion of sound waves and shear diffusion (first two set of modes in Eq.\eqref{HBEDIS}) are shown in Fig. \ref{F2F4THF1} and they are in perfect agreement with the formulae above.
\begin{figure}[]
\centering
     \includegraphics[width=6.3cm]{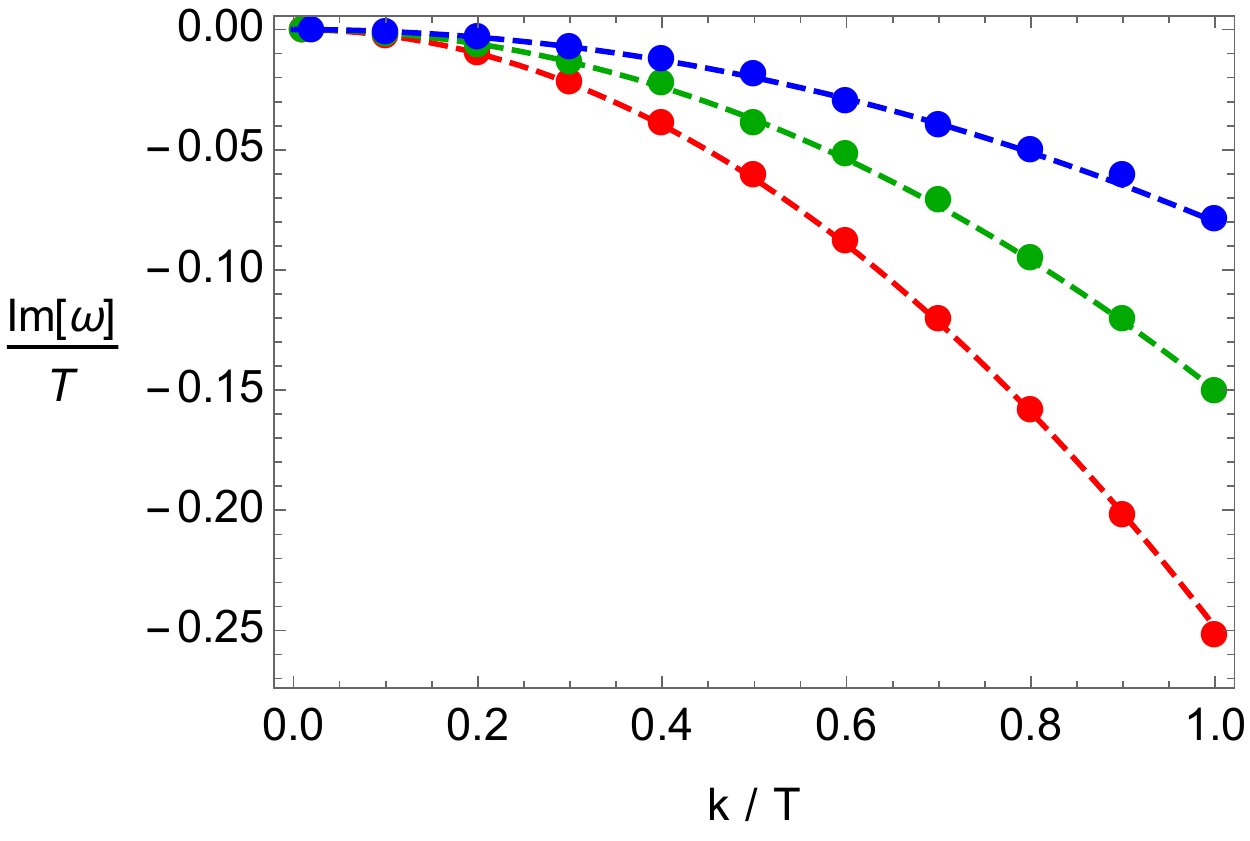} 
     \includegraphics[width=6.3cm]{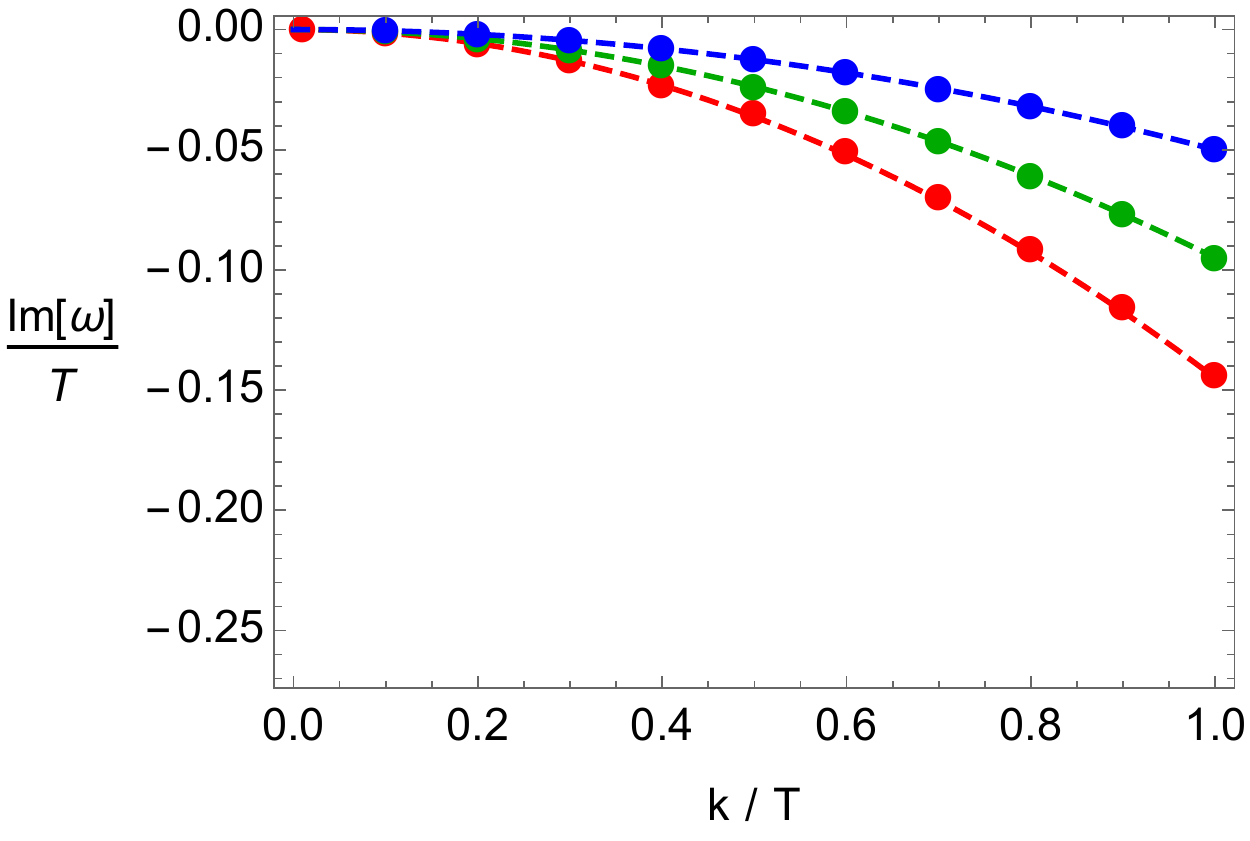} 
 \caption{\textbf{Left:} $F^2$ theory with Neumann boundary conditions; \textbf{right:} $F^4$ theory with Dirichlet boundary conditions. Diffusion mode for $B/T^2$ = $10^{-6}$, 40, 150 (red, green, blue). The dashed lines here are not analytic results from magnetohydrodynamics but simply numerical fits to the quasinormal modes data.}\label{F2F4THF4}
\end{figure}
Let us emphasize that in the limit of small magnetic field, the formulae presented in Eq.\eqref{HBEDIS} can be consistently derived from hydrodynamics by taking the $\lambda \rightarrow \infty$ limit. In particular, in that regime, we find from hydrodynamics
\begin{align}\label{NEUHYDRO1}
\begin{split}
v_{\text{ms}}^2 \,=\, \frac{1}{2} \,+\, \frac{\partial \chi_{BB}/\partial{T}}{\partial{\bar{P}}/\partial{T}} \frac{B^2}{2} \,+\, \mathcal{O}(B)^3 \,, \qquad \Gamma_{\text{ms}} \,=\,  \frac{\eta}{\epsilon+\bar{P}} + \frac{\chi_{BB}^2}{\sigma (\epsilon+\bar{P})} \frac{B^2}{2} \,+\, \mathcal{O}(B)^3 \,,
\end{split}
\end{align} 
which agree with Eq.\eqref{HBEDIS} at small $B$.
Similarly, from hydrodynamics, the diffusion constants of shear and magnetic diffusion, in the limit of $\lambda\rightarrow\infty$ and small magnetic field, are given by
\begin{align}\label{NEUHYDRO2}
\begin{split}
D_{\text{shear}} \,=\,  \frac{\eta}{\epsilon+\bar{P}}  \,, \qquad D_{\text{mag}} \,=\, - \, \frac{\chi_{BB}}{\sigma} \,-\,  \frac{\chi_{BB}^2}{\sigma (\epsilon+\bar{P})} B^2 \,+\, \mathcal{O}(B)^3 \,.
\end{split}
\end{align} 

Notice that, using Eq.\eqref{SUSEMU2} in the $\lambda=\infty$ limit, the magnetic susceptibility is given by:
\begin{equation}
    \chi_{BB}\sim -\frac{1}{\mu_m} \,,
\end{equation}
and it is negative. Then, $D_{\text{mag}}>0$.

For the magnetic diffusion mode, we find agreement between the hydrodynamic predictions and the numerical data only in the low-$B$ limit and for the $F^2$ model (see left panel in Fig. \ref{F2F4THF5}). We do not believe that the failure of the magnetohydrodynamic theory for the $F^4$ theory in the small $B$ limit is meaningful. On the contrary, that is just a signal of our failure in correctly identifying the hydrodynamic transport coefficients, such as the conductivity $\sigma$, in the $F^4$ theory. We plan to revisit these results and the transport dynamics of the $F^4$ theory in more detail in the future.\\
In summary, this analysis once more shows that the magnetohydrodynamic theory only fails in the EM sector and only in the concomitant limit of large (in this case infinite) EM coupling and large magnetic field. Moreover, it shows that modifying the bulk action with a higher-derivative kinetic term is equivalent to considering the standard kinetic term with alternative boundary conditions. This is in close analogy with the case of holographic models with broken translations \cite{Baggioli:2021xuv,Baggioli:2022pyb}.

\subsection{Further comments on magnetic diffusion}

\begin{figure}[]
\centering
     \includegraphics[width=6.3cm]{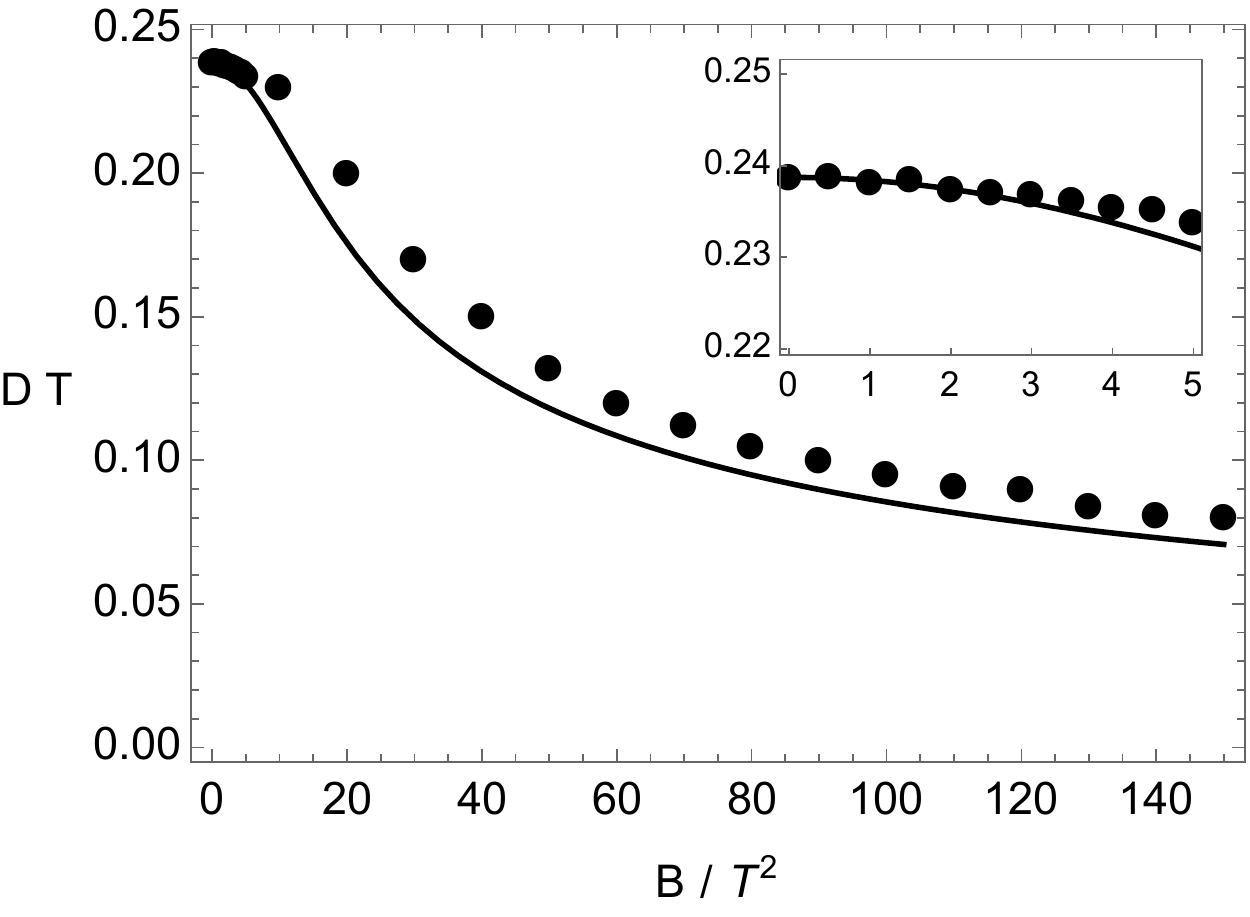} \qquad 
     \includegraphics[width=6.3cm]{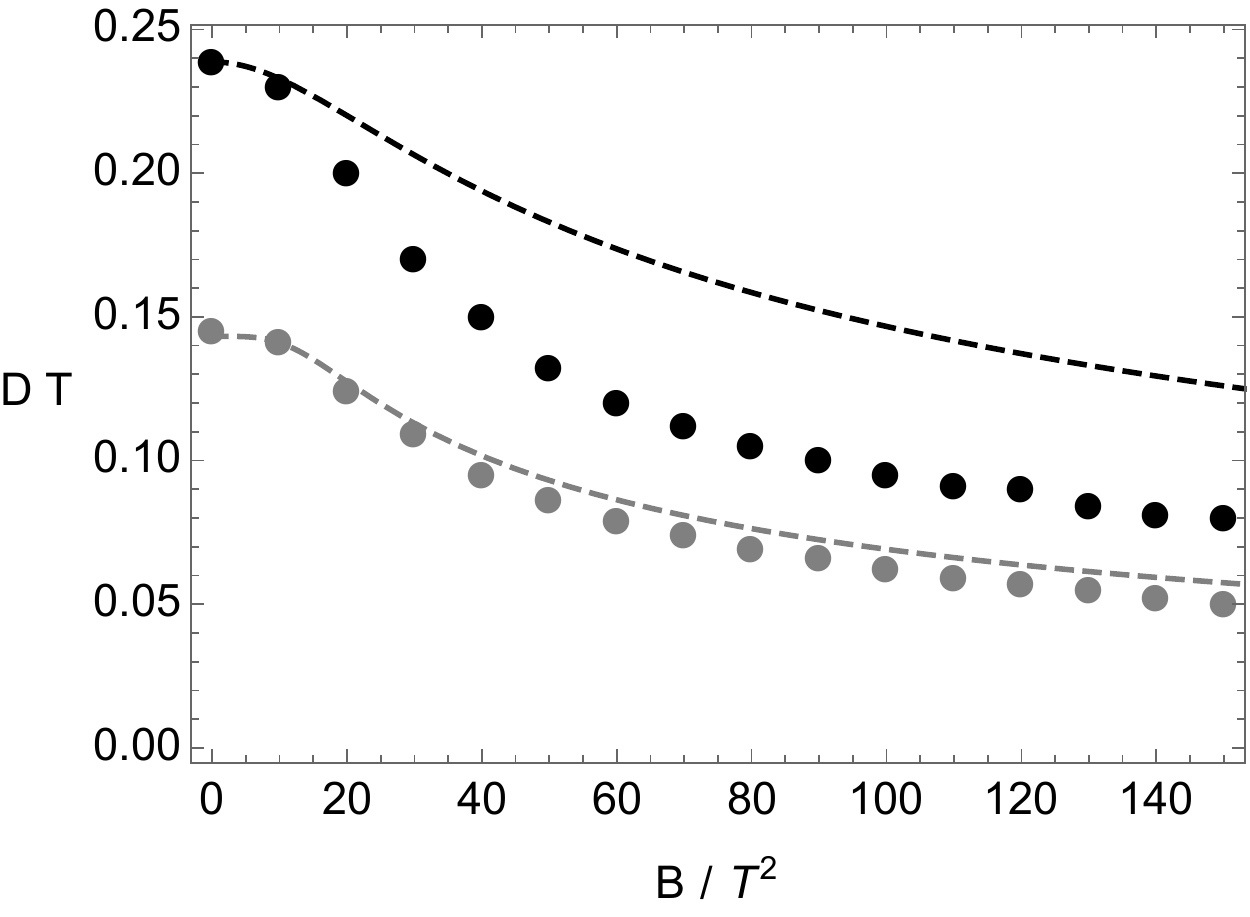} 
 \caption{The diffusion constant of the diffusive mode. \textbf{Left:} the result for the $F^2$ model with Neumann boundary conditions. The solid line is the hydrodynamic prediction which at small $B$ is given in Eq.\eqref{NEUHYDRO2}. \textbf{Right: } Black dots are for $F^2$ model and gray dots are for $F^4$ model with Dirichelt boundary conditions. The dashed lines are the analytic results from the perturbative bulk computation, Eq.\eqref{DB0RES}, valid at $B/T^2\rightarrow0$. The zoom shows the validity of the hydrodynamic formula for the $F^2$ theory with alternative b.c.s. in the limit of small magnetic field.}\label{F2F4THF5}
\end{figure}
As already mentioned, we have not been able to match the magnetic diffusion constant for the $F^4$ model using our magnetohydrodynamic theory because we could not robustly derive the transport coefficients needed. In particular, in order to achieve this, one would need to understand how to extract the electric conductivity $\sigma$ and the magnetic susceptibility $\chi_{BB}$ in the higher-derivative $F^4$ model. Nevertheless, one can gain further insights on the diffusion constant $D$ by performing a perturbative bulk analysis.
From the equation for the fluctuations, one can check that such a diffusive mode originates from the gauge fluctuation sector $\delta a_y$ which couples to the metric fluctuation sector at finite $B$.
In the limit of a vanishing $B$, one can find that the gauge sector decouples so that one can study the dynamics of the gauge fields on a fixed Schwarzschild background.

In this decoupling limit the equation of motion for $\delta a_y$ reads
\begin{align}\label{DCEOM}
\begin{split}
\delta a_y'' \,+\, \delta a_y' \left( \frac{f'}{f} \,+\, \frac{4(1-\frac{N}{2})}{r} \right) + \delta a_y \left( \frac{\omega^2}{f^2} + \frac{k^2 (1-N)}{r^2 f} \right) = 0  \,,
\end{split}
\end{align} 
where $f(r)$ is given by Eq.\eqref{BGFN} in the limit of $B=0$.
Implementing standard perturbative techniques, we are able to solve the above equation analytically and obtain the Green's function for the operator dual to $\delta a_y$. By looking at the poles structure of the latter, we can identify the presence of a mode whose dispersion is given by
\begin{align}\label{DB0RES}
\begin{split}
\omega = -i D_{B\rightarrow0} \, k^2  \,, \qquad D_{B\rightarrow0} = \frac{{N}-1}{2{N}-3}\,r_{h}^{-1} +\mathcal{O}\left(B\right)\,.
\end{split}
\end{align} 

As shown in Fig. \ref{F2F4THF5}, the analytic result above is consistent with the numerical results in the limit of $B/T^2\rightarrow0$.

\subsection{On the existence of a free boundary photon in the alternative quantization scheme}
For the case of alternative boundary conditions, a propagating photon $\omega=\pm k$ was identified in the neutral AdS$_3$ case ~\cite{Gao:2012yw}.
More precisely, the emergent photon was found by imposing the vanishing of the subleading term of the gauge field in the gauge-invariant way as
\begin{equation}\label{NBCLW33}
\begin{split}
Z_{A_1}^{(S)}  = 0 \,.
\end{split}
\end{equation}
In order to understand the photon dispersion from \eqref{NBCLW33}, it is useful to re-express \eqref{NBCLW33} as
\begin{equation}\label{NBCLW44}
\begin{split}
\omega \, \delta a_x^{(S)} + k \, \delta a_t^{(S)}  = 0 \,.
\end{split}
\end{equation}
Using the AdS boundary expansion, one can also find the following relation
\begin{equation}\label{NBCLW55}
\begin{split}
\delta a_t^{(S)} = -\frac{k}{\omega}\delta a_x^{(S)}  \,,
\end{split}
\end{equation}
which is nothing else that the conservation of the current.
Putting Eqs.\eqref{NBCLW44}-\eqref{NBCLW55} together, one immediately obtains
\begin{equation}\label{NBCLW66}
\begin{split}
\left( \omega^2 - k^2 \right) \delta a_x^{(S)} = 0 \,,
\end{split}
\end{equation}
which has a trivial solution at $\omega=\pm k$. This is exactly the propagating photon observed in~\cite{Gao:2012yw}.

However, if one considers as boundary conditions the vanishing of the external current, as we do, the situation is different. Instead of Eq.\eqref{NBCLW33}, one has to impose
\begin{equation}\label{NBCLW77}
\begin{split}
\frac{1}{\omega^2 - k^2} Z_{A_1}^{(S)}  = 0 \,.
\end{split}
\end{equation}
The frequency dependent pre-factor cancels out and the emergent photon does not appear anymore.

In order to justify our findings, let us have another look at the standard Maxwell equation for electromagnetic waves in matter, given by:
\begin{equation}
    \omega\left(\omega+i \frac{\sigma}{\epsilon_\text{e}}\right)\,=\,\frac{k^2}{\epsilon_\text{e}\,\mu_\text{m}}\,.
\end{equation}
Solving this equation together with \eqref{HOLOPER} in the $\lambda\rightarrow \infty$ limit gives two modes with dispersion:
\begin{equation}
    \omega=-i \lambda \sigma\,,\qquad \omega=-i\frac{|\chi_{BB}|}{\sigma}\,k^2\,,
\end{equation}
where the diffusive mode is the magnetic diffusion given in \eqref{NEUHYDRO2} at $B=0$.
This means that in such a limit the photon disappears. The only way that a photon could emerge in the limit of $\lambda \rightarrow \infty$ would be if $\sigma/\epsilon_e \rightarrow 0$. As proved numerically in Fig. \ref{PFSLDEP}, this is certainly not the case. 
{In summary, in the limit of infinite EM coupling, $\lambda \rightarrow \infty$, we do not find any propagating photon.}

%

%
\section{Conclusions}\label{SECMHDDIS444}
In this work, we have studied the low-energy dynamics of bottom-up holographic models at finite (free) charge density and magnetic field in presence of dynamical electromagnetism at the boundary. We have achieved the presence of a local U(1) symmetry in the boundary field theory by appropriately modifying the boundary conditions for the bulk gauge fields. We have then compared the numerical results from the holographic models with the predictions of magnetohydrodynamic theory in $2+1$ dimensions. We have found perfect agreement between the two results. This proves that modified mixed boundary conditions for the bulk gauge fields provide the correct magnetohydrodynamic phenomenology in the dual field theory.

Importantly, our work proves that the dual higher-form bulk description (e.g., \cite{Grozdanov:2017kyl}) is not necessary to obtain dynamical electromagnetism in the boundary field theory of bottom-up holographic models. This is somehow not surprising given that one could derive a precise duality between higher-form models and standard Maxwell model using different mixed boundary conditions \cite{DeWolfe:2020uzb}. 

{Interestingly, we numerically observe the breaking down of magnetohydrodynamics only in the concomitant limit of large EM coupling, $\lambda/T \gg 1$, and large magnetic field, $B/T^2\gg 1$. On the contrary, we find that, as far as the electromagnetic coupling is small, the predictions from magnetohydrodynamics at small frequencies and wave-vectors are in good agreement with the numerical data even in the limit of large magnetic field. Despite the magnetic field is treated in the ``strong field" limit, where $B\sim \mathcal{O}(1)$, this is somehow surprising. A few possible explanations arise. (I) This is a pure coincidence valid only for the model considered. (II) We have not been able to probe very large values for the magnetic field $B$ where maybe the predictions from magnetohydrodynamics would fail. (III) We are witnessing another case in favor of ``\textit{unreasonable effectiveness}" of hydrodymamics. (IV) A solid argument behind this observation exists but we have not found it yet. We find this aspect particularly interesting and we leave further investigation of this open question for the near future.}

More in general, our results provide a good playground to describe holographic models with finite electromagnetic interactions in view of possible applications to plasma physics, astrophysical objects and condensed matter systems. A set of additional open questions is left for future studies.
\begin{itemize}
    \item What is the emergent physics at infinite electromagnetic coupling in $2+1$ dimensions and how can that be described (see \cite{Gao:2012yw} for earlier discussions on this point)?
     \item {Can we find a way to compute the electric susceptibility $\chi_{EE}$ from holography and improve our understanding at large EM coupling? It would be interesting to understand whether the models and analyses of \cite{Karch:2010kt,Horowitz:2013mia,Withers:2016lft} could shed light on this point.}
    \item Can we understand better the large $B$ limit and in particular test the recent claims made in \cite{Vardhan:2022wxz} about magnetic diffusion?
    \item Can the numbers of gapless (hydrodynamic) modes be understood in terms of symmetries? In this sense, is the plasma frequency related to the explicit breaking of any symmetry? If the photon can be identified as a Goldstone mode, is the plasmon the manifestation of a pseudo-Goldstone mode?
    \item Is there an emergent photon in the strong $B$ regime? And, why?
    \item What is the correct dual field theory interpretation of the higher-derivative $F^{2N}$ bulk model? Which are the corresponding transport properties?
    \item Are the modified boundary conditions giving the correct phenomenology of superconductors once the U(1) symmetry is spontaneously broken~\cite{wipYW3} (see for example \cite{Natsuume:2022kic})?
\end{itemize}
Finally, it would be instructive to re-do our computations in a four dimensional boundary theory in which magnetohydrodynamics displays a richer, and angle dependent, spectrum with for example Alfv\'en waves and fast/slow magnetosonic waves \cite{Grozdanov:2017kyl}. Also, it would be interesting to extend our results in presence of a chiral anomaly, as done in \cite{Ammon:2020rvg} for the case of external gauge fields. We plan to report on some of these issues in the near future.

\acknowledgments

We would like to thank U.~Gran, L.~Li, A.~Amoretti, E.~Nilsson, J.~Zaanen, D.~Brattan, S.~Grozdanov, S.~Grieninger, N.~Iqbal and N.~Poovuttikul  for valuable discussions and correspondence.
This work was supported by the National Key R$\&$D Program of China (Grant No. 2018FYA0305800), Project 12035016 supported by National Natural Science Foundation of China, the Strategic Priority Research Program of Chinese Academy of Sciences, Grant No. XDB28000000, Basic Science Research Program through the National Research Foundation of Korea (NRF) funded by the Ministry of Science, ICT $\&$ Future Planning (NRF- 2021R1A2C1006791) and GIST Research Institute (GRI) grant funded by the GIST in 2022.
K.-B. Huh was also supported by Basic Science Research Program through 
the National Research Foundation of Korea (NRF) funded by 
the Ministry of Education (Grants No. NRF-2020R1I1A2054376)
M.B. acknowledges the support of the Shanghai Municipal Science and Technology Major Project (Grant No.2019SHZDZX01) and the sponsorship from the Yangyang Development Fund. M.B. would like to thank IFT Madrid, NORDITA and GIST for the warm hospitality during the completion of this work and acknowledges the support of the NORDITA distinguished visitor program and GIST visitor program.
H.-S Jeong would like to thank GIST for the warm hospitality during the completion of this work. K-Y Kim acknowledges the hospitality at APCTP where part of this work was done.

\appendix
\section{The unreasonable effectiveness of first-order magnetohydrodynamics}\label{lala}
In this Appendix, we provide a few more details about the discussion of Section \ref{nn}.

\begin{figure}[h]
\centering
     \includegraphics[width=8.3cm]{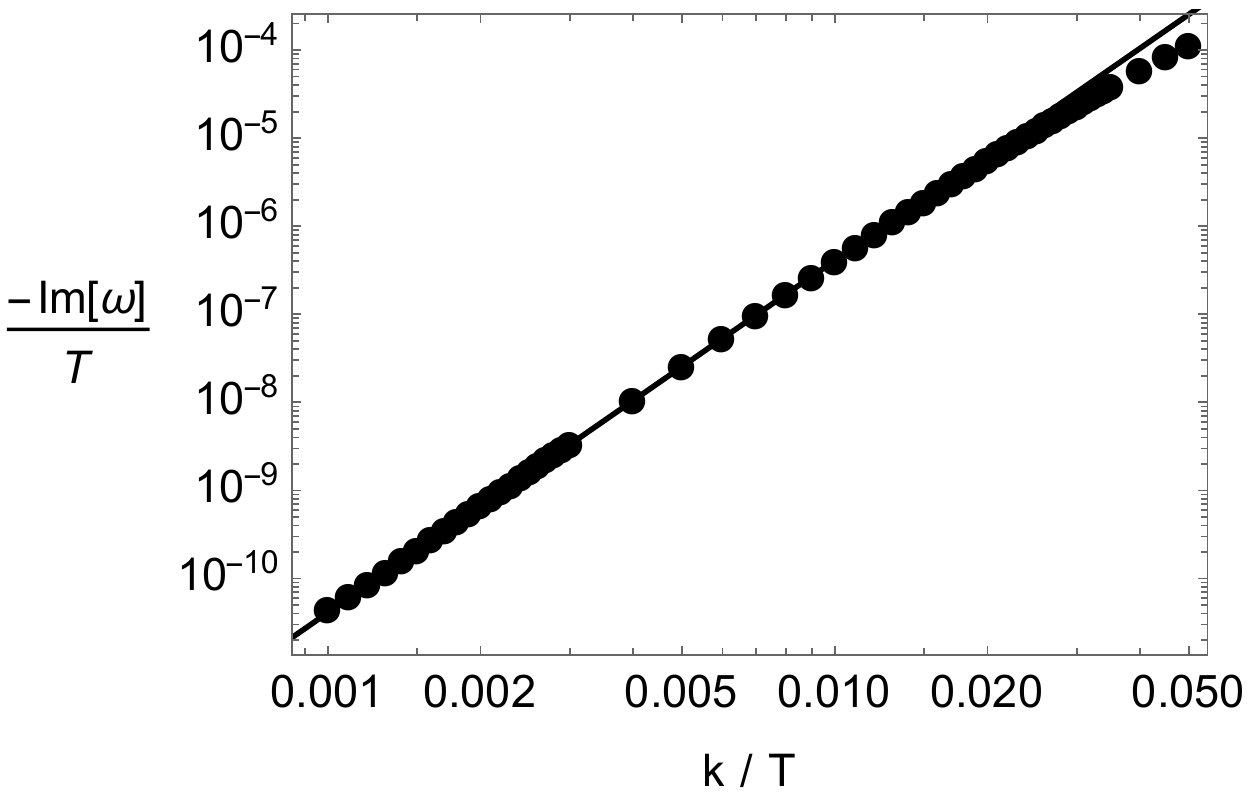}
 \caption{The dispersion relation of the subdiffusive mode at finite density $(\mu/T = 0.5)$ and $B/T^2=0$. The solid line is the prediction from first-order magnetohydrodynamics presented in the main text.}
 \label{che}
\end{figure}

First, we analyze in more detail the dispersion relation of the subdiffusive mode, $\omega=-i D_{\text{subdiff}}\,k^4$ and the validity of the predictions from first-order magnetohydrodynamics presented in the main text. As already argued in Section \ref{nn}, taking a overly pessimistic attitude, one could expect that the corrections from second-order hydrodynamics could modify the dispersion relation of the subdiffusive mode in Eq.\eqref{FINITEDEN1} at order $k^3$. Fortunately, this is not the case. In Fig. \ref{che}, we present an accurate analysis of the dispersion relation of the subdiffusive mode at low wave-vector. As evident from there, the dispersion displays a $k^4$ scaling up to low wave-vector, indicating the $k^3$ correction is not present. Additionally, the prediction from first-order hydrodynamics of the $k^4$ coefficient matches perfectly the data up to $k/T \approx 0.04$. This indicates that the dispersion relation extracted from first order hydrodynamics is reliable and, at least up to $k^4$ order, no corrections appear.

\begin{figure}[h]
\centering
     \includegraphics[width=7.3cm]{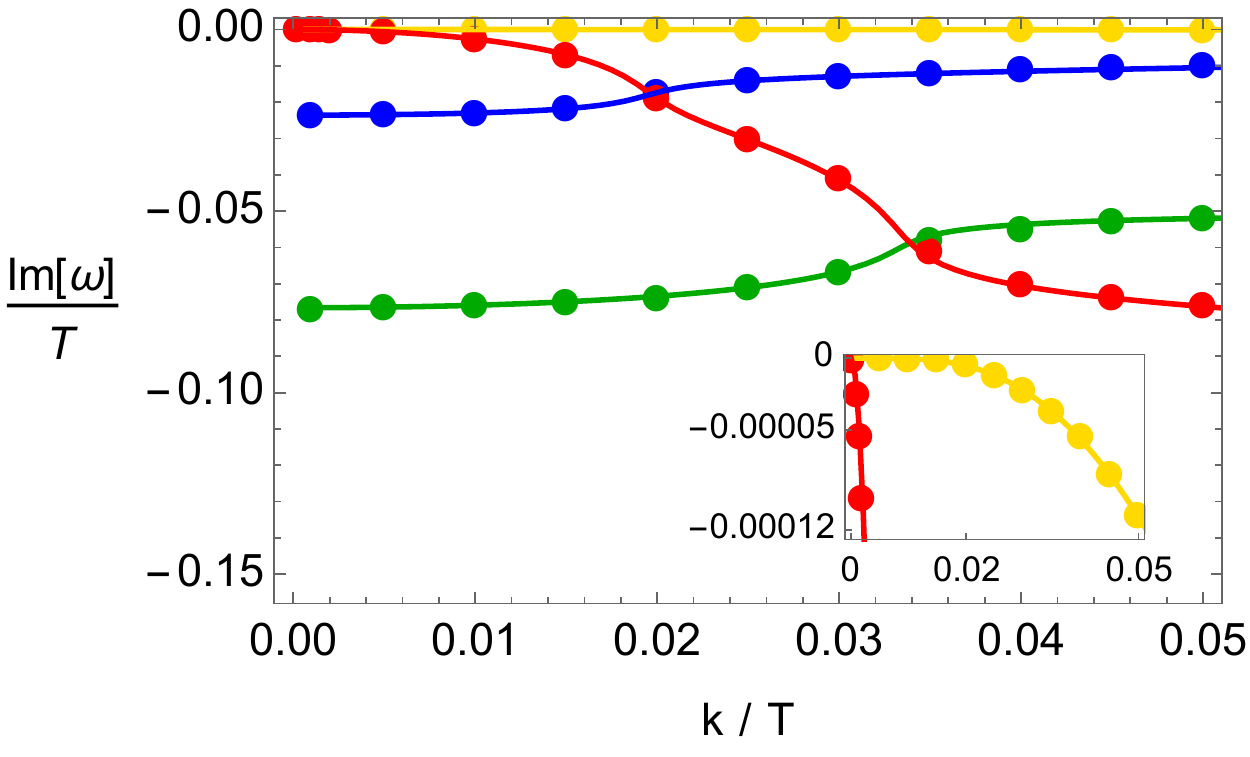}
     \includegraphics[width=7.3cm]{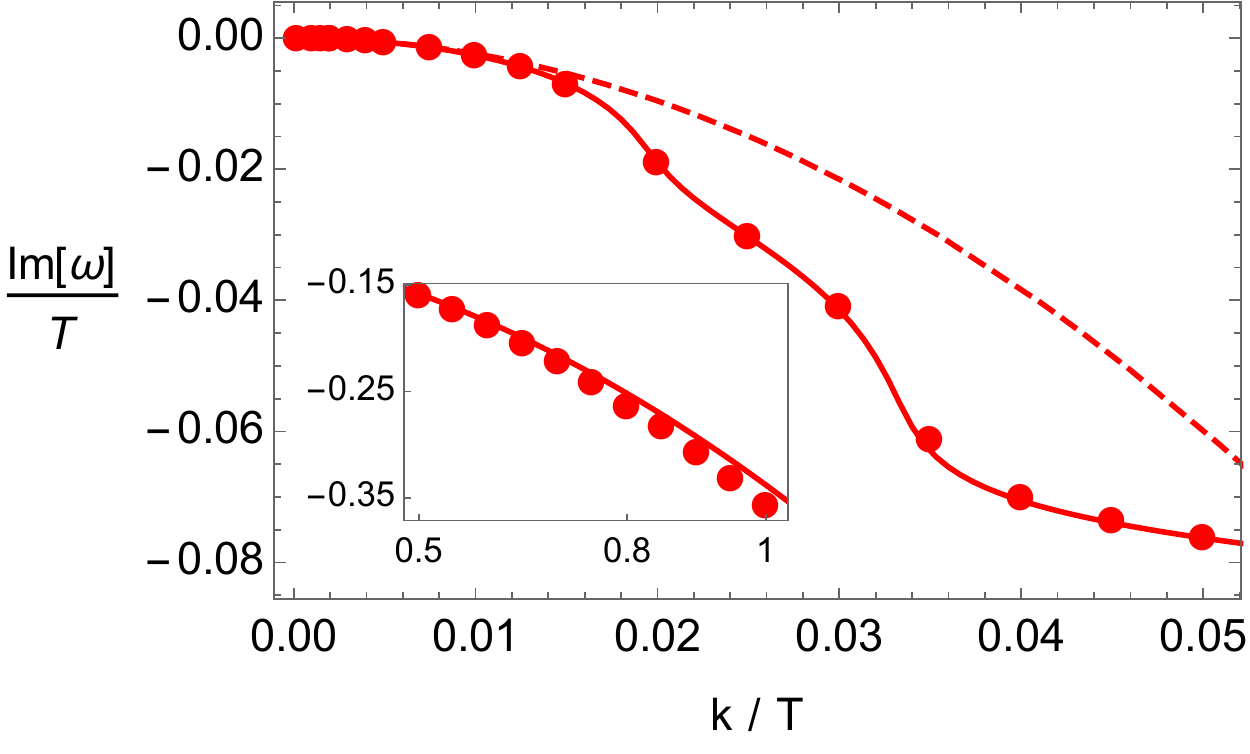}
 \caption{Dispersion relations of the lowest QNMs at finite density $(\mu/T = 0.5)$ and $B/T^2=0.5$ corresponding to Fig. \ref{finden1} in the main text. The solid lines are the hydrodynamic formulas without truncating the solution while the dashed line correspond to the controlled hydrodynamic predictions shown in the text. See Section \ref{nn} for more details.}\label{bella}
\end{figure}

Finally, we discuss the validity of the first order hydrodynamic formalism used in the main text. As explained in detail in Section \ref{nn}, apart from the subdiffusive mode in Eq.\eqref{FINITEDEN1}, in the main text we take a conservative attitude and we consider the results from the first-order formalism only to the order at which we are sure they cannot be affected by second-order corrections, $\sim k^3$. Here, we want to relax this attitude and consider the dispersion relations from first-order hydrodynamics without expanding the solutions of $\mathrm{det}(\mathcal{M}(\omega,k))=0$ at small wave-vector. Since $\mathcal{M}$ is a $6\times 6$ square matrix and every entry is at most order $k^2$, we do expect the final polynomial to be order six in frequency and order twelve in wave-vector. To perform this analysis we consider only the QNMs  shown in the bottom panel of Fig. \ref{finden1} in the main text. We show the results in Fig. \ref{bella}: 
we also provide the data for all other cases (corresponding to Fig. \ref{zeroden1}-Fig. \ref{bb} in the main text) in the GitHub repository available \href{https://github.com/sicobysico/MHD_HOLO}{here}.

First, in the left panel, we show that this unjustified relaxed attitude significantly enlarges the validity of the hydrodynamic predictions. The latter are now in perfect agreement with the numerical data up to $k/T \sim 0.05$. This has to be contrasted with the results shown in the bottom panel of Fig. \ref{finden1} in which the first-order hydrodynamic predictions fail already around $k/T \sim 0.013$. In order to make this more evident, in the right panel of Fig. \ref{bella}, we show both predictions for a single mode. In dashed red line we display the conservative predictions used in the main text while in solid red line the enlarged attitude described in this Appendix. The difference is evident. What is this suggesting us? These results are telling us that, at least for the system at hand, the higher order corrections which come from expanding the constitutive relations at higher-order are subleading for a quite large range of wave-vector. This might certainly not be the case in general, but it is nevertheless a nice and interesting observation. Let us conclude this appendix saying that, even considering the $\sim k^{12}$ solution from $\mathrm{det}(\mathcal{M}(\omega,k))=0$, the hydrodynamic predictions will not match the data at arbitrarily large values of $k$. Increasing the range of $k$ further, one would see deviations as well.

%

%
%

\clearpage
\newpage

\begin{thebibliography}{100}

\bibitem{Witten:1998qj}
E.~Witten, \emph{{Anti-de Sitter space and holography}}, {\emph{Adv. Theor.
  Math. Phys.} {\bf 2} (1998) 253--291},
  [\href{http://arxiv.org/abs/hep-th/9802150}{{\tt hep-th/9802150}}].

\bibitem{Gubser:1998bc}
S.~S. Gubser, I.~R. Klebanov and A.~M. Polyakov, \emph{{Gauge theory
  correlators from non-critical string theory}},
  \href{http://dx.doi.org/10.1016/S0370-2693(98)00377-3}{\emph{Phys. Lett.}
  {\bf B428} (1998) 105--114}, [\href{http://arxiv.org/abs/hep-th/9802109}{{\tt
  hep-th/9802109}}].

\bibitem{Hartnoll:2016apf}
S.~A. Hartnoll, A.~Lucas and S.~Sachdev, \emph{{Holographic quantum matter}},
  \href{http://arxiv.org/abs/1612.07324}{{\tt 1612.07324}}.

\bibitem{Witten:2003ya}
E.~Witten, \emph{{SL(2,Z) action on three-dimensional conformal field theories
  with Abelian symmetry}},  in \emph{{From Fields to Strings: Circumnavigating
  Theoretical Physics: A Conference in Tribute to Ian Kogan}}, pp.~1173--1200,
  7, 2003.
\newblock \href{http://arxiv.org/abs/hep-th/0307041}{{\tt hep-th/0307041}}.

\bibitem{Klebanov:1999tb}
I.~R. Klebanov and E.~Witten, \emph{{AdS / CFT correspondence and symmetry
  breaking}},
  \href{http://dx.doi.org/10.1016/S0550-3213(99)00387-9}{\emph{Nucl. Phys. B}
  {\bf 556} (1999) 89--114}, [\href{http://arxiv.org/abs/hep-th/9905104}{{\tt
  hep-th/9905104}}].

\bibitem{Leigh:2003ez}
R.~G. Leigh and A.~C. Petkou, \emph{{SL(2,Z) action on three-dimensional CFTs
  and holography}},
  \href{http://dx.doi.org/10.1088/1126-6708/2003/12/020}{\emph{JHEP} {\bf 12}
  (2003) 020}, [\href{http://arxiv.org/abs/hep-th/0309177}{{\tt
  hep-th/0309177}}].

\bibitem{Yee:2004ju}
H.-U. Yee, \emph{{A Note on AdS / CFT dual of SL(2,Z) action on 3-D conformal
  field theories with U(1) symmetry}},
  \href{http://dx.doi.org/10.1016/j.physletb.2004.05.082}{\emph{Phys. Lett. B}
  {\bf 598} (2004) 139--148}, [\href{http://arxiv.org/abs/hep-th/0402115}{{\tt
  hep-th/0402115}}].

\bibitem{Breitenlohner:1982jf}
P.~Breitenlohner and D.~Z. Freedman, \emph{{Stability in Gauged Extended
  Supergravity}},
  \href{http://dx.doi.org/10.1016/0003-4916(82)90116-6}{\emph{Annals Phys.}
  {\bf 144} (1982) 249}.

\bibitem{Marolf:2006nd}
D.~Marolf and S.~F. Ross, \emph{{Boundary Conditions and New Dualities: Vector
  Fields in AdS/CFT}},
  \href{http://dx.doi.org/10.1088/1126-6708/2006/11/085}{\emph{JHEP} {\bf 11}
  (2006) 085}, [\href{http://arxiv.org/abs/hep-th/0606113}{{\tt
  hep-th/0606113}}].

\bibitem{Cottrell:2017gkb}
W.~Cottrell, A.~Hashimoto, A.~Loveridge and D.~Pettengill, \emph{{Stability and
  boundedness in AdS/CFT with double trace deformations II: Vector Fields}},
  \href{http://arxiv.org/abs/1711.01257}{{\tt 1711.01257}}.

\bibitem{Montull:2009fe}
M.~Montull, A.~Pomarol and P.~J. Silva, \emph{{The Holographic Superconductor
  Vortex}}, \href{http://dx.doi.org/10.1103/PhysRevLett.103.091601}{\emph{Phys.
  Rev. Lett.} {\bf 103} (2009) 091601},
  [\href{http://arxiv.org/abs/0906.2396}{{\tt 0906.2396}}].

\bibitem{Maeda:2010br}
K.~Maeda, M.~Natsuume and T.~Okamura, \emph{{On two pieces of folklore in the
  AdS/CFT duality}},
  \href{http://dx.doi.org/10.1103/PhysRevD.82.046002}{\emph{Phys. Rev. D} {\bf
  82} (2010) 046002}, [\href{http://arxiv.org/abs/1005.2431}{{\tt 1005.2431}}].

\bibitem{Domenech:2010nf}
O.~Domenech, M.~Montull, A.~Pomarol, A.~Salvio and P.~J. Silva, \emph{{Emergent
  Gauge Fields in Holographic Superconductors}},
  \href{http://dx.doi.org/10.1007/JHEP08(2010)033}{\emph{JHEP} {\bf 08} (2010)
  033}, [\href{http://arxiv.org/abs/1005.1776}{{\tt 1005.1776}}].

\bibitem{Hartnoll:2008kx}
S.~A. Hartnoll, C.~P. Herzog and G.~T. Horowitz, \emph{{Holographic
  Superconductors}},
  \href{http://dx.doi.org/10.1088/1126-6708/2008/12/015}{\emph{JHEP} {\bf 0812}
  (2008) 015}, [\href{http://arxiv.org/abs/0810.1563}{{\tt 0810.1563}}].

\bibitem{Compere:2008us}
G.~Compere and D.~Marolf, \emph{{Setting the boundary free in AdS/CFT}},
  \href{http://dx.doi.org/10.1088/0264-9381/25/19/195014}{\emph{Class. Quant.
  Grav.} {\bf 25} (2008) 195014}, [\href{http://arxiv.org/abs/0805.1902}{{\tt
  0805.1902}}].

\bibitem{Ecker:2021cvz}
C.~Ecker, W.~van~der Schee, D.~Mateos and J.~Casalderrey-Solana,
  \emph{{Holographic evolution with dynamical boundary gravity}},
  \href{http://dx.doi.org/10.1007/JHEP03(2022)137}{\emph{JHEP} {\bf 03} (2022)
  137}, [\href{http://arxiv.org/abs/2109.10355}{{\tt 2109.10355}}].

\bibitem{Witten:2001ua}
E.~Witten, \emph{{Multitrace operators, boundary conditions, and AdS / CFT
  correspondence}},  \href{http://arxiv.org/abs/hep-th/0112258}{{\tt
  hep-th/0112258}}.

\bibitem{Berkooz:2002ug}
M.~Berkooz, A.~Sever and A.~Shomer, \emph{{'Double trace' deformations,
  boundary conditions and space-time singularities}},
  \href{http://dx.doi.org/10.1088/1126-6708/2002/05/034}{\emph{JHEP} {\bf 05}
  (2002) 034}, [\href{http://arxiv.org/abs/hep-th/0112264}{{\tt
  hep-th/0112264}}].

\bibitem{Hartman:2006dy}
T.~Hartman and L.~Rastelli, \emph{{Double-trace deformations, mixed boundary
  conditions and functional determinants in AdS/CFT}},
  \href{http://dx.doi.org/10.1088/1126-6708/2008/01/019}{\emph{JHEP} {\bf 01}
  (2008) 019}, [\href{http://arxiv.org/abs/hep-th/0602106}{{\tt
  hep-th/0602106}}].

\bibitem{Silva:2011zzc}
P.~J. Silva, \emph{{Dynamical gauge fields in holographic superconductors}},
  \href{http://dx.doi.org/10.1002/prop.201100016}{\emph{Fortsch. Phys.} {\bf
  59} (2011) 756--761}.

\bibitem{Albash:2009iq}
T.~Albash and C.~V. Johnson, \emph{{Vortex and Droplet Engineering in
  Holographic Superconductors}},
  \href{http://dx.doi.org/10.1103/PhysRevD.80.126009}{\emph{Phys. Rev.} {\bf
  D80} (2009) 126009}, [\href{http://arxiv.org/abs/0906.1795}{{\tt
  0906.1795}}].

\bibitem{Rozali:2012ry}
M.~Rozali, D.~Smyth and E.~Sorkin, \emph{{Holographic Higgs Phases}},
  \href{http://dx.doi.org/10.1007/JHEP08(2012)118}{\emph{JHEP} {\bf 08} (2012)
  118}, [\href{http://arxiv.org/abs/1202.5271}{{\tt 1202.5271}}].

\bibitem{Gao:2012yw}
X.~Gao, M.~Kaminski, H.-B. Zeng and H.-Q. Zhang, \emph{{Non-Equilibrium Field
  Dynamics of an Honest Holographic Superconductor}},
  \href{http://dx.doi.org/10.1007/JHEP11(2012)112}{\emph{JHEP} {\bf 11} (2012)
  112}, [\href{http://arxiv.org/abs/1204.3103}{{\tt 1204.3103}}].

\bibitem{Salvio:2012at}
A.~Salvio, \emph{{Holographic Superfluids and Superconductors in
  Dilaton-Gravity}},
  \href{http://dx.doi.org/10.1007/JHEP09(2012)134}{\emph{JHEP} {\bf 09} (2012)
  134}, [\href{http://arxiv.org/abs/1207.3800}{{\tt 1207.3800}}].

\bibitem{Salvio:2013ja}
A.~Salvio, \emph{{Superconductivity, Superfluidity and Holography}},
  \href{http://dx.doi.org/10.1088/1742-6596/442/1/012040}{\emph{J. Phys. Conf.
  Ser.} {\bf 442} (2013) 012040}, [\href{http://arxiv.org/abs/1301.0201}{{\tt
  1301.0201}}].

\bibitem{Dias:2013bwa}
O.~J.~C. Dias, G.~T. Horowitz, N.~Iqbal and J.~E. Santos, \emph{{Vortices in
  holographic superfluids and superconductors as conformal defects}},
  \href{http://dx.doi.org/10.1007/JHEP04(2014)096}{\emph{JHEP} {\bf 04} (2014)
  096}, [\href{http://arxiv.org/abs/1311.3673}{{\tt 1311.3673}}].

\bibitem{Montull:2011im}
M.~Montull, O.~Pujolas, A.~Salvio and P.~J. Silva, \emph{{Flux Periodicities
  and Quantum Hair on Holographic Superconductors}},
  \href{http://dx.doi.org/10.1103/PhysRevLett.107.181601}{\emph{Phys. Rev.
  Lett.} {\bf 107} (2011) 181601}, [\href{http://arxiv.org/abs/1105.5392}{{\tt
  1105.5392}}].

\bibitem{delCampo:2021rak}
A.~del Campo, F.~J. G\'omez-Ruiz, Z.-H. Li, C.-Y. Xia, H.-B. Zeng and H.-Q.
  Zhang, \emph{{Universal statistics of vortices in a newborn holographic
  superconductor: beyond the Kibble-Zurek mechanism}},
  \href{http://dx.doi.org/10.1007/JHEP06(2021)061}{\emph{JHEP} {\bf 06} (2021)
  061}, [\href{http://arxiv.org/abs/2101.02171}{{\tt 2101.02171}}].

\bibitem{Zeng:2019yhi}
H.-B. Zeng, C.-Y. Xia and H.-Q. Zhang, \emph{{Topological defects as relics of
  spontaneous symmetry breaking from black hole physics}},
  \href{http://dx.doi.org/10.1007/JHEP03(2021)136}{\emph{JHEP} {\bf 03} (2021)
  136}, [\href{http://arxiv.org/abs/1912.08332}{{\tt 1912.08332}}].

\bibitem{Natsuume:2022kic}
M.~Natsuume and T.~Okamura, \emph{{Holographic Meissner Effect}},
  \href{http://arxiv.org/abs/2207.07182}{{\tt 2207.07182}}.

\bibitem{Jokela:2013hta}
N.~Jokela, G.~Lifschytz and M.~Lippert, \emph{{Holographic anyonic
  superfluidity}}, \href{http://dx.doi.org/10.1007/JHEP10(2013)014}{\emph{JHEP}
  {\bf 10} (2013) 014}, [\href{http://arxiv.org/abs/1307.6336}{{\tt
  1307.6336}}].

\bibitem{Brattan:2013wya}
D.~K. Brattan and G.~Lifschytz, \emph{{Holographic plasma and anyonic fluids}},
  \href{http://dx.doi.org/10.1007/JHEP02(2014)090}{\emph{JHEP} {\bf 02} (2014)
  090}, [\href{http://arxiv.org/abs/1310.2610}{{\tt 1310.2610}}].

\bibitem{Brattan:2014moa}
D.~K. Brattan, \emph{{A strongly coupled anyon material}},
  \href{http://dx.doi.org/10.1007/JHEP11(2015)214}{\emph{JHEP} {\bf 11} (2015)
  214}, [\href{http://arxiv.org/abs/1412.1489}{{\tt 1412.1489}}].

\bibitem{Gran:2017jht}
U.~Gran, M.~Torns\"o and T.~Zingg, \emph{{Holographic Plasmons}},
  \href{http://dx.doi.org/10.1007/JHEP11(2018)176}{\emph{JHEP} {\bf 11} (2018)
  176}, [\href{http://arxiv.org/abs/1712.05672}{{\tt 1712.05672}}].

\bibitem{Gran:2018iie}
U.~Gran, M.~Torns\"o and T.~Zingg, \emph{{Plasmons in Holographic Graphene}},
  \href{http://dx.doi.org/10.21468/SciPostPhys.8.6.093}{\emph{SciPost Phys.}
  {\bf 8} (2020) 093}, [\href{http://arxiv.org/abs/1804.02284}{{\tt
  1804.02284}}].

\bibitem{Gran:2018vdn}
U.~Gran, M.~Torns\"o and T.~Zingg, \emph{{Exotic Holographic Dispersion}},
  \href{http://dx.doi.org/10.1007/JHEP02(2019)032}{\emph{JHEP} {\bf 02} (2019)
  032}, [\href{http://arxiv.org/abs/1808.05867}{{\tt 1808.05867}}].

\bibitem{Gran:2018jnt}
U.~Gran, M.~Torns\"o and T.~Zingg, \emph{{Holographic Response of Electron
  Clouds}}, \href{http://dx.doi.org/10.1007/JHEP03(2019)019}{\emph{JHEP} {\bf
  03} (2019) 019}, [\href{http://arxiv.org/abs/1810.11416}{{\tt 1810.11416}}].

\bibitem{Baggioli:2019aqf}
M.~Baggioli, U.~Gran, A.~J. Alba, M.~Torns and T.~Zingg, \emph{{Holographic
  Plasmon Relaxation with and without Broken Translations}},
  \href{http://arxiv.org/abs/1905.00804}{{\tt 1905.00804}}.

\bibitem{Gran:2019djz}
U.~Gran, N.~Jokela, D.~Musso, A.~V. Ramallo and M.~Torns\"o, \emph{{Holographic
  fundamental matter in multilayered media}},
  \href{http://dx.doi.org/10.1007/JHEP12(2019)038}{\emph{JHEP} {\bf 12} (2019)
  038}, [\href{http://arxiv.org/abs/1909.01864}{{\tt 1909.01864}}].

\bibitem{Baggioli:2019sio}
M.~Baggioli, U.~Gran and M.~Torns{\"o}, \emph{{Transverse Collective Modes in
  Interacting Holographic Plasmas}},
  \href{http://dx.doi.org/10.1007/JHEP04(2020)106}{\emph{JHEP} {\bf 04} (2020)
  106}, [\href{http://arxiv.org/abs/1912.07321}{{\tt 1912.07321}}].

\bibitem{Baggioli:2021ujk}
M.~Baggioli, U.~Gran and M.~Torns\"o, \emph{{Collective modes of polarizable
  holographic media in magnetic fields}},
  \href{http://dx.doi.org/10.1007/JHEP06(2021)014}{\emph{JHEP} {\bf 06} (2021)
  014}, [\href{http://arxiv.org/abs/2102.09969}{{\tt 2102.09969}}].

\bibitem{Romero-Bermudez:2019lzz}
A.~Romero-Berm\'udez, \emph{{Density response of holographic metallic IR fixed
  points with translational pseudo-spontaneous symmetry breaking}},
  \href{http://dx.doi.org/10.1007/JHEP07(2019)153}{\emph{JHEP} {\bf 07} (2019)
  153}, [\href{http://arxiv.org/abs/1904.06237}{{\tt 1904.06237}}].

\bibitem{pines2018theory}
D.~Pines, \emph{Theory of Quantum Liquids: Normal Fermi Liquids}.
\newblock CRC Press, 2018.

\bibitem{Mauri:2018pzq}
E.~Mauri and H.~T.~C. Stoof, \emph{{Screening of Coulomb interactions in
  Holography}}, \href{http://dx.doi.org/10.1007/JHEP04(2019)035}{\emph{JHEP}
  {\bf 04} (2019) 035}, [\href{http://arxiv.org/abs/1811.11795}{{\tt
  1811.11795}}].

\bibitem{Romero-Bermudez:2018etn}
A.~Romero-Berm\'udez, A.~Krikun, K.~Schalm and J.~Zaanen, \emph{{Anomalous
  attenuation of plasmons in strange metals and holography}},
  \href{http://dx.doi.org/10.1103/PhysRevB.99.235149}{\emph{Phys. Rev. B} {\bf
  99} (2019) 235149}, [\href{http://arxiv.org/abs/1812.03968}{{\tt
  1812.03968}}].

\bibitem{Grozdanov:2016tdf}
S.~Grozdanov, D.~M. Hofman and N.~Iqbal, \emph{{Generalized global symmetries
  and dissipative magnetohydrodynamics}},
  \href{http://dx.doi.org/10.1103/PhysRevD.95.096003}{\emph{Phys. Rev. D} {\bf
  95} (2017) 096003}, [\href{http://arxiv.org/abs/1610.07392}{{\tt
  1610.07392}}].

\bibitem{Grozdanov:2017kyl}
S.~Grozdanov and N.~Poovuttikul, \emph{{Generalised global symmetries in
  holography: magnetohydrodynamic waves in a strongly interacting plasma}},
  \href{http://dx.doi.org/10.1007/JHEP04(2019)141}{\emph{JHEP} {\bf 04} (2019)
  141}, [\href{http://arxiv.org/abs/1707.04182}{{\tt 1707.04182}}].

\bibitem{Poovuttikul:2021fdi}
N.~Poovuttikul and A.~Rajagopal, \emph{{Operator lifetime and the force-free
  electrodynamic limit of magnetised holographic plasma}},
  \href{http://dx.doi.org/10.1007/JHEP09(2021)091}{\emph{JHEP} {\bf 09} (2021)
  091}, [\href{http://arxiv.org/abs/2101.12540}{{\tt 2101.12540}}].

\bibitem{Das:2022auy}
A.~Das, R.~Gregory and N.~Iqbal, \emph{{Higher-form symmetries, anomalous
  magnetohydrodynamics, and holography}},
  \href{http://arxiv.org/abs/2205.03619}{{\tt 2205.03619}}.

\bibitem{Hernandez:2017mch}
J.~Hernandez and P.~Kovtun, \emph{{Relativistic magnetohydrodynamics}},
  \href{http://dx.doi.org/10.1007/JHEP05(2017)001}{\emph{JHEP} {\bf 05} (2017)
  001}, [\href{http://arxiv.org/abs/1703.08757}{{\tt 1703.08757}}].

\bibitem{Grozdanov:2018fic}
S.~Grozdanov, A.~Lucas and N.~Poovuttikul, \emph{{Holography and hydrodynamics
  with weakly broken symmetries}},
  \href{http://dx.doi.org/10.1103/PhysRevD.99.086012}{\emph{Phys. Rev. D} {\bf
  99} (2019) 086012}, [\href{http://arxiv.org/abs/1810.10016}{{\tt
  1810.10016}}].

\bibitem{Armas:2018ibg}
J.~Armas, J.~Gath, A.~Jain and A.~V. Pedersen, \emph{{Dissipative hydrodynamics
  with higher-form symmetry}},
  \href{http://dx.doi.org/10.1007/JHEP05(2018)192}{\emph{JHEP} {\bf 05} (2018)
  192}, [\href{http://arxiv.org/abs/1803.00991}{{\tt 1803.00991}}].

\bibitem{Benenowski:2019ule}
B.~Benenowski and N.~Poovuttikul, \emph{{Classification of magnetohydrodynamic
  transport at strong magnetic field}},
  \href{http://arxiv.org/abs/1911.05554}{{\tt 1911.05554}}.

\bibitem{Hofman:2017vwr}
D.~M. Hofman and N.~Iqbal, \emph{{Generalized global symmetries and
  holography}},
  \href{http://dx.doi.org/10.21468/SciPostPhys.4.1.005}{\emph{SciPost Phys.}
  {\bf 4} (2018) 005}, [\href{http://arxiv.org/abs/1707.08577}{{\tt
  1707.08577}}].

\bibitem{Hofman:2018lfz}
D.~M. Hofman and N.~Iqbal, \emph{{Goldstone modes and photonization for higher
  form symmetries}},
  \href{http://dx.doi.org/10.21468/SciPostPhys.6.1.006}{\emph{SciPost Phys.}
  {\bf 6} (2019) 006}, [\href{http://arxiv.org/abs/1802.09512}{{\tt
  1802.09512}}].

\bibitem{DeWolfe:2020uzb}
O.~DeWolfe and K.~Higginbotham, \emph{{Generalized symmetries and 2-groups via
  electromagnetic duality in $AdS/CFT$}},
  \href{http://dx.doi.org/10.1103/PhysRevD.103.026011}{\emph{Phys. Rev. D} {\bf
  103} (2021) 026011}, [\href{http://arxiv.org/abs/2010.06594}{{\tt
  2010.06594}}].

\bibitem{Grozdanov:2018ewh}
S.~Grozdanov and N.~Poovuttikul, \emph{{Generalized global symmetries in states
  with dynamical defects: The case of the transverse sound in field theory and
  holography}}, \href{http://dx.doi.org/10.1103/PhysRevD.97.106005}{\emph{Phys.
  Rev. D} {\bf 97} (2018) 106005}, [\href{http://arxiv.org/abs/1801.03199}{{\tt
  1801.03199}}].

\bibitem{Armas:2019sbe}
J.~Armas and A.~Jain, \emph{{Viscoelastic hydrodynamics and holography}},
  \href{http://dx.doi.org/10.1007/JHEP01(2020)126}{\emph{JHEP} {\bf 01} (2020)
  126}, [\href{http://arxiv.org/abs/1908.01175}{{\tt 1908.01175}}].

\bibitem{Hartnoll:2007ih}
S.~A. Hartnoll, P.~K. Kovtun, M.~Muller and S.~Sachdev, \emph{{Theory of the
  Nernst effect near quantum phase transitions in condensed matter, and in
  dyonic black holes}},
  \href{http://dx.doi.org/10.1103/PhysRevB.76.144502}{\emph{Phys.Rev.} {\bf
  B76} (2007) 144502}, [\href{http://arxiv.org/abs/0706.3215}{{\tt
  0706.3215}}].

\bibitem{Jensen:2011xb}
K.~Jensen, M.~Kaminski, P.~Kovtun, R.~Meyer, A.~Ritz and A.~Yarom,
  \emph{{Parity-Violating Hydrodynamics in 2+1 Dimensions}},
  \href{http://dx.doi.org/10.1007/JHEP05(2012)102}{\emph{JHEP} {\bf 05} (2012)
  102}, [\href{http://arxiv.org/abs/1112.4498}{{\tt 1112.4498}}].

\bibitem{Kovtun:2016lfw}
P.~Kovtun, \emph{{Thermodynamics of polarized relativistic matter}},
  \href{http://dx.doi.org/10.1007/JHEP07(2016)028}{\emph{JHEP} {\bf 07} (2016)
  028}, [\href{http://arxiv.org/abs/1606.01226}{{\tt 1606.01226}}].

\bibitem{Jeong:2022luo}
H.-S. Jeong, K.-Y. Kim and Y.-W. Sun, \emph{{Quasi-normal modes of dyonic black
  holes and magneto-hydrodynamics}},
  \href{http://dx.doi.org/10.1007/JHEP07(2022)065}{\emph{JHEP} {\bf 07} (2022)
  065}, [\href{http://arxiv.org/abs/2203.02642}{{\tt 2203.02642}}].

\bibitem{Hartnoll:2007ip}
S.~A. Hartnoll and C.~P. Herzog, \emph{{Ohm's Law at strong coupling: S duality
  and the cyclotron resonance}},
  \href{http://dx.doi.org/10.1103/PhysRevD.76.106012}{\emph{Phys.Rev.} {\bf
  D76} (2007) 106012}, [\href{http://arxiv.org/abs/0706.3228}{{\tt
  0706.3228}}].

\bibitem{Amoretti:2021fch}
A.~Amoretti, D.~Arean, D.~K. Brattan and N.~Magnoli, \emph{{Hydrodynamic
  magneto-transport in charge density wave states}},
  \href{http://dx.doi.org/10.1007/JHEP05(2021)027}{\emph{JHEP} {\bf 05} (2021)
  027}, [\href{http://arxiv.org/abs/2101.05343}{{\tt 2101.05343}}].

\bibitem{Amoretti:2020mkp}
A.~Amoretti, D.~K. Brattan, N.~Magnoli and M.~Scanavino, \emph{{Magneto-thermal
  transport implies an incoherent Hall conductivity}},
  \href{http://dx.doi.org/10.1007/JHEP08(2020)097}{\emph{JHEP} {\bf 08} (2020)
  097}, [\href{http://arxiv.org/abs/2005.09662}{{\tt 2005.09662}}].

\bibitem{Baggioli:2020edn}
M.~Baggioli, S.~Grieninger and L.~Li, \emph{{Magnetophonons \& type-B
  Goldstones from Hydrodynamics to Holography}},
  \href{http://arxiv.org/abs/2005.01725}{{\tt 2005.01725}}.

\bibitem{Blake:2015hxa}
M.~Blake, \emph{{Magnetotransport from the fluid/gravity correspondence}},
  \href{http://dx.doi.org/10.1007/JHEP10(2015)078}{\emph{JHEP} {\bf 10} (2015)
  078}, [\href{http://arxiv.org/abs/1507.04870}{{\tt 1507.04870}}].

\bibitem{donos2016dc}
A.~Donos, J.~P. Gauntlett, T.~Griffin and L.~Melgar, \emph{Dc conductivity of
  magnetised holographic matter}, {\emph{Journal of High Energy Physics} {\bf
  2016} (2016) 1--37}.

\bibitem{Amoretti:2014zha}
A.~Amoretti, A.~Braggio, N.~Maggiore, N.~Magnoli and D.~Musso,
  \emph{{Thermo-electric transport in gauge/gravity models with momentum
  dissipation}},  \href{http://arxiv.org/abs/1406.4134}{{\tt 1406.4134}}.

\bibitem{Amoretti:2015gna}
A.~Amoretti and D.~Musso, \emph{{Magneto-transport from momentum dissipating
  holography}}, \href{http://dx.doi.org/10.1007/JHEP09(2015)094}{\emph{JHEP}
  {\bf 09} (2015) 094}, [\href{http://arxiv.org/abs/1502.02631}{{\tt
  1502.02631}}].

\bibitem{Ammon:2020rvg}
M.~Ammon, S.~Grieninger, J.~Hernandez, M.~Kaminski, R.~Koirala, J.~Leiber
  et~al., \emph{{Chiral hydrodynamics in strong external magnetic fields}},
  \href{http://dx.doi.org/10.1007/JHEP04(2021)078}{\emph{JHEP} {\bf 04} (2021)
  078}, [\href{http://arxiv.org/abs/2012.09183}{{\tt 2012.09183}}].

\bibitem{Baggioli:2021xuv}
M.~Baggioli, K.-Y. Kim, L.~Li and W.-J. Li, \emph{{Holographic Axion Model: a
  simple gravitational tool for quantum matter}},
  \href{http://dx.doi.org/10.1007/s11433-021-1681-8}{\emph{Sci. China Phys.
  Mech. Astron.} {\bf 64} (2021) 270001},
  [\href{http://arxiv.org/abs/2101.01892}{{\tt 2101.01892}}].

\bibitem{Baggioli:2014roa}
M.~Baggioli and O.~Pujolas, \emph{{Holographic Polarons, the Metal-Insulator
  Transition and Massive Gravity}},  \href{http://arxiv.org/abs/1411.1003}{{\tt
  1411.1003}}.

\bibitem{Alberte:2017oqx}
L.~Alberte, M.~Ammon, A.~Jim\'enez-Alba, M.~Baggioli and O.~Pujol\`as,
  \emph{{Holographic Phonons}},
  \href{http://dx.doi.org/10.1103/PhysRevLett.120.171602}{\emph{Phys. Rev.
  Lett.} {\bf 120} (2018) 171602}, [\href{http://arxiv.org/abs/1711.03100}{{\tt
  1711.03100}}].

\bibitem{Ammon:2020xyv}
M.~Ammon, M.~Baggioli, S.~Gray, S.~Grieninger and A.~Jain, \emph{{On the
  Hydrodynamic Description of Holographic Viscoelastic Models}},
  \href{http://dx.doi.org/10.1016/j.physletb.2020.135691}{\emph{Phys. Lett. B}
  {\bf 808} (2020) 135691}, [\href{http://arxiv.org/abs/2001.05737}{{\tt
  2001.05737}}].

\bibitem{Baggioli:2022pyb}
M.~Baggioli and B.~Gout\'eraux, \emph{{Colloquium: Hydrodynamics and holography
  of charge density wave phases}},  \href{http://arxiv.org/abs/2203.03298}{{\tt
  2203.03298}}.

\bibitem{Hattori:2022hyo}
K.~Hattori, M.~Hongo and X.-G. Huang, \emph{{New developments in relativistic
  magnetohydrodynamics}},
  \href{http://dx.doi.org/10.3390/sym14091851}{\emph{Symmetry} {\bf 14} (2022)
  1851}, [\href{http://arxiv.org/abs/2207.12794}{{\tt 2207.12794}}].

\bibitem{andreanew}
A.~Amoretti and D.~K. Brattan, \emph{On the hydrodynamics of (2 +
  1)-dimensional strongly coupled relativistic theories in an external magnetic
  field}, \href{http://dx.doi.org/10.1142/S0217732322300105}{\emph{Modern
  Physics Letters A} {\bf 37} (2022) 2230010}.

\bibitem{griffiths2014introduction}
D.~Griffiths, \emph{Introduction to Electrodynamics}.
\newblock Pearson Education, 2014.

\bibitem{doi:10.1080/14786447608639176}
O.~Heaviside, \emph{Xix. on the extra current},
  \href{http://dx.doi.org/10.1080/14786447608639176}{\emph{The London,
  Edinburgh, and Dublin Philosophical Magazine and Journal of Science} {\bf 2}
  (1876) 135--145}.

\bibitem{Baggioli:2019jcm}
M.~Baggioli, V.~V. Brazhkin, K.~Trachenko and M.~Vasin, \emph{{Gapped momentum
  states}}, \href{http://dx.doi.org/10.1016/j.physrep.2020.04.002}{\emph{Phys.
  Rept.} {\bf 865} (2020) 1--44}, [\href{http://arxiv.org/abs/1904.01419}{{\tt
  1904.01419}}].

\bibitem{Kovtun:2012rj}
P.~Kovtun, \emph{{Lectures on hydrodynamic fluctuations in relativistic
  theories}}, \href{http://dx.doi.org/10.1088/1751-8113/45/47/473001}{\emph{J.
  Phys. A} {\bf 45} (2012) 473001}, [\href{http://arxiv.org/abs/1205.5040}{{\tt
  1205.5040}}].

\bibitem{Kim:2015wba}
K.-Y. Kim, K.~K. Kim, Y.~Seo and S.-J. Sin, \emph{{Thermoelectric
  Conductivities at Finite Magnetic Field and the Nernst Effect}},
  \href{http://dx.doi.org/10.1007/JHEP07(2015)027}{\emph{JHEP} {\bf 07} (2015)
  027}, [\href{http://arxiv.org/abs/1502.05386}{{\tt 1502.05386}}].

\bibitem{El-Showk:2011xbs}
S.~El-Showk, Y.~Nakayama and S.~Rychkov, \emph{{What Maxwell Theory in
  D\ensuremath{<}\ensuremath{>}4 teaches us about scale and conformal
  invariance}},
  \href{http://dx.doi.org/10.1016/j.nuclphysb.2011.03.008}{\emph{Nucl. Phys. B}
  {\bf 848} (2011) 578--593}, [\href{http://arxiv.org/abs/1101.5385}{{\tt
  1101.5385}}].

\bibitem{Nakayama:2013is}
Y.~Nakayama, \emph{{Scale invariance vs conformal invariance}},
  \href{http://dx.doi.org/10.1016/j.physrep.2014.12.003}{\emph{Phys. Rept.}
  {\bf 569} (2015) 1--93}, [\href{http://arxiv.org/abs/1302.0884}{{\tt
  1302.0884}}].

\bibitem{Amoretti:2019buu}
A.~Amoretti, M.~Meinero, D.~K. Brattan, F.~Caglieris, E.~Giannini, M.~Affronte
  et~al., \emph{{Hydrodynamical description for magneto-transport in the
  strange metal phase of Bi-2201}},
  \href{http://dx.doi.org/10.1103/PhysRevResearch.2.023387}{\emph{Phys. Rev.
  Res.} {\bf 2} (2020) 023387}, [\href{http://arxiv.org/abs/1909.07991}{{\tt
  1909.07991}}].

\bibitem{Kovtun:2004de}
P.~Kovtun, D.~T. Son and A.~O. Starinets, \emph{{Viscosity in strongly
  interacting quantum field theories from black hole physics}},
  \href{http://dx.doi.org/10.1103/PhysRevLett.94.111601}{\emph{Phys. Rev.
  Lett.} {\bf 94} (2005) 111601},
  [\href{http://arxiv.org/abs/hep-th/0405231}{{\tt hep-th/0405231}}].

\bibitem{Kovtun:2003wp}
P.~Kovtun, D.~T. Son and A.~O. Starinets, \emph{{Holography and hydrodynamics:
  Diffusion on stretched horizons}}, {\emph{JHEP} {\bf 0310} (2003) 064},
  [\href{http://arxiv.org/abs/hep-th/0309213}{{\tt hep-th/0309213}}].

\bibitem{Jain:2015txa}
S.~Jain, R.~Samanta and S.~P. Trivedi, \emph{{The Shear Viscosity in
  Anisotropic Phases}},
  \href{http://dx.doi.org/10.1007/JHEP10(2015)028}{\emph{JHEP} {\bf 10} (2015)
  028}, [\href{http://arxiv.org/abs/1506.01899}{{\tt 1506.01899}}].

\bibitem{Finazzo:2016mhm}
S.~I. Finazzo, R.~Critelli, R.~Rougemont and J.~Noronha, \emph{{Momentum
  transport in strongly coupled anisotropic plasmas in the presence of strong
  magnetic fields}},
  \href{http://dx.doi.org/10.1103/PhysRevD.94.054020}{\emph{Phys. Rev. D} {\bf
  94} (2016) 054020}, [\href{http://arxiv.org/abs/1605.06061}{{\tt
  1605.06061}}].

\bibitem{Rebhan:2011vd}
A.~Rebhan and D.~Steineder, \emph{{Violation of the Holographic Viscosity Bound
  in a Strongly Coupled Anisotropic Plasma}},
  \href{http://dx.doi.org/10.1103/PhysRevLett.108.021601}{\emph{Phys. Rev.
  Lett.} {\bf 108} (2012) 021601}, [\href{http://arxiv.org/abs/1110.6825}{{\tt
  1110.6825}}].

\bibitem{Giataganas:2013lga}
D.~Giataganas, \emph{{Observables in Strongly Coupled Anisotropic Theories}},
  \href{http://dx.doi.org/10.22323/1.177.0122}{\emph{PoS} {\bf Corfu2012}
  (2013) 122}, [\href{http://arxiv.org/abs/1306.1404}{{\tt 1306.1404}}].

\bibitem{Mamo:2012sy}
K.~A. Mamo, \emph{{Holographic RG flow of the shear viscosity to entropy
  density ratio in strongly coupled anisotropic plasma}},
  \href{http://dx.doi.org/10.1007/JHEP10(2012)070}{\emph{JHEP} {\bf 10} (2012)
  070}, [\href{http://arxiv.org/abs/1205.1797}{{\tt 1205.1797}}].

\bibitem{Kaminski:2009dh}
M.~Kaminski, K.~Landsteiner, J.~Mas, J.~P. Shock and J.~Tarrio,
  \emph{{Holographic Operator Mixing and Quasinormal Modes on the Brane}},
  \href{http://dx.doi.org/10.1007/JHEP02(2010)021}{\emph{JHEP} {\bf 1002}
  (2010) 021}, [\href{http://arxiv.org/abs/0911.3610}{{\tt 0911.3610}}].

\bibitem{Davison:2011uk}
R.~A. Davison and N.~K. Kaplis, \emph{{Bosonic excitations of the $AdS_4$
  Reissner-Nordstrom black hole}},
  \href{http://dx.doi.org/10.1007/JHEP12(2011)037}{\emph{JHEP} {\bf 12} (2011)
  037}, [\href{http://arxiv.org/abs/1111.0660}{{\tt 1111.0660}}].

\bibitem{Denef:2009yy}
F.~Denef, S.~A. Hartnoll and S.~Sachdev, \emph{{Quantum oscillations and black
  hole ringing}},
  \href{http://dx.doi.org/10.1103/PhysRevD.80.126016}{\emph{Phys. Rev.} {\bf
  D80} (2009) 126016}, [\href{http://arxiv.org/abs/0908.1788}{{\tt
  0908.1788}}].

\bibitem{Donos:2012yu}
A.~Donos, J.~P. Gauntlett, J.~Sonner and B.~Withers, \emph{{Competing orders in
  M-theory: superfluids, stripes and metamagnetism}},
  \href{http://dx.doi.org/10.1007/JHEP03(2013)108}{\emph{JHEP} {\bf 03} (2013)
  108}, [\href{http://arxiv.org/abs/1212.0871}{{\tt 1212.0871}}].

\bibitem{Jeong:2021zhz}
H.-S. Jeong, K.-Y. Kim and Y.-W. Sun, \emph{{Bound of diffusion constants from
  pole-skipping points: spontaneous symmetry breaking and magnetic field}},
  \href{http://dx.doi.org/10.1007/JHEP07(2021)105}{\emph{JHEP} {\bf 07} (2021)
  105}, [\href{http://arxiv.org/abs/2104.13084}{{\tt 2104.13084}}].

\bibitem{Karch:2010kt}
A.~Karch and S.~L. Sondhi, \emph{{Non-linear, Finite Frequency Quantum Critical
  Transport from AdS/CFT}},
  \href{http://dx.doi.org/10.1007/JHEP01(2011)149}{\emph{JHEP} {\bf 01} (2011)
  149}, [\href{http://arxiv.org/abs/1008.4134}{{\tt 1008.4134}}].

\bibitem{Horowitz:2013mia}
G.~T. Horowitz, N.~Iqbal and J.~E. Santos, \emph{{Simple holographic model of
  nonlinear conductivity}},
  \href{http://dx.doi.org/10.1103/PhysRevD.88.126002}{\emph{Phys. Rev. D} {\bf
  88} (2013) 126002}, [\href{http://arxiv.org/abs/1309.5088}{{\tt 1309.5088}}].

\bibitem{Withers:2016lft}
B.~Withers, \emph{{Nonlinear conductivity and the ringdown of currents in
  metallic holography}},
  \href{http://dx.doi.org/10.1007/JHEP10(2016)008}{\emph{JHEP} {\bf 10} (2016)
  008}, [\href{http://arxiv.org/abs/1606.03457}{{\tt 1606.03457}}].

\bibitem{Vardhan:2022wxz}
S.~Vardhan, S.~Grozdanov, S.~Leutheusser and H.~Liu, \emph{{A new formulation
  of strong-field magnetohydrodynamics for neutron stars}},
  \href{http://arxiv.org/abs/2207.01636}{{\tt 2207.01636}}.

\bibitem{wipYW3}
M.~Baggioli, H.-S. Jeong, K.-Y. Kim and Y.-W. Sun, \emph{Collective excitations
  in holographic superconductors}, {\emph{work in progress} (2022) to appear
  soon}.

\end{thebibliography}
\bibliographystyle{JHEP}

\providecommand{\href}[2]{#2}\begingroup\raggedright\endgroup

\end{document}